\documentclass[a4paper,twoside]{book}
\usepackage{epsfig}
\usepackage{amssymb}
\begin{document}
\begin{center}
\vspace{5cm}
\Huge
Transverse Momentum Distributions in $B$ Decays
\vspace{5cm}
\\
\Large
by Roberto Sghedoni
\\
\vspace{3cm}
\begin{figure}[h]
\center\epsfig{file=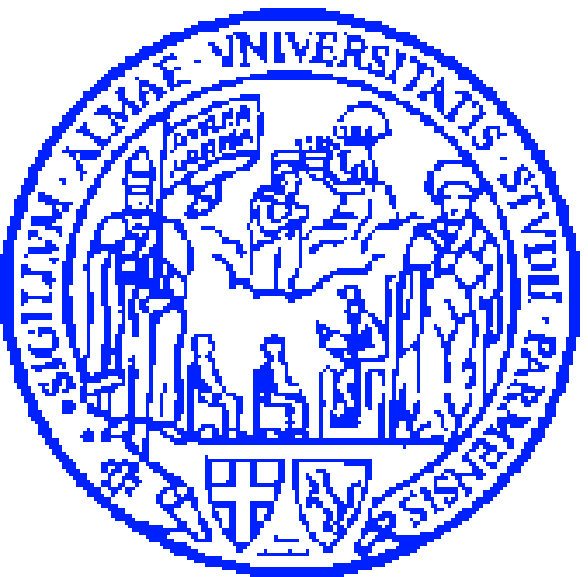, width=3cm}
\end{figure}
Ph.D. Thesis, Universit\'a di Parma
\end{center}

\chapter*{}
This thesis is the result of my collaboration with my supervisor,
Prof. L.Trentadue, and Dr.U.Aglietti. I would like to thank
Prof.Trentadue to have been my supervisor in these years and
Dr.Aglietti for his essential help: his suggestions and teachings
have been fundamental for me and his passion for perturbative QCD
has been a great source of improvement and ideas.
\\
\\
I would like to thank all the friends which have shared with me
this experience: all the members of the Stilton Social Club, Gigi,
Roberto, Fra\& Fra, Gio, Alberto and in particular my proof reader
Franco (and why not the most french among italians Federico), and
the Ph.D. students of Parma, Enzo, Carlo, Andrea, Alessia,
Alberto, Nicola and Cristian.

\tableofcontents
\chapter{Introduction}\large
One of the most recurrent philosophical questions in human thought
has surely been: "\it What are things made of?\rm".
\\
People tried to answer this question during the times and found
several and sometimes eccentric solutions to the problem: from
Greek philosophers to modern scientists many centuries of
improvements and progress passed and the original question changed
horizon time after time.
\\
Since the introduction of the galilean method of inquiry in
science the search for the laws governing the world of the
infinitely small building blocks of nature has been a great source
of discoveries. The attempt to answer our initial question has
increased dramatically the knowledge of mankind and has brought
incredible applications in many sectors, up to be applied in
everyday life of ordinary people (even if many people do not
realize this, quantum mechanics has brilliant applications in many
devices we use every day). This kind of physics can look abstract
and far away from ordinary needs, so that  in non scientific
environments the question about the sense of the construction of
very expensive accelerators  makes sometimes its appearance.
However if we look at the past we can recognize that very
important inventions were introduced on the basis of physical laws
or phenomena that physicists discovered without any practical aim
(no one worked out Quantum Mechanics to understand how to build a
transistor nor anybody studied radiations, at the beginning, to
radiograph someone else's broken leg\dots). The discoveries about
the microscopic world that we are inquiring today will likely have
a practical application in the future.
\\
Anyhow every new discovery about the world around us, at
microscopic or macroscopic or cosmic level, increases our
knowledge of the universe and the nature and this constitutes a
progress for our consciousness about what are things we can or
cannot see around us, from the largest scales to the smallest
distances.
\\
What is considered "small" is obviously a function of time: in the
19th century scientists began to introduce the concept of atom, an
indivisible particle, building block of every state of matter.
Just a few decades later, at the beginning of the 20th century
scientists as Rutherford and Thompson showed that the indivisible
atom was not so indivisible, but had its own inner components,
nuclei and electrons. Moreover this microscopic world was
described by new and unexpected laws, the Quantum Theory,
sometimes in deep contradiction with our usual way of thinking.
\\
In the 30s even nuclei began to show an inner structure, a bound
state of neutrons and protons: little by little a new branch of
physics was born, the Elementary Particle Physics. To study the
new features of elementary particles larger and larger energies
needed to be reached and this was made possible by bigger and
bigger instruments, accelerators and more and more advanced and
refined detectors were needed too. At the beginning of this
adventure, a powerful accelerator could be safely laid on a
laboratory desk, nowadays people build accelerators of several
kilometers of diameter wide to reach the astonishingly high
energies necessary to go further in the exploration of small
distances.
\\
As time went by the number and the content of discoveries
concerning the world of extremely small distances began to
separate from everyday life: before the discovery of the muons
this branch of physics was studying particles involved in the
matter we can touch every day. After that High Energy Physics
began to study a world which exist over scales of time of the
order of few microseconds at first, then nanoseconds and now even
smaller times. This is a world with does not exist in ordinary
life.
\\
A larger and larger number of short-living particles appear: this
was a great puzzle for physicists trying to understand why nature
was composed by so many different building blocks. Even if
sometimes they do not admit, theoretical physicists are attracted
by the idea that an ultimate theory of nature (both by elementary
constituents and fundamental interactions) should be \it
beautiful\rm, though this assumption is surely difficult to define
within a scientific framework. We use to think that \it beautiful
\rm in science is something simple, symmetric, which needs the
smallest number of assumptions and \it ingredients\rm. And a large
number of building blocks is not a \it beautiful \rm feature for a
theory!
\\
The modern history of High Energy Physics begins in the 50s of the
last century with Hofstadter's experiments at Stanford, which
demonstrated that protons are extended objects and measured their
form factors. In the following years a linear collider was built
at Stanford and the experiments performed at SLAC (Stanford Linear
Accelerator Center) brought to the resolution of proton components
and the introduction of quarks. The hypothesis that hadrons are
composed by point-like building blocks was introduced by Gell-Mann
and Zweig, from spectroscopical observations, and by Feynman and
Bjorken, to explain the so called Bjorken scaling. In the same
years  the Standard Model was codified by Glashow, Weinberg and
Salam. The latter is surely one of the most advanced goal reached
by scientists in the 20th century: it was able to describe
unexplained effects, to predict new phenomena. In practice all the
quantitative predictions in High Energy Physics made by the
Standard Model are correct within the limit of experimental
errors.
\\
Is this the final answer? The existence of three families of
particles (even if only the first one is involved in ordinary
matter) and the known fundamental forces? Many problems are still
open as it is discussed in the following chapter.
\\
Physicists are not fully satisfied with this answer: they would
like to include gravity in this picture, to understand where
particle masses come from, why they interact in that way, why the
families are just three,\dots.
\\
This is the reason why the main problem of this kind of physics
nowadays is how to go beyond the Standard Model: however the still
correct predictions of the theory complicates the game, there are
not data in sensational disagreement with the theoretical
prediction, even if the discovery of a mass for neutrinos could
open some door.
\\
In the last years, due to the progress made in understanding the
strong and weak interactions, a new topic arose in High Energy
Physics: the high precision study of the properties of the
$beauty$ quark, $b$ physics. This quark was discovered in 1977 and
its role inside the Standard Model is peculiar: for contingent
phenomenological reasons, in $b$ physics many parameters very
important to test the Standard Model with a high level of
precision are involved. Moreover $b$ physics can be very sensitive
to effects due to phenomena that cannot be described by the
Standard Model, the so called New Physics, a theory beyond the
Standard Model which has not yet a precise form.
\\
Finally in the decays of the $b$ quark the perturbation theory for
the strong interactions can be applied, as we have done, since the
property of asymptotic freedom, discovered by Wilczek, Politzer
and Gross in 1974, implies that $\frac{\alpha_S}{2\pi}\ll 1$,
while for the decays of lightest quarks in general this is not
true.
\\
A large program of measurements has begun in the past years, the
final goal being a determination of the parameters concerning the
physics of the $b$ quark with high precision, to constraint the
Standard Model and conclude if its predictions are fully
compatible with experiments.
\\
This program is mainly based on the construction of the so called
$b$ factories, Belle and BaBar, accelerators built specifically to
study the properties of the $b$ quark. Also CLEO gave important
results concerning this topic. In these years they were dedicated
to the measurements of decay rates of heavy hadrons, lifetimes,
branching ratios, parameters of the Cabibbo-Kobayashi-Maskawa
(CKM) matrix. In particular the experiments gave the most precise
measure of the elements involved in the unitary triangle and the
first experimental observation of CP violation in $b$ physics
(with a measurement of the associated phase of the CKM matrix).
\\
At the same time new theoretical tools, based on the Standard
Model, were introduced to face in the proper way the physics of
this quark: in particular an effective theory, called Heavy Quark
Effective Theory (HQET), found a large application in theoretical
predictions, being able both to simplify the full theory and to
reproduce its dynamics.
\\
This is the horizon where we decided to move our studies and
inquiries: the specifical motivations about the calculations and
the analysis performed are in chapter \ref{introtmd}, here let us
just introduce the topic of this thesis.
\\
In the framework of the $b$ physics we decided to study, from a
theoretical point of view, a rare process of decay of this quark,
the transition $b\rightarrow s\gamma$.  This process has been
widely studied in the past years because physicists hoped to see
clear signals of new physics: this hope was frustrated by the good
agreement of the theoretical predictions with experimental data
(even if for a few times a disagreement was observed). We focused
our calculations on the transverse momentum distribution of the
\it strange \rm quark with respect to the direction defined by
photon flight. The kinematics of the process is introduced in
chapter \ref{introtmd}.
\\
The main topic of this thesis is an application of the technique
of resummation to this particular channel and an accurate and
complete evaluation of the strong corrections. In particular a
full $O(\alpha_S)$ calculation and the structure of large
logarithms are the crucial ingredients to reach this goal.
\\
The resummation of large logarithms is a technique introduced in
QCD about 25 years ago, to improve the perturbative expansion in
regions of the phase space for processes where strong interactions
are involved, for example strong radiative corrections to
$e^+e^-\rightarrow \ {\rm 2 \ jets}$ or in semi-inclusive
distributions in Drell-Yan.
\\
This technique has been developed and applied to processes where
only light quarks were involved, since at the time accelerators
were not dedicated to the intensive study of heavy quarks (they
were just discovered). Here we will apply this technique to a
process were an heavy quark is present and this will change some
of the dynamical features.
\\
Another important point we will argue is the reliability of
perturbative QCD at relatively low energies: in fact perturbative
QCD has had brilliant confirmations in high energy processes at
the scale of the intermediate boson masses, while in this case the
energy scale is a factor 20 smaller ($M_Z \sim 20 m_b$) and the
strong coupling constant is about twice ($\alpha_S(M_Z)\approx
0.12 \sim 2\alpha_S(m_b)$).
\\
These problems will be discussed during the text, where the
explicit calculations are shown.
\\
This thesis is organized as follows.
\\
In Chapter \ref{basic} aspects of the Standard Model are briefly
described and in particular the most important features of QCD,
involved in our following discussions, are outlined.
\\
In Chapter \ref{capbsgamma} a brief review about the specifical
process $b\rightarrow s\gamma$ is presented, above all a
description of the effective hamiltonian used to disentangle the
dynamics of this transition is given.
\\
Chapter \ref{introtmd} contains the motivations for our work, the
reasons why we decided to apply the technique of resummation to
this particular process and against what background is set our
analysis. In this chapter the kinematics of the process is also
outlined.
\\
From Chapter \ref{resumtmd} on the explicit calculations are
shown: there the large logarithms appearing near the limit of the
phase space are resummed in the impact parameter space, according
to the method of resumming transverse momentum in an auxiliary
space.
\\
In following Chapter \ref{compare} we take into account the
singularities of the resummed distribution and we compare them
with the analogous ones of another distribution concerning this
process, the threshold distribution. This provides informations
about non perturbative effects.
\\
Chapter \ref{fixedorder} completes the calculation, with the
evaluation of constants or regular terms which are not resummed
and which do not show a logarithmic enhancement.
\\
Chapter \ref{masseffects} contains preliminary results about the
effects of the introduction of a mass in the final state, for
example for the \it strange \rm quark: this is a necessary step to
deal with with another important transition of the $b$ quark,
namely $b\rightarrow c$.
\\
Finally Chapter \ref{conclusionith} a brief summary of the
contents of thesis is presented with an outlook of future
improvements.
\\
Appendices, which deepen some topic met during the text, are given
at the end.
\\
\\
Some of the results obtained during the work for my thesis have
been already presented in \cite{noi1,noi2}.

\newpage

\part{Physics of the $b$ Quark and the Standard Model}

\large
\chapter{Standard Model}\label{basic}

\section{The Standard Model of Elementary Particles}
The Standard Model is nowadays the framework describing
interactions among elementary particles in processes occurring in
a range of energies up to about 1 TeV.
\\
It is a non abelian and renormalizable gauge theory based on the
symmetry group
\begin{equation}\label{gruppo}
{\cal{G}}_{SM}=SU(3)_C \otimes SU(2)_L \otimes U(1)_Y.
\end{equation}
The sector corresponding to the group $SU(3)_C$ describes strong
interactions through Quantum Chromodynamics (QCD), which will be
discussed in this chapter from section (\ref{qcd}) on, while the
sector corresponding to the group $SU(2)_L \otimes U(1)_Y$
describes the unification of weak and electromagnetic interactions
through the Glashow-\-Weinberg-Salam model
\cite{glashow,weinberg}, briefly described in section (\ref{gws}).
\\
The original symmetry (\ref{gruppo}) is broken by  a non vanishing
vacuum expectation value in the theory for a scalar field, which
gives mass to the gauge bosons of the electroweak sector and to
the fermions.
\\
The main details of theory are discussed in the next section, here
we will just make some observations about the model:
\begin{itemize}
\item
up to now experiments performed in the laboratories all over the
world confirm the predictions of the Standard Model: this
statement could seem too optimistic and strong, but, as a matter
of fact, there are not observations or calculations which are
within the errors violated \cite{pdg};
\item
the recent results obtained in agreement with hypothesis of
neutrinos oscillations seem to be a highly non trivial check of
the Standard Model: we are not currently able to state wether this
phenomenon can be brought back to the Standard Model or is a
signal of new physics, leading to a theory \it beyond\rm. In
principle a mass for the neutrinos could be introduced in the
Standard Model, as discussed later. Other searches of new physics
(studying rare decays of the $beauty$ quark, measuring CKM matrix
elements and so on) have not given any result.  Up to now no
signals of supersymmetry, extra dimensions, technicolor have been
found;
\item
the Higgs boson has not yet been detected: one of the most
important goals of future accelerators is its discovery. If the
Higgs boson does exist it will be possibly detected at LHC: the
most disappointing eventuality would be the discovery of a single
Higgs particle as predicted by the Standard Model, since this will
not give any information about extensions of the Standard Model;
\item
though its success the Standard Model is not believed to be a
fundamental theory for several (and good) reasons: the most
relevant one is that it does not describe gravitational
interactions. At present a quantum description of gravity is
lacking;
\item
moreover it depends on many free parameters (34) which are fixed
by experiments and cannot be calculated from the theory: a final
theory should be able to explain why the mass of particles have
their values, why coupling behave that way and so on and the
answer to this question is still lacking;
\item
the interactions described in the Standard Model are not really
unified, because they depends on different and not related
couplings. Maybe they are unified at energies much larger than the
ones reachable by present machines. However without a
supersymmetric model the couplings of the interactions in the
Standard Model do not unify;
\item
furthermore the Standard Model does not explain correctly the
asymmetry between matter and anti-matter. Inside the theory the CP
violation (in the last years confirmed by Belle and BaBar) could
introduce such an asymmetry, but the quantitative prevision is
largely incorrect;
\item
other issues (such as the \it hierarchy problem\rm) suggest that
the theory should be valid up to a new physics scale $\Lambda$,
where supersymmetry should make its appearance. This is the hope
of many high energy physicists: in my opinion, lacking strong
experimental evidences that violate the theory, it is very
complicated to put forward the correct guess.
\end{itemize}
By the light of these observations we can conclude that the
Standard Model is an excellent parametrization of particle physics
and has been able to explain high energy processes studied in
experiments performed up to now.
\\
The theoretical issues, as the ones mentioned above, suggest that
this is not the final theory of particle physics and fundamental
interactions: a theory beyond the Standard Model is the Holy Graal
of modern particle physicists, but its elaboration collide with
the lack of any signal of new physics.
\\
In this chapter the main features of the Standard Model will be
recalled, with a particular emphasis to its strong sector.

\section{Glashow-Weinberg-Salam Model for Electroweak Interactions}\label{gws}
In this section the main properties of the Glashow-Weinberg-Salam
model \cite{glashow,weinberg}, the theory of electroweak
interactions, are briefly recalled.
\\
This theory was put forward in the 60s, as a description of the
unification between weak and electromagnetic interactions: several
attempts were tried to introduce a theory of massive bosons for
weak interactions, but this kind of models turn out to be non
renormalizable. Moreover physicists realized that also the photon
had to be included in this description, to make the theory
predictive\footnote{The renormalizability of the theory was shown
by 't-Hooft and Veltman in \cite{thooft}}.
\\
Experimental observations suggested that weak interactions does
not conserve parity \cite{parity} and in particular that only left
handed fermionic fields interacts weakly: this drove to build a
lagrangian with massless fermionic fields, with independent
left-handed and right-handed components:
\begin{eqnarray}
\Psi_L=\frac{1-\gamma_5}{2}\Psi,\nonumber\\
\Psi_R=\frac{1+\gamma_5}{2}\Psi.
\end{eqnarray}
The model is based on the hypothesis that fermionic left-handed
fields are arranged in isospin doublets, while fermionic
right-handed fields are singlets.
\\
Fermionic elementary fields, leptons and quarks, are rearranged
into three families, each containing a lepton doublet and a quark
doublet of left-handed fields:
\begin{equation}
\begin{array}{ccc}
\left(\begin{array}{c}\nu_e\\e\end{array}\right)&
\left(\begin{array}{c} \nu_\mu\\ \mu\end{array}\right)&
\left(\begin{array}{c} \nu_\tau \\ \tau\end{array}\right)\\ \\
\left(\begin{array}{c} u \\ d\end{array}\right) &
\left(\begin{array}{c} c \\ s\end{array}\right) &
\left(\begin{array}{c} t \\ b\end{array}\right).
\end{array}
\end{equation}
Right-handed fields are singlet of isospin:
\begin{equation}
e_R, \ \ \ \mu_R,\ \ \ u_R,\ \ \ d_R, \ \ \ \dots
\end{equation}
In order to consider also electromagnetic interactions in the
model a new quantum number, the hypercharge $Y$, was introduced.
The hypercharge of a particle is defined as twice the difference
of its electric charge and the third component of its isospin:
\begin{equation}
Y=2(Q-T_3).
\end{equation}
The gauge group of symmetry of the model is:
\begin{equation}
{\cal{G}}_{GWS}=SU(2)_L\otimes U(1)_Y.
\end{equation}
Such a group has 4 generators, corresponding to the gauge bosons
of the theory, $A^a_\mu(x)$, with $a=1,2,3$, and $B_\mu(x)$: in
order to get the invariance under the action of the group, a
covariant derivative is introduced as
\begin{equation}\label{derivative}
D^\mu=\partial^\mu-igT^a\cdot A^\mu-ig^\prime B^\mu.
\end{equation}
The kinematical term for the gauge bosons is introduced through
the strength tensor:
\begin{equation}
{\cal{L}}_{kin}=-\frac{1}{4}G_{\mu\nu}^a
G^{a\mu\nu}-\frac{1}{4}F_{\mu\nu}F^{\mu\nu},
\end{equation}
defined as
\begin{eqnarray}
G^a_{\mu\nu}&=&\partial_\mu A^a_\nu-\partial_\nu
A^a_\mu-ig\left[A^a_\mu,A^a_\nu\right],\nonumber\\
F_{\mu\nu}&=&\partial_\mu B_\nu- \partial_\nu B_\mu.
\end{eqnarray}
The kinematical part of the lagrangian contains no mass terms for
the gauge bosons: the experimental evidence that weak interactions
has a small range of action , about $10^{-17}m$, suggest that
these interactions are mediated by a massive boson, whose mass is
about $10^2 {\rm GeV}$. As it was demonstrated, a mass term
introduced by hand would lead to a non renormalizable theory. The
mass is given to the gauge bosons (and to the fermions) by the
mechanism of spontaneous symmetry breaking, through the
introduction of a scalar field $\Phi$ which acquire a non zero
vacuum expectation value, $<\Phi>_0\not=0$.
\\
One introduces the scalar doublet \cite{higgs}
\begin{equation}
\phi(x)=\left(\begin{array}{r@{\quad}l}\phi^+ \\ \phi_0
\end{array} \right),
\end{equation}
and in the lagrangian the term
\begin{equation}
{\cal{L}}_{Higgs}=|D_\mu\phi|^2+\frac{1}{2}\mu^2\phi^\ast\phi-\lambda
(\phi^\ast\phi)^2.
\end{equation}
In this way the non vanishing vacuum expectation value reads
\begin{equation}
<\phi>_0=\frac{v}{\sqrt{2}}=\frac{1}{\sqrt{2}}\sqrt{\frac{\mu^2}{\lambda}}\not=0
\end{equation}
and, by using the gauge symmetry of the theory it can be
rearranged as
\begin{equation}
\phi(x)=\frac{1}{\sqrt{2}}\left(\begin{array}{c} 0 \\
v+\eta(x)\end{array} \right).
\end{equation}
In this way the scalar field can be expanded around its minimum
$<\phi>=v/\sqrt{2}$, introducing the scalar field $\eta(x)$ which
represents the Higgs boson.
\\
The covariant derivative in ${\cal{L}}_{Higgs}$ produces terms of
interactions between the gauge bosons and the scalar doublet: once
the symmetry is broken, by giving a non vanishing vacuum
expectation value to $\phi$, mass terms for the gauge bosons
arise:
\begin{equation}\label{massboson}
\frac{v^2}{8}(gA_3^\mu-g^\prime B^\mu)^2+\frac{v^2}{4}g^2
(A_1+iA_2)^\mu (A_1-iA_2)_\mu.
\end{equation}
beside terms of interactions between the gauge bosons and the
Higgs boson, $\eta(x)$.
\\
The terms in (\ref{massboson}) can be diagonalized in order to
interpret them as mass terms in the lagrangian for the gauge
bosons: this can be accomplished defining
\begin{eqnarray}
W^\pm&=&\frac{1}{\sqrt{2}}(A_1\pm i A_2)\nonumber\\
Z&=& -B\sin\theta_W+A_3\cos\theta_W\nonumber\\
\gamma&=&B\cos\theta_W+A_3\sin\theta_W,
\end{eqnarray}
where
\begin{equation}
\sin\theta_W=\frac{g}{\sqrt{g^2+g^{\prime 2}}}.
\end{equation}
This implies that bosons $W^\pm$ and $Z^0$ acquire a mass, while
the field $\gamma$, which can be interpreted as the photon,
remains massless:
\begin{equation}
m_\gamma=0, \ \ \ m_W=\frac{gv}{2}, \ \ \
m_Z=\frac{\sqrt{g^2+g^{\prime 2}}}{2}.
\end{equation}
The electric charge can be identified with the following
combination of parameters of the theory
\begin{equation}
e=g\sin\theta_W.
\end{equation}
Moreover, from a comparison with the Fermi theory of weak
interactions, one can state that
\begin{equation}
\frac{g^2}{8m^2_W}=\frac{G_F}{\sqrt{2}}.
\end{equation}
At the end the original symmetry is broken into
\begin{equation}
SU(2)_L\otimes U(1)_Y\rightarrow U(1)_{em}.
\end{equation}
The interaction of the fermionic fields (leptons and quarks) is
introduced as usual in the construction of a gauge theory,
considering the Dirac lagrangian and imposing the invariance under
the symmetry group: this implies that the ordinary derivative have
to be replaced by the covariant derivative (\ref{derivative}).
\\
The mass of fermions cannot be introduced by hand, because the
massive term in the Dirac lagrangian would violate the symmetry
$SU(2)_L$, so that
\begin{equation}
{\cal{L}}_{fermi}=\overline{\Psi}_L(i\gamma^\mu
D_\mu)\Psi_L+\overline{\Psi}_R(i\gamma^\mu D_\mu)\Psi_R.
\end{equation}
Fermions acquire a mass through the symmetry breaking, by
introducing in the lagrangian a Yukawa type interactions between
(left-handed and right-handed) fermion fields and the scalar
field, that is terms such as
\begin{equation}
{\cal{L}}_{Yukawa}=g_f(\overline{L}e_R\phi+\overline{e}_R
L\overline{\phi}),
\end{equation}
After symmetry breaking a mass for fermions arise:
\begin{equation}
{\cal{L}}_{mf}=\sum_f \frac{g_f v}{\sqrt{2}}\overline{f}f, \ \ \
m_f=\frac{g_f v}{\sqrt{2}}.
\end{equation}
where the summation runs over the different fermions (possibly
except neutrinos).
\\
Moreover terms of interaction among the Higgs boson and the
fermions
\begin{equation}
{\cal{L}}_{hf}=\sum_f \frac{g_f v}{\sqrt{2}}\eta\overline{f}f,
\end{equation}
showing that the Higgs boson couples to fermions with a factor
proportional to their mass.
\\
Unfortunately the parameters $g_f$ cannot be calculated from the
theory, this fact also indicating that the Standard Model cannot
be considered a complete description of the particle world.
\\
A different discussion for neutrino masses can be made: one should
introduce right-handed neutrinos, but being the neutrino a
chargeless particle it interacts neither through electromagnetic
interactions, and the right-handed part nor through weak
interactions, so that the introduction of a right-handed neutrino
seems not to be a necessary step in the Standard Model. Recent
measurements indicates that neutrinos do have a mass: this implies
that or a right-handed neutrino is introduced in the theory, as
done for other fermions, or other mechanisms of generation of a
mass are necessary, which would eventually point to new physics
effects. At present the problem of neutrino masses and its
implications in the search of new physics is still open.
\\
The last point to deal with to get the lagrangian of the Standard
Model is the mixing between quarks. An analogous phenomenon for
charged leptons can be introduced, but it has been poorly studied
and is at the moment less relevant, while neutrino mixing is one
of the major topics, related to the measurement of their masses
through oscillations. Writing down the lagrangian of the Standard
Model one realizes that quark fields interacting with gauge bosons
and quark fields mass eigenvectors are not forced to be the same,
because they appear in different contexts. This allows to
introduce a unitary matrix $V_{CKM}$, the
Cabibbo-Kobayashi-Maskawa matrix\footnote{Cabibbo at first
introduced a mixing between the two first generations of quarks.
Kobayashi and Maskawa extended it to the heaviest one.}
\cite{ckm}, which describes the relation between the eigenvectors
of the interaction hamiltonian $q^\prime$ and the mass
eigenvectors $q$:
\begin{equation}
\left( \begin{array}{c} d^\prime \\ s^\prime \\ b^\prime
\end{array} \right)= V_{CKM}\left( \begin{array}{c} d \\ s \\ b
\end{array} \right).
\end{equation}
The choice of down type quarks for the mixing is purely a matter
of convention: a quark mixing using up type quarks could be
considered and physical prediction would not be affected by this
choice.
\\
The existence of this matrix explains transitions between
different families, which would not be allowed without it: CKM
matrix elements are in practice a measure of the strength of
interactions between two quarks of different flavour.
\\
The CKM matrix can be parametrized and written in different ways,
for example:
\begin{equation}\label{ckm}
V_{CKM}=\left( \begin{array}{ccc} V_{ud} & V_{us} & V_{ub}\\
V_{cd} & V_{cs}& V_{cb} \\ V_{td} & V_{ts} & V_{tb} \end{array}
\right)\backsimeq \left(\begin{array}{ccc} 1-\frac{\lambda^2}{2} & \lambda & A\lambda^3(\rho-i\eta)\\
-\lambda & 1-\frac{\lambda^2}{2}& A\lambda^3 \\
A\lambda^3(1-\rho-i\eta) & -A\lambda^3 & 1
\end{array} \right).
\end{equation}
The last term in (\ref{ckm}) is the Wolfenstein parametrization
\cite{wolf}: it is an approximated form, valid up to
$O(\lambda^4)$, where $\lambda$ is the Cabibbo angle. $\eta$
parametrizes the complex part of the matrix and it is a measure of
$CP$ violation in the Standard Model.
\begin{figure}[t]
\begin{center}
\mbox{\epsfig{file=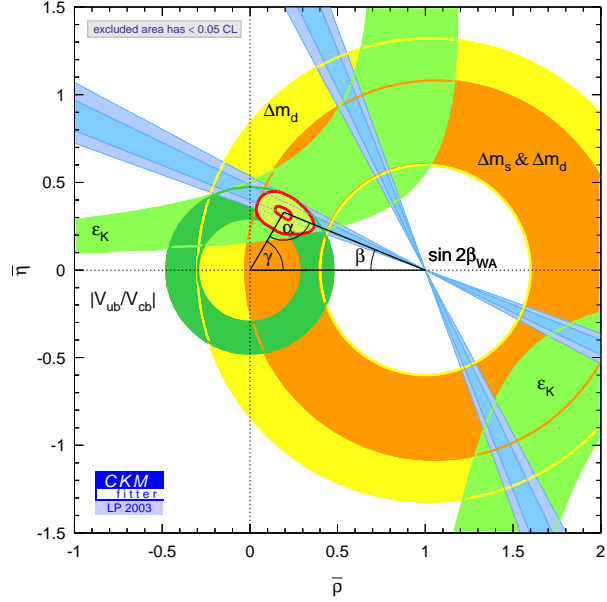,
height=8cm}}\caption{\label{triangolo}Unitarity triangle obtained
by data from Lepton-Photon 2003 \cite{ckm2}.}
\end{center}
\end{figure}
In the last years, at the b-factories, a large effort has been
made to measure, with high precision, CKM matrix elements, above
all the smallest ones, in order to verify the unitarity of the
matrix. A violation of unitarity could be a possible signal that
something is incomplete in the theory. In particular most of the
experiments tried to verify the relation
\begin{equation}\label{ut}
V_{ud}V_{ub}^*+V_{cd}V_{cb}^*+V_{td}V_{tb}^*=0
\end{equation}
also known as unitarity triangle\footnote{Unitarity triangles are
in general three and corresponds to the constraint of unitarity
for the elements of the matrix $V_{CKM}$. However, for
phenomenological reasons, the other two triangles are more
difficult to test, so that the unitarity triangle for antonomasia
is the one in (\ref{ut}).}. Latest experimental results give the
following values for the parameter in the Wolfenstein
parametrization:
\begin{eqnarray}
\lambda&=&0.2279\pm0.0032\\
A&=& 0.768-0.824\\
\overline{\rho}&=&0.118-0.273\\
\overline{\eta}&=&0.305-0.393.
\end{eqnarray}
These values are consistent with the unitarity of the CKM matrix,
as shown in the figure (\ref{triangolo}) \cite{ckm2}. Once again
the Standard Model seems to withstand experimental tests.

\section{Quantum Chromodynamics}\label{qcd}
Hadrons are particles interacting through strong interactions:
since the 30s-40s, from Yukawa's theory of strong forces on, they
were studied to find a coherent theory of this kind of forces.
Theoretical attempts have clashed with the large coupling constant
associated to these interactions which would have precluded any
perturbative expansion.
\\
This difficulty was overcome in the 70s with the discovery of
Quantum Chromodynamics (QCD). QCD is a non abelian gauge theory
introducing a new quantum number, called \it color, \rm which
exists in three different forms: \it red, blue \rm and \it green.
\rm The fundamental fields of the theory are the \it quarks \rm,
which carry the charge of color and interact with each other by
exchanging the carriers of the strong force, called \it gluons\rm.
\\
Experiments suggest that colored particles cannot be observed as
free states in nature and to explain this evidence an additional
property, the \it confinement\rm, has been postulated.
\\
In this picture, colored particles are confined into hadrons and
they are combined in such a way that the color is screened outside
the hadron and does not show up.
\\
A description of the confinement is not yet possible: in
particular we cannot solve completely the theory, starting from
the lagrangian of QCD, written in term of fundamental fields
(quarks and gluons), to explain the spectrum of the bound states
observed in nature, i.e. the hadrons.
\\
In spite of the lacking of this complete solution, QCD can explain
in great detail strong interactions in high energy experiments.
\\
Although strong interactions involve a large coupling constant at
low energies, the application of renormalization group equations
shows that, at higher energies, the coupling constant becomes
rather small and a perturbative approach is viable.
\\
This occurs because
\begin{equation}
\alpha_S(Q^2)\ll 1 \ \ \rm for \ \  Q^2\gg\Lambda^2_{QCD}
\end{equation}
where $\Lambda_{QCD}$ is the typical scale of strong interactions,
$\Lambda_{QCD}\sim 200 \ \rm MeV$ and $Q$ is the \it hard \rm
scale of the process, that is the highest energy scale where the
process takes place.
\\
At energies comparable with $\Lambda_{QCD}$ the coupling constant
blows up and any perturbative approach is barred. This behaviour
can qualitatively explain the origin of confinement: when quarks
separate (large distances correspond to small exchanged momenta),
the interaction between them becomes stronger and stronger, in
such a way that they can never be further than a distance
corresponding to the typical scale of QCD.
\\
During the times different techniques were developed in field
theory to deal with the non perturbative region of strong
interactions, such as for example QCD sum rules and lattice QCD.
\\
On the contrary the perturbative part has been widely studied and
tested in the last twenty five years and perturbative QCD has
become a precision tool to deal with strong interactions in high
energy physics.
\\
In the next sections some of the most important features of QCD
will be briefly recalled, above all those properties which are
needed further down.

\section{The Parton Model}
In the 60s experiments performed at the Stanford Linear
Accelerator Center (SLAC), showing the so called \it Bjorken
scaling \rm \cite{bjorken}, and spectroscopical observations
suggest that mesons (bosonic hadrons) and baryons (fermionic
hadrons) are composed by building blocks which, at high energies,
interact with external currents as free particles.
\\
These constituents were called \it partons, \rm the name
introduced by Feynman, or \it quarks, \rm as they were called by
Gell-Mann \cite{quark}. On theoretical basis also Zweig
\cite{zweig} introduced a similar entity\footnote{Zweig introduced
the name \it aces. \rm His paper was not published.}.
\\
Experiments at SLAC, pointing out the partonic behaviour at high
energies, concerned Deep Inelastic Scattering, that is the
scattering of an electron over a proton with a large exchanged
momentum (that is a momentum much larger than the proton mass). In
terms of the Bjorken variable
\begin{equation}
x=-\frac{q^2}{2P\cdot q}
\end{equation}
where $q^\mu$ is the exchanged momentum and $P^\mu$ the proton
momentum (usually taken in its rest frame, where $P^\mu=(M,0)$).
In first approximation the structure functions $f_a(x)$,
representing the probability to find a parton of type $a$ in the
hadron with a value $x$ of the Bjorken variable, measured in the
process do not depend on $x$ (Bjorken scaling). As it was noticed,
this is the typical situation occurring in point-like elastic
scatterings: this behaviour can be explained assuming that the
electron is scattered by a point-like constituent inside the
proton, suggesting that at high energies hadrons can be thought as
composed by partons, which behave as free particles.
\\
At the beginning of the 60s, to explain hadron spectroscopy, three
types of quarks were introduced: \it up, down \rm and \it strange.
\rm In the following years other three quarks were discovered: the
\it charm \rm quark in 1974, \it bottom \rm or \it beauty \rm in
1977 and , finally, the \it top \rm quark in 1994.
\\
The first three quarks have masses negligible at high energies and
for this reason they are usually referred as \it light \rm quarks.
The three quarks discovered more recently are referred as \it
heavy \rm quark, because their mass is in general relevant.
\begin{eqnarray}
m_c&\simeq&1.5 {\rm GeV},\nonumber\\
m_b&\simeq&5 {\rm GeV},\nonumber\\
m_t&\simeq&175 {\rm GeV}.
\end{eqnarray}
This is true for the \it top \rm and, in many processes (and in
particular the ones we are interested in), for the \it beauty\rm.
The mass of the \it charm \rm quark is relevant in many processes,
but the application of perturbative QCD at charm mass energies is
doubtful, since the coupling constant is quite large
$\alpha_S(m_c)=g^2/4\pi\simeq 0.35$.
\\
The parton model was introduced by Feynman and Bjorken for light
quarks, assuming that each parton in the hadron carries a fraction
of the total quadrimomentum of the hadron itself. Defining
$f_a(z)$ the probability to find a parton of type $a$, carrying a
momentum fraction $z$, the cross section $d\sigma$ for the Deep
Inelastic Scattering can be written as
\begin{equation}\label{dis}
d\sigma= \sum_a \int_0^1 dz \ f_a(z) \ d\hat{\sigma}_a(z)
\end{equation}
where $d\hat{\sigma}$ is the partonic cross section, that is the
cross section for the scattering of an electron over a quark $a$.
\\
At first the parton model was a na\"{i}ve picture which did not
take into account strong corrections: an improved parton model can
be consistently introduced considering QCD corrections, which
explain for example Bjorken scaling violations.
\\
Eq.(\ref{dis}) shows a paradigmatic situation in QCD, the
interplay between non perturbative and perturbative terms, which
contribute together to the result, but can be obtained in
different ways: $d\hat{\sigma}_a(z)$ being a point-like object can
be in principle computed in perturbation theory, while the \it
parton distribution functions \rm (pdf), being log distance
objects, cannot be calculated and have to be extracted by
experimental data. However the theory allows to calculate
perturbatively the evolution of the pdf's as a function of the
scale.

\section{The Lagrangian of Quantum Chromodynamics}
QCD is a non abelian gauge theory, whose group of symmetry is
$SU(3)_c$. The index $c$ indicates that the quantum number is the
color.
\\
The Lagrangian of QCD is obtained according to Yang-Mills
theories, by requiring a local invariance under the group
$SU(3)_c$\footnote{Since now the suffix $c$ will be neglected.}.
The construction of the Lagrangian involves the definition of a
covariant derivative
\begin{equation}
(D^\mu)_{ab}=\partial^\mu\delta_{ab}+ig_s \ (t^c A^{\mu c})_{ab}
\end{equation}
where $A^\mu(x)$ are gluons, the gauge fields of the theory, and
$t^a$ are matrices of the fundamental representation of $SU(3)$,
having the properties
\begin{equation}
\left[t^A,t^B\right]=if^{ABC}t^C
\end{equation}
being $f^{ABC}$ the structure constants of $SU(3)$.
\\
The most common choice is provided by the eight Gell-Mann
matrices, hermitean and traceless, normalized as
\begin{equation}
{\rm Tr} \ t^At^B=\frac{1}{2}\delta^{AB}
\end{equation}
\\
Important relations satisfied by the colour matrices are
\begin{equation}
\sum_A t^A_{ab}t^A_{bc}=C_F\delta_{ac} \ \ \ {\rm{with}} \ \ \
C_F=\frac{4}{3}.
\end{equation}
For the adjoint representation it holds
\begin{equation}
{\rm{Tr}} \ \mathit T^AT^B=C_A\delta_{AB} \ \ \ {\rm{with}} \ \ \
C_A=3
\end{equation}
The lagrangian density describing the interaction between
fermionic fields $q(x)$ (quark) and gauge bosons (gluons), locally
invariant under $SU(3)$ turns out to be:
\begin{equation}
\mathcal{L}_{QCD}=-\frac{1}{4}F^a_{\mu\nu}F^{a
\mu\nu}+\sum_{flavours}\overline q_a (i D\!\!\!\!/-m)_{ab}q_b
\end{equation}
where $F^a_{\mu\nu}$ is the strength tensor defined as
\begin{equation}
F^a_{\mu\nu}=\left[\partial_\mu A^a_\nu- \partial_\nu A^a_\mu -
g_sf^{abc} A^b_\mu A^c_\nu\right].
\end{equation}
The indices $a,b,c$ run over the eight degrees of freedom of the
gluon field.
\\
Feynman rules deriving from this lagrangian are contained in
appendix (\ref{feynrules}).

\section{Evolution of the Coupling Constant and Asymptotic Freedom}\label{running}
Let us consider an observable $\mathcal{O}(Q^2)$, where $Q$ is an
energy scale much larger than every other energy parameter
involved in the process, for example the masses of the quarks,
$Q^2\gg m^2_i$. In general a prediction about the value of this
observable can be performed in perturbation theory, that is as an
expansion in the coupling constant
\begin{equation}
\alpha_s=\frac{g^2_s}{4\pi},
\end{equation}
provided that $\alpha_S$ is small enough to justify this approach.
Obviously let us consider an observable where only strong
corrections need to be calculated.
\\
In order to remove ultraviolet divergences, arising to every order
of the perturbative expansion, the theory of renormalization has
to be applied to the observable $\mathcal{O}(Q^2)$, by introducing
a substraction point $\mu$. Now, having introduced a second energy
scale, the observable $\mathcal{O}$ will depend on the ratio
$Q^2/\mu^2$.
\\
However the renormalization scale $\mu$ is arbitrary and,
according to the theory of renormalization, the observable
$\mathcal{O}(Q^2/\mu^2)$ has to satisfy the Callan-Symanzik
equation\footnote{We are neglecting the masses and this will lead
to simpler equations.} (renormalization group equation):
\begin{equation}\label{rge}
\mu^2\frac{d}{d\mu^2}\mathcal{O}(\frac{Q^2}{\mu^2},\alpha_S)\equiv
\left[
\mu^2\frac{\partial}{\partial\mu^2}+\mu^2\frac{\partial\alpha_S}{\partial\mu^2}\frac{\partial}{\partial\alpha_S}\right]\mathcal{O}(\frac{Q^2}{\mu^2},\alpha_S)=0.
\end{equation}
Defining the $\beta$ function as
\begin{equation}\label{beta}
\beta(\alpha_S)=\mu^2\frac{\partial\alpha_S}{\partial\mu^2}
\end{equation}
and defining $t=\log(Q^2/\mu^2)$, the equation (\ref{rge}) can be
written as
\begin{equation}
\left[ -\frac{\partial}{\partial
t}+\beta(\alpha_S)\frac{\partial}{\partial
\alpha_S}\right]\mathcal{O}(e^t,\alpha_S)=0.
\end{equation}
An implicit solution of this equation is given by
\begin{equation}
\mathcal{O}(\mu^2=Q^2,\alpha_S(Q^2))=\mathcal{O}(1,\alpha_S(Q^2)),
\end{equation}
where the new function $\alpha_S(Q^2)$, the \it running coupling
constant \rm, is defined according to the property
\begin{equation}
\int_{\alpha_S(\mu^2)}^{\alpha_S(Q^2)} \
\frac{d\alpha}{\beta(\alpha)}=\log\left(\frac{Q^2}{\mu^2}\right).
\end{equation}
Let us remark that in this way the whole dependence on the scale
is absorbed in the running coupling constant.
\\
The function $\beta(\alpha_S)$ admits a perturbative expansion in
the form
\begin{equation}\label{betaexp}
\beta(\alpha_S)=-\beta_0\alpha_S^2-\beta_1\alpha_S^3-\beta_2\alpha_S^4-\dots
\end{equation}
and substituting (\ref{betaexp}) into (\ref{beta}),  the running
of the coupling can be calculated in perturbation theory.
\\
The coefficients $\beta_n$ can be calculated from higher order
corrections to the bare vertices of the theory: at present they
are known up to fourth order. For our purposes only $\beta_0$ and
$\beta_1$ will be relevant:
\begin{eqnarray}\label{b}
\beta_0&=&\frac{33-2n_f}{12\pi},\nonumber\\
\beta_1&=&\frac{153-19n_f}{12\pi^2},
\end{eqnarray}
where $n_f$ is the number of the active flavours at the energies
where the process takes place.
\\
The leading order solution of the equation (\ref{beta}) reads
\begin{equation}\label{alfarun1}
\alpha_S(Q^2)=\frac{\alpha_S(Q_0^2)}{1+4\pi
\beta_0\log(Q^2/Q_0^2)}.
\end{equation}
In this way we can extract the value of the coupling constant at a
scale $Q$, known its value at $Q_0$. The most common choice is to
consider as a reference $\alpha_S(M^2_{Z^0})$ which is a parameter
of the Standard Model known with very high precision.
\begin{equation}
\alpha_S(M^2_{Z^0})=0.11720\pm 0.00017
\end{equation}
Another choice to express the solution of (\ref{beta}) is
possible: one can introduce a parameter $\Lambda_{QCD}$ in such a
way that (\ref{alfarun1}) takes the form
\begin{equation}\label{alfarun0}
\alpha_S(Q^2)=\frac{4\pi}{\beta_0\log
\left(Q^2/\Lambda^2_{QCD}\right)}.
\end{equation}
In this way the running coupling constant is expressed in terms of
one single parameter and it is evident that it blows up when the
energy scale is set equal to $\Lambda_{QCD}$.
\\
The exact value of $\Lambda_{QCD}$ strongly depends on its precise
definition, but it can be considered of the order of
$\Lambda_{QCD}\sim 200 \ \rm{MeV}$. Roughly speaking, for energies
of the order of a few hundred MeV, the perturbative approach for
strong interactions is not reliable because $\alpha_S$ becomes too
large. Let us notice that this is the typical scale of masses of
light hadrons and this suggests that the large growth of the
running coupling at low energies is an ingredient necessary to
explain phenomena such as the confinement of the quarks inside the
hadrons.
\\
On the contrary when the scale of the process becomes very large,
the running coupling constant vanishes:
\begin{equation}
\alpha_S(Q^2)\rightarrow 0 \ \ \ {\rm for} \ \ \ Q^2\rightarrow
\infty.
\end{equation}
This property is known as \it asymptotic freedom \rm and was
discovered by Wilczek, Politzer and Gross in 1974
\cite{asymptoticfreedom}. Asymptotic freedom is a fundamental
property of QCD:
\begin{itemize}
\item
it justifies the perturbative approach for energies much larger
than $\Lambda_{QCD}$: since $\alpha_S$ becomes small enough, one
can expand any observable in powers of this parameter;
\item
it explains why the parton model is a successful picture of the
structure of an hadron: for large $Q^2$ the partons in the hadrons
behave as non interacting particles in first approximation,
because the coupling constant is small. Through radiative
corrections one can calculate violations to this behaviour.
\end{itemize}
Let us observe that the existence of asymptotic freedom is related
to the sign of $\beta_0$: as long as the number of flavour is
$n_f<33/2$ (as it happens in nature), the first coefficient
$\beta_0$ is positive, that is the first term in the expansion
(\ref{betaexp}) is negative. A step-by-step solution of equation
(\ref{rge}) would show that $\alpha_S(Q^2)$ becomes smaller and
smaller as the energy becomes larger because of the sign of
$\beta_0$.
\\
This behaviour is opposite with respect to QED, where
$\beta_0=-\frac{1}{3\pi}$ (according to the conventions of eq.
(\ref{betaexp})) \cite{peskin}.
\\
The physical origin of this difference arises from the non abelian
nature of QCD: being a non abelian theory gauge bosons carry the
quantum number of the symmetry group, the color, contrarily to
what follows in QED, where photons do not carry any electric
charge. This implies that gluons can interact with each other, as
shown by Feynman rules in appendix \ref{feynrules} . In QED the
coupling constant decreases at large distances (small energies)
and this effect is na\"{i}vely explained as a consequence of
vacuum polarization by electron-positron pairs, which screen the
electric charge.
\\
In QCD this effect related to quark interactions is overwhelmed by
an anti-screen effect due to gluons self-interactions, which
causes a growth of the coupling constant at large distances.
\\
Just to conclude this section let us notice that in next
applications, instead of (\ref{alfarun0}), we will use the
following expression for $\alpha_S(Q^2)$:
\begin{equation}\label{alfarun}
\alpha_S(Q^2)=\frac{4\pi}{\beta_0\log
\left(Q^2/\Lambda^2_{QCD}\right)}\left[1-
\frac{\beta_1}{\beta_0}\frac{\log\log\left(Q^2/\Lambda^2_{QCD}\right)}{\log\left(Q^2/\Lambda^2_{QCD}\right)}\right],
\end{equation}
which represents the running coupling constant calculated with
next-to-leading accuracy.

\section{Infrared Divergencies}
Infrared divergencies and their cancellation, which is a crucial
test of the consistency of a massless theory as QCD and QED, will
be one of the main topics of this thesis.
\\
Let us consider a generic process involving at least a quark as an
external leg and assume that the quark mass can be neglected with
respect to the other relevant energy scales.
\\
Strong corrections to this process involve gluon bremsstrahlung:
for example, for one real gluon emission, a straightforward
calculation shows that the emission rate turns out to be
proportional to the term
\begin{equation}\label{brem}
\Gamma\propto\int_0^1 \ \frac{d\omega}{\omega} \ \int_0^1 \
\frac{d\theta}{(1-\cos\theta)},
\end{equation}
where $\omega$ is an adimensional and unitary variable, defined as
\begin{equation}
\omega=\frac{E_{gluon}}{E^{MAX}_{gluon}}
\end{equation}
and $\theta$ is the angle between the gluon and the quark.
\\
The expression in (\ref{brem}) is manifestly divergent and, in
particular, it shows two types of singularities:
\begin{itemize}
\item
when the gluon momentum vanishes
\begin{equation}
\omega\rightarrow 0
\end{equation}
the so called \it soft \rm singularity arises. It originates from
the massless nature of the gluon and can be regularized by giving
to it a fictious mass $\lambda$;
\item
when the angle of emission vanishes
\begin{equation}
\theta\rightarrow 0
\end{equation}
the singularity is called \it collinear\rm. It can be regularized
by giving a mass $m$ to the quark.
\end{itemize}
Soft and collinear singularities will be referred with the
comprehensive term of \it infrared \rm divergencies (or
singularities).
\\
In general infrared singularities arise in theories when a
massless field is present (soft divergence), the gluon in QCD, or
when it couples to another massless field (collinear divergence),
for example to a quark \cite{muta}.
\\
With the regularization indicated above, (\ref{brem})
reads\footnote{For simplicity let $m$ be an adimensional variable
proportional to the quark mass.}:
\begin{equation}\label{brem1}
\Gamma\propto\int_\lambda^1 \ \frac{d\omega}{\omega} \
\int_{m^2}^1 \ \frac{d\theta}{(1-\cos\theta)}.
\end{equation}
After two simple integrations one gets that the singularities are
parametrized in terms such as $\log \lambda$ and $\log m$.
\\
This regularization is not the one we will use in explicit
calculation: we will use dimensional regularization
\cite{dimensional}, which is more elegant and theoretically
useful, even if it hides the origin of the singularities discussed
above. In practice one calculates integrals in $N=4+\epsilon$
dimensions, instead of 4 dimensions: in this number of dimensions
infrared divergent integrals becomes regular and singularities are
parametrized as poles in the regulator $\epsilon$. Once
divergencies are removed in such a way, one can consider the limit
$\epsilon\rightarrow 0$, coming back to a physical number of
dimensions.
\\
In this regularization scheme the equation (\ref{brem}) takes the
form
\begin{equation}
\Gamma\propto\int_0^1 \ \frac{d\omega}{\omega^{1-\epsilon}} \
\int_0^1 \
\frac{d\theta}{(1-\cos\theta)^{1-\epsilon/2}}=\frac{2}{\epsilon^2}.
\end{equation}
More details about this topic will be discussed in chapter
\ref{fixedorder} where explicit calculation will be shown.

\section{Cancellation of Infrared Divergencies}
Obviously meaningful physical results cannot be affected by
singularities: somehow infrared divergencies have to be cancelled.
\\
At first let us consider the case of inclusive quantities, such as
for example the total rate of a scattering or the total width of a
decay.
\\
The non abelian nature of QCD complicates the question: let us
recall at first the solution for QED, which is an abelian theory.
\\
In QED it is known from a very long time that infrared
divergencies arise in photon brehmsstrahlung processes and in
radiative corrections in general. Their cancellation is assured by
Bloch-Nordsieck theorem \cite{bn}: it states that infrared
singularities cancel if we sum over all the degenerate final
states. In practice, order by order in perturbation theory, one
has to sum all the real and virtual diagrams describing the same
process: for example, for one photon emission one has to take into
account photon brehmasstrahlung and virtual one loop diagrams with
no photon emission.
\\
From a physical point of view, the explanation of this theorem is
rather intuitive: for soft emission , for example, one should
consider that every real detector has a finite resolution and
cannot distinguish between an electron and an electron plus a soft
photon\footnote{In practice a photon with an energy lower than the
energy resolution of the detector.}. In practice the bare state of
electron does not exist: a physical electron is always surrounded
by a cloud of soft photons.
\\
Analogous considerations can be made for collinear singularities:
every detector cannot resolve particles in a arbitrarily small
angular cone, so that a summation over the state of electron and
electron plus a cloud of collinear photon is required.
\\
However in non abelian theories the Bloch-Nordsieck theorem is in
general violated, as a direct consequence of the non abelian
nature of the theory \cite{comelliciafaloni2,dft}.
\\
The cancellation of infrared singularities in non abelian theories
is assured, on the basis of general arguments, by the
Kinoshita-Lee-Nauenberg theorem, which provides a generalization
of Bloch-Nordsieck assertions. The main results of this important
theorem are recalled in the next section.

\subsection{Kinoshita-Lee-Nauenberg Theorem}
\begin{figure}[h]
\begin{center}
\mbox{\epsfig{file=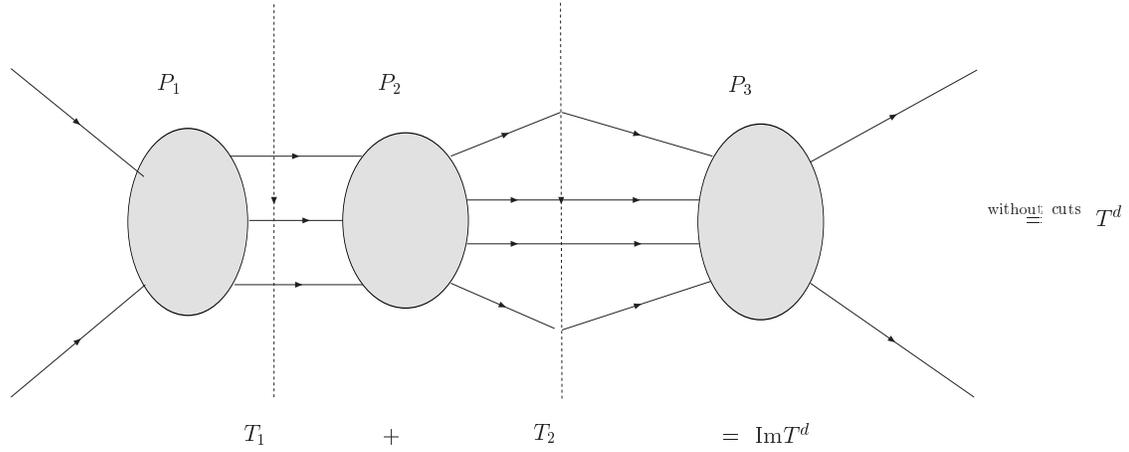, width=15cm}}
\caption{\label{kino2}\footnotesize{Example of cut diagrams: $T_1$
and $T_2$ are two different cuts of the same diagram $T^d$.}}
\end{center}
\end{figure}

\begin{figure}[h]
\begin{center}
\mbox{\epsfig{file=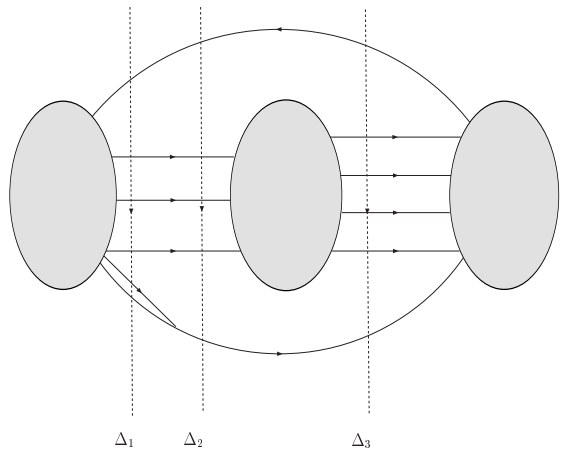, height=6cm}}
\caption{\label{kino3}\footnotesize{Example of double cut diagram:
the amplitudes $\Delta_1$,$\Delta_2$ and $\Delta_3$ belong to the
same set $\Delta$. }}
\end{center}
\end{figure}

According to the Kinoshita-Lee-Nauenberg theorem
\cite{kino,leenauen} the cancellation is attained when a summation
over final degenerate states (like in Bloch-Nordsieck theorem) and
initial degenerate states (due to the non abelian nature of the
theory) is performed.
\begin{figure}[h]
\begin{center}
\mbox{\epsfig{file=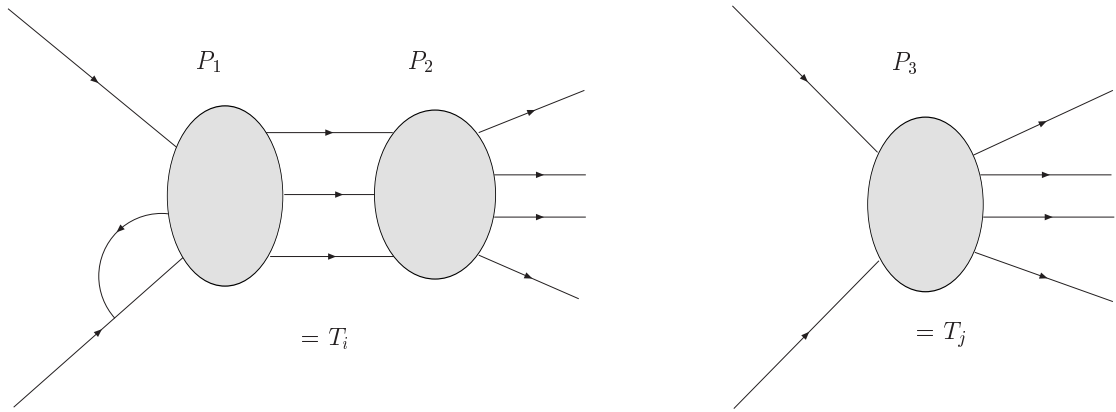, width=15cm}}
\caption{\label{kino1}\footnotesize{Example of two diagrams having
the same final states.}}
\end{center}
\end{figure}
In particular, in Kinoshita's work \cite{kino}, the relation
between the Feynman diagrams involved in the process and the
cancellation of mass (another usual name for collinear
divergencies) and soft singularities is widely discussed.
\\
Let us recall the main results of \cite{kino}: let us consider a
process at a given order of perturbation theory; let us call $T_i$
the corresponding Feynman amplitude (see Fig.(\ref{kino1})), the
total transition probability is proportional to $\sum_{ij}
T^\dagger_iT_j$, where the sum over indices is performed
considering diagrams with the same final states. $T^\dagger_i$ is
represented by a diagram obtained by $T_i$ by reversing time, that
is by the exchange of initial and final states.
\\
In this way $T^\dagger_i T_j$ may be represented joining the final
states of $T_j$ and the initial states of $T^\dagger_i$ (that is
the final states of $T_i$): in order to distinguish it from a
Feynman diagram let us draw a line intersecting the final states
of $T_i$ and $T_j$.
\\
Let us call $T^d$ the Feynman diagram obtained removing the
cutting line and consider all the $T^\dagger_i T_j$ which reduces
to $T^d$ when the line is removed: by the optical theorem
$\sum_{ij}T^\dagger_i T_j$, with the sum over the diagrams which
reduce to $T^d$ when the cutting line is removed, is the
absorbitive part $A^d$ of $T^d$ and it's called \it cut diagram
\rm(see Fig.(\ref{kino2})).
The total transition probability is the sum of cut diagrams
involved in the process.
\\
Let us now connect the initial states of $T^d$ with its final
states (remember that they carry the same momentum), that is the
initial states of $T_i$ and $T_j$ and let us represent this
junction with a second cutting line.
\\
Removing the cutting lines we obtain a vacuum-to-vacuum transition
$T^v$: let us consider the set $\Delta$ of diagrams $\Delta_i$
which reduce to $T^v$ by removing the cutting lines (see
Fig.(\ref{kino3})).
\\
$\Delta$ is called a \it double cut diagram \rm and the total
transition probability is a sum of $\Delta$s.
Kinoshita's paper \cite{kino} shows that not only the total
transition probability is free from infrared divergences but also
every $\Delta$ doesn't show neither soft singularities nor
collinear (or mass) ones.

\subsection{Cancellation of Infrared Singularities in QCD}
In principle, in QCD the summation over initial states would be
required, according to what stated above.
\\
In many processes however Bloch-Nordsieck theorem can be applied,
provided that a summation over initial and final color is
performed; actually this is first of all a phenomenological
requirement: because of confinement colored particles cannot be
observed, so that a summation over all the possible color is
natural, since only this superposition is observable.
\\
However some counterexamples have been found, where the only
summation over colors is not sufficient: for examples in
\cite{dft} has been found that in Drell-Yan
\begin{equation}
q\overline{q}\rightarrow \gamma X,
\end{equation}
where $X$ is every hadronic state, subleading soft divergencies
arise at two loops and they are not cancelled by soft gluon
emissions. If the degeneracy of the initial quark with the state
of a quark plus a soft gluon is taken into account, this
divergence is cancelled. This is nothing but an application of the
Kinoshita-Lee-Nauenberg theorem.
\\
Let us underline that this is not necessary for our purposes and
in our explicit calculation the simple summation over colors and
final states will be enough.

\subsection{An Example: Incomplete Cancellation of Infrared Logarithms in the Electroweak Sector}
In the last few years, some authors
\cite{comelliciafaloni,comelliciafaloni2} showed that
Bloch-Nordsieck violations can be observed also in the electroweak
sector of the Standard Model. This happens since the
Glashow-Weinberg-Salam model is based on a non abelian gauge
theory. It assumes particular characteristics because of some
peculiarity of the theory.
\\
Apart the photon, whose behaviour in QED is discussed above and it
is not interesting here, the gauge bosons of electroweak
interactions, $W^\pm$ and $Z^0$, are massive. The mass provides a
natural cut off: in this case infrared singularities are screened,
but residual large logarithms, such as
\begin{equation}\label{largel}
\frac{\alpha_W}{\pi}\log^2 \left(\frac{M^2_W}{Q^2}\right),
\end{equation}
arise as in (\ref{brem1}), though here the cut off mass is
physical and not fictious.
\\
In \cite{comelliciafaloni,comelliciafaloni2} authors noticed that
these logarithmic terms do not cancel even in inclusive
quantities: in QED and QCD large infrared logarithms appear in
semi-inclusive distributions (they will be introduced in section
(\ref{seclogs})), but cancel in inclusive quantities, such as
total rates.
\\
The appearance of residual logarithms in the Electroweak Model is
a violation of the Bloch-Nordsieck theorem, due to the non abelian
nature of the theory: in this case a summation over initial weak
charges is not required, contrarily to what performed in QCD with
initial colors, because particles carrying weak isospin can exist
as asymptotic states, for example as electrons or neutrinos.
\\
This implies that in processes such as
\begin{equation}
e^+e^-\rightarrow \ \rm everything
\end{equation}
terms like (\ref{largel}) remain.
\\
A cancellation of these terms would be achieved by summing the
rate $\sigma(e^+e^-\rightarrow \ \rm everything)$ to the rate
$\sigma(\overline{\nu}_e+e^-\rightarrow \ \rm everything)$: this
corresponds to a summation over initial state weak charges and,
according to KLN theorem, provides the cancellation of logarithms.
Obviously there is not a compelling physical reason to do that.
\\
The presence of terms (\ref{largel}) does not spoil the theory:
after all they are finite and not singular, even if they are a
sort of shadow of the infrared singularity.
\\
However, at asymptotic energies, these terms can become large and
enhance electroweak corrections, even if $\alpha_W$ is quite
small, making electroweak corrections comparable to strong ones.
If this effect does exist, it will be surely detectable at future
Linear Colliders.

\section{Evolution Equations}\label{evolution}
Evolutions equations are an important property of QCD dynamics
\cite{altarelliparisi,dgl}.
\\
Let us consider multiple branching of partons from another one,
for example multiple emissions of gluons from a quark or splitting
of a gluon into a quark-antiquark pair and so on.
\begin{figure}[h]
\begin{center}
\mbox{\epsfig{file=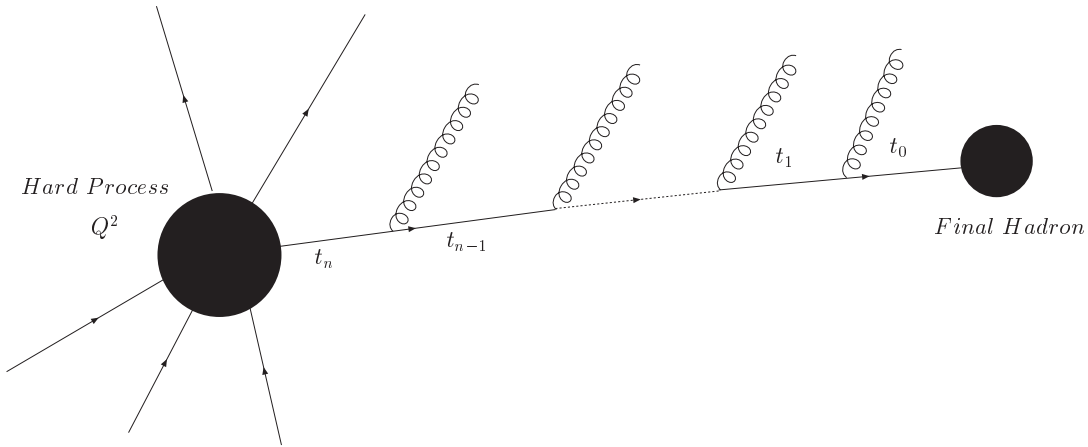, height=6cm}}
\caption{\label{multibranch}Evolution of a light quark outgoing
from an hard process at the energy $Q^2$. Through multiple gluon
emissions it evolves from the virtuality $t_n$ to the virtuality
$t_0$.}
\end{center}
\end{figure}
As shown in picture (\ref{multibranch}), the parent parton,
participating to an hard process at a scale $Q^2$, evolves
emitting other partons (gluons or quark) and increasing its
virtual mass-squared, as it approaches the hard process. In order
to be plain we can consider an outgoing quark from the hard
process, having virtual mass-squared $-t_n$, carrying a momentum
fraction $x_n$, which evolves to less virtual masses and momentum
fraction, by emitting multiple gluons at small angles.
\\
The total rate of the hard process will depend on the momentum
fraction distribution of the partons $f_i(x,Q^2)$ inside the
hadron, as seen by an external probe (for example the virtual
photon in the deep inelastic scattering).
\\
GLAPD (Gribov-Lipatov-Altarelli-Parisi-Dokshitzer)
\cite{altarelliparisi,dgl} evolution equations describe the
evolution of the distribution $f_i(x,Q^2)$ changing the scale of
the hard process: they are a set of coupled integro-differential
equations
\begin{equation}\label{dglap}
t\frac{\partial}{\partial t}f_i(x,t)=\sum_j\int_x^1 \frac{dz}{z}
\frac{\alpha_S}{2\pi}P_{ij}(z)f_j(x/z,t).
\end{equation}
The index $j$ runs over the different types of partons, that is
quark ($q$), antiquark ($\overline{q}$) and gluon ($g$).
\\
The kernels in the equations, $P_{ij}(z)$ are called the parton
splitting functions, taking into account different types of
branching: pair production from a gluon $P_{qg}$, gluon emission
from a quark $P_{qq}$ and gluon splitting into gluons $P_{gg}$.
\\
The charge conjugation invariance implies
\begin{eqnarray}
P_{q_iq_j}&=&P_{\overline{q}_i\overline{q}_j}\nonumber\\
P_{q_i\overline{q}_j}&=&P_{\overline{q}_iq_j}\nonumber\\
P_{\overline{q}_ig}&=&P_{q_ig}\equiv P_{qg}\nonumber\\
P_{g\overline{q}_i}&=&P_{gq_i}\equiv P_{gq}
\end{eqnarray}
They can expanded in perturbation theory as
\begin{equation}
P_{ij}(z)=\frac{\alpha_S}{2\pi}P^{(0)}_{ij}(z)+\left(\frac{\alpha_S}{2\pi}\right)^2P^{(1)}_{ij}(z)+\dots
\end{equation}
At the lowest order they take the look \cite{altarelliparisi}
\begin{eqnarray}
P^{(0)}_{qq}(z)&=&C_F\left[\frac{1+z^2}{(1-z)_+}+\frac{3}{2}\delta(1-z)\right],\label{apkernel}\\
P^{(0)}_{qg}(z)&=&n_f\frac{z^2+(1-z)^2}{2},\\
P^{(0)}_{gq}(z)&=&C_F\left[\frac{1+(1-z)^2}{z}\right],\\
P^{(0)}_{gg}(z)&=&2C_A\left[\frac{z}{(1-z)_+}+\frac{1-z}{z}+z(1-z)\right]+\frac{11C_A-2n_f}{6}\delta(1-z).
\end{eqnarray}
The next-to-leading order kernels, $P^{(1)}_{ij}$ have been
calculated in \cite{cfp}.
\\
The symbol "+" denotes plus distributions, defined through the
relation
\begin{equation}
\int_0^1 dz \ \frac{g(z)}{(1-z)_+}\equiv \int dz \
\frac{g(z)-g(1)}{1-z}.
\end{equation}
Let us notice that the plus distribution removes the infrared
singularities for $z\rightarrow 1$: in fact the "+" distribution
reproduces the summation between real and virtual diagrams, which
assures the cancellation of singularities. Let us finally remark
that the singularity for $z\rightarrow 0$ is outside the domain of
integration.
\\
The parton splitting function can be na\"{i}vely interpreted as
the probability to emit a parton with momentum fraction $z$ by the
parent parton: the interpretation as probability implies that
\begin{eqnarray}
\int_0^1 dx P_{qq}^{(0)}(x)=0,\nonumber\\
\int_0^1 dx \ x
\left[P_{qq}^{(0)}(x)+P_{gq}^{(0)}(x)\right]=0,\nonumber\\
\int_0^1 dx \ x \left[2n_f
P_{qg}^{(0)}(x)+P_{gg}^{(0)}(x)\right]=0.
\end{eqnarray}
Actually the interpretation as probability is formally incorrect,
because it does not handle carefully the infrared singularity for
$z\rightarrow 1$.
\\
A more correct interpretation can be introduced, by defining the
\it Sudakov form factor \rm
\begin{equation}
\Delta_i(Q,Q_0)\equiv
\exp{\left[-\sum_j\int_{Q_0}^{Q}\frac{dQ^\prime}{Q}\int dz
\frac{\alpha_S}{2\pi} P_{ji}(z)\right]}
\end{equation}
which turns out to be the probability to evolve from the scale
$Q_0$ to $Q$ without any branching \cite{webber}.
\\
A final observation is necessary for our purposes: as long as a
single logarithmic accuracy is required, instead of the bare
coupling one should consider the running coupling, evaluated at
the transverse momentum squared, as suggested in \cite{abcmv},
\begin{equation}
\alpha_S \rightarrow \alpha_S\left[z(1-z)Q\right].
\end{equation}
The effect of this redefinition of the scale where to evaluate the
coupling constant, is necessary at next-to-leading order: in fact,
expanding the running coupling one has
\begin{equation}
\alpha_S(z(1-z)Q)=\alpha_S(Q)+\beta_0\log\left[z(1-z)\right]\frac{\alpha_S^2(Q)}{4\pi}+O(\alpha_S^3).
\end{equation}
At two loops, the logarithmic term, combined with the soft
singularity $1/(1-z)$, gives next-to-leading terms: since we have
performed calculation with next-to-leading accuracy, these terms
due to the running of the coupling cannot be neglected.

\section{Semi-inclusive Observables and Infrared Logarithms}\label{seclogs}
Another approach to study the properties of strong interactions is
to define quantities which are not fully inclusive, but select
particular region of the phase space.
\\
Among these distributions are the so called \it shape variables
\rm, which permits to characterized the shape of the final
hadronic jet: an example is the \it thrust \rm, defined as
\cite{thrust}
\begin{equation}
T= \ {\rm max}_n \ \frac{\sum_i |\vec{p}_i \cdot \vec{n}|}{\sum_i
|\vec{p}_i|},
\end{equation}
where $\vec{n}$ is an arbitrary an unitary vector.
\\
In practice one wants to define a variable $X$, which in our
calculation will be the transverse momentum, and calculate the
differential distribution
\begin{equation}
d(X)=\frac{1}{\Gamma_0} \ \frac{d\Gamma}{dX},
\end{equation}
or the partially (cumulative) integrated distribution
\begin{equation}\label{distrint}
D(X)=\frac{1}{\Gamma_0} \ \int_{X_0}^X \
dX^\prime\frac{d\Gamma}{dX^\prime},
\end{equation}
where $X=X_0$ is the elastic point, that is the Born (lowest
order) value of the observable. Let us consider the case where
\begin{equation}
\frac{1}{\Gamma_0} \ \frac{d\Gamma}{dX}=\delta(X-X_0)
\end{equation}
and
\begin{equation}
\frac{1}{\Gamma_0} \ \int_{X_0}^X \
dX^\prime\frac{d\Gamma}{dX^\prime}=1.
\end{equation}
Radiative corrections spread the spectrum in general and
introduced theoretical problems to face: in order to be calculable
in perturbation theory, the considered quantity has to be infrared
safe. Higher order corrections show infrared singularities for
real and virtual emission separately: in infrared safe quantities
these singularities cancel in the sum of real and virtual diagrams
as stated in the sections above.
\\
Another way to state this property is requiring that the
observable is insensitive to the emission of soft or collinear
gluons, that is it is invariant under the branching
\begin{equation}
\vec{p_i}\rightarrow \vec{p_j}+\vec{p_k}
\end{equation}
where the final momenta are collinear or one of them is very
small.
\\
At the order $O(\alpha_S)$ the calculation can be performed
inserting in the phase space a kinematical constraint, describing
the distribution we want to calculate, in terms of the variable of
the system:
\begin{equation}
\frac{1}{\Gamma_0} \ \frac{d\Gamma}{dX}=\frac{1}{\Gamma_0}
\int_\Omega \ \frac{d\Gamma}{d\Omega} \ \delta(X-o(\Omega)) \
d\Omega,
\end{equation}
where $\Omega$ represents generically the set of variables which
parametrize the phase space. $o(\Omega)$ is a function of the
kinematical variables of the system and, in order to have an
infrared safe distribution $o(\Omega)$ has to vanish for soft
emissions and for collinear emissions:
\[ o(\Omega)\rightarrow 0 \left\{ \begin{array}{l@{\quad}l}
{\rm for} \ E_{gluon}\rightarrow 0\\
{\rm for} \ \theta_{emission}\rightarrow 0
\end{array}\right. \]
In this kind of observables the cancellation of infrared
singularities occurs, but large logarithms arise in every order of
the perturbation theory. In fact the structure of  the calculation
at a finite order (fixed order), is
\begin{equation}
D(X)=\sum_{m=0}^{2n} \ G_{mn}\log^m |X-X_0|.
\end{equation}
These logarithms will be referred as \it infrared \rm or \it large
\rm logarithms, because they become large when the variable
approach the infrared region of the phase space. This occurs for
$X\rightarrow X_0$, that is near the elastic point of the process
and corresponds to the emission of soft or collinear gluons.
\subsection{General Structure for Single Gluon Emission}
As a remarkable example let us consider the rate for one real
gluon emission. The contribution to the distribution in a generic
variable $u$ is
\begin{equation}
f_R(u)=\int_0^1 d\omega \int_0^1 dt \ |{\overline{\cal
M}}|^2(\omega,t) \ \delta(u-o(\omega,t)),
\end{equation}
where $|{\overline{\cal M}}|^2(\omega,t)$ is the squared amplitude
for the emission of one real gluon, integrated over the azimuthal
angle and averaged/summed over the helicities and the colors of
the initial/final partons. We defined the unitary variables
$\omega$ and $t$
\begin{equation}
{\omega =}\frac{E_{g}}{E_{g}^{\max }}  \label{Eunit}
\end{equation}
\begin{equation}
{t=}\frac{1-\cos \theta }{2}.
\end{equation}
The function $o\left( \omega ,t\right) $ is the kinematical
constraint which specifies the states of the infrared gluon which
are included in the distribution and it is rescaled in such a way
that
\begin{equation}
0\leq o\left( \omega ,t\right) \leq 1.
\end{equation}
For a collinear-safe observable
\begin{equation}
o\left( \omega ,t=0\right) =0  \label{collsafe}
\end{equation}
while for a soft safe observable we have
\begin{equation}
o\left( \omega =0,t\right) =0.  \label{softsafe}
\end{equation}
An observable is therefore infrared safe when both conditions
above are satisfied, or, in other words when it is unaffected by
the emission of a soft and/or collinear gluon. If conditions
(\ref{collsafe}) or (\ref {softsafe}) holds we have the
cancellation of the two delta functions on the r.h.s. of equation
(\ref{generale}) for a collinear or soft gluon respectively . As a
consequence, the corresponding singularity is screened. The
distribution can be therefore computed in perturbation theory in
the limit of vanishing parton masses and virtualities.
\\
We know that the amplitude shows a pole in $\omega = 0$ (soft
singularity), due to the vanishing mass of the gluon, and a pole
in $t = 0$ (collinear singularity), due to the vanishing mass of
the quark.
\\
Therefore we can write \cite{noi2}
\begin{equation}\label{decomposizionematrel}
|{\overline{\cal M}}|^2(\omega,t) = \frac{\,M(\varepsilon
,t)}{\omega \,t} =\frac{\alpha _{S}A_{1}}{\omega \,t}
+\frac{\alpha _{S}S_{1}( t ) }{\omega }+\frac{\alpha_{S}C_{1}(
\omega ) }{t}+\alpha_{S} F_{1}( \omega ,t) , \label{matrel}
\end{equation}
where the functions $A_{1},\,S_{1}( t ) ,\,C_{1}( \omega )
,\,F_{1} ( \omega ,t) $ are defined as follows:
\begin{eqnarray}
A_{1} &\equiv &M(0,0),   \\ S_{1}\left( t\right)
&\equiv &\frac{M(0,t)-M(0,0)}{t},    \\
C_{1}\left( \omega \right) &\equiv &\frac{M(\omega ,0)-M(0,0)}{
\omega },   \\ F_{1}\left( \omega ,t\right) &\equiv
&\frac{M(\omega ,t)-M(0,t)-M(\omega ,0)+M(0,0)}{\omega \,t}.
\end{eqnarray}
From their definition it is clear that they are finite in the soft
limit
\begin{equation}
\omega \rightarrow 0,
\end{equation}
as well as in the collinear one
\begin{equation}
t\rightarrow 0.
\end{equation}
By integrating the amplitude in $\omega$ and $t$ at the border of
the phase space, large logarithms arise from the terms which are
singular in (\ref{matrel}):
\begin{equation}\label{oneloop}
D(X)=1+\frac{\alpha_S}{\pi}\left(\tilde{A}_1 \log^2 X +B_1 \log X
+\kappa_1 + r(X)\right).
\end{equation}
The double logarithm represents the leading term, while the single
logarithm is referred as the next-to-leading term, $\kappa_1$ is a
constant and $r(X)$ is a regular function in the whole phase
space, vanishing for $X\rightarrow X_0$. $\tilde{A}_1$ is strictly
related to $A_1$, typically $A_1$ times some coefficient coming
from the variable we are considering. $B_1$ is a combination of
$C_1$ and $S_1$.
\\
The leading term is the well known Sudakov double logarithm
\cite{sudakov}, whose coefficient is $A_1$ and have both soft and
collinear nature. The next-to-leading term is given by single
logarithms of soft or
collinear nature. $F_1(\omega,t)$ contains the terms which are not singular in the infrared limit.\\
Let us recall that $f_R$ contains infrared divergences which
cancel in the sum with virtual diagrams:
\begin{equation}\label{frv}
f(u)=f_R(u) + f_V(u),
\end{equation}
giving rise to "+" distributions, as defined above. $f(u)$ is the
physical distribution taking into account real and virtual
emissions.
\\
$f(u)$ at the border of the phase space, omitting the finite term
$F_1(\omega,t)$, which does not product any large logarithm, may
be written to $O(\alpha_{S})$ as:
\begin{equation}
f(u)=\delta (1-u)+\alpha_{S}\int_{0}^{1} d\omega \int_{0}^{1}
dt\left[ {\frac{A_{1}}{\omega t}}+{\frac{S_{1}(t)}{ \omega
}}+{\frac{C_{1}(\omega )}{t}}\right] \left\{ \delta \left[
u-o\left( \omega ,t\right) \right] -\delta \left[ u\right]
\right\}.  \label{generale}
\end{equation}
\\
According to the conditions stated the integral of the
distribution over the whole kinematical domain is one, i.e. the
distribution is normalized:
\begin{equation}
\int_{0}^{1} f(u)\,du=1.
\end{equation}
This means that the total rate is unchanged with respect to its
value by the emission of an infrared gluon.
\\
Obviously the functions $A_{1}$, $S_{1}\left( t\right) $ e $
C_{1}\left( \omega \right) $ may be obtained with an explicit
evaluation of the Feynman diagrams for the emission of a real
gluon. It is however possible to give a general derivation of
these functions by using the properties of the amplitudes in the
collinear and soft limits, as we will do in the two following
paragraphes.

\subsection{Terms of Collinear Origin}\label{collinearlogs}
The functions $A_{1}$ and $C_{1}\left( \omega \right) $ may be
obtained by considering the perturbative evolution of a light
quark in QCD. \newline Let us define
\begin{equation}
z=1-\omega ,
\end{equation}
$z$ is the energy fraction carried by the light quark, so that
\begin{equation}
o \left( \omega ,t\right) =\omega .  \label{framm}
\end{equation}
As we can see from condition (\ref{collsafe}), this observable is
not collinear safe\footnote{ The condition (\ref{framm}) is
instead soft-safe.}: we regulate the collinear divergence with a
small quark mass $\mu \neq 0.$ The denominator of the light quark
propagator is therefore modified in $\left( p^{2}=\mu ^{2}\right)
$:
\begin{equation}
\frac{1}{\left( p+k\right) ^{2}-\mu
^{2}}\cong\frac{1}{Q^{2}}\,\,\frac{1}{ \omega \left( t+\mu
^{2}/Q^{2}\right) }.
\end{equation}
where we have defined as usual $$ Q = 2p^0 \ \ , \ \ \omega =
{2k^0 \over Q}$$ The integration over the polar angle is $$
\int_0^1 dt \ {1\over t+{\mu^2 \over Q^2}} = \int_{\mu^2\over
Q^2}^1 dt^\prime {1 \over t^\prime} $$ This result is the same of
the massless case $ \left( \mu =0\right) $, with the angular
restriction:
\begin{equation}\label{deadcone}
t> \frac{\mu^{2}}{Q^{2}}.
\end{equation}
known as dead cone effect.
\\
The fragmentation function may be written as:
\begin{eqnarray}
D(z;Q^{2}) &=&\delta (1-z)+\alpha _{S}\int_{\mu
^{2}/Q^{2}}^{1}\frac{dt}{t} \int_{0}^{1}d\omega \left[
{\frac{A_{1}}{\omega }}+C_{1}(\omega ) \right] \left[ \delta
(1-z-\omega )-\delta (1-z)\right]  \nonumber\\ &=&\delta
(1-z)+\left[ \alpha _{S}A_{1}\left( \frac{1}{1-z}\right)
_{+}+\alpha _{S}\, C_{1}(1-z)_{+}\right] \log \frac{Q^{2}}{\mu
^{2}}.
\end{eqnarray}
Let us compare with the fragmentation function to the order
$\alpha _{S}$, containing the leading Altarelli-Parisi kernel:
\begin{eqnarray}
D(z;Q^{2}) &=&\delta (1-z)+{\frac{C_{F}\alpha _{S}}{2\pi }}\left(
{\frac{ 1+z^{2}}{1-z}}\right) _{+}\log \frac{Q^{2}}{\mu
^{2}}+\ldots   \\ &=&\delta (1-z)+\left[ {\frac{C_{F}\alpha
_{S}}{\pi }}\left( {\frac{1}{1-z}} \right) _{+}-\frac{C_{F}\alpha
_{S}}{2\pi }\left( 1+z\right) _{+}\right] \log \frac{Q^{2}}{\mu
^{2}}.
\end{eqnarray}
The plus-distribution in the infrared regular kernel may be made
explicit as:
\begin{equation}
\left( 1+z\right) _{+}{ =1+z-}\frac{3}{2}{\delta }\left(
1-z\right) .
\end{equation}
This decomposition allows the identification
\begin{eqnarray}
A_{1} &=&\frac{C_{F}}{\pi },  \\ C_{1}{(1-z)}
&=&{-}\frac{C_{F}}{2\pi }\left( 1+z\right)
\end{eqnarray}
implying that
\begin{equation}
C_{1}{ \equiv }\int_{0}^{1}\ dz\ {C}_{1}{(1-z)=}-\frac{
3}{4}\frac{C_{F}}{\pi }.
\end{equation}
This is the usual collinear coefficient appearing at the
$\alpha_S$ order in processes involving light quarks (DIS,
Drell-Yan, \dots).

\subsection{Soft Emissions and Eikonal
Approximation}\label{eikonal} In processes involving light quarks
only, single logarithms coming from soft emissions do not appear.
This means that in these processes soft gluons can be emitted only
at small angles, at least in the approximation where only
next-to-leading terms are taken into account. It can be proved
that, in such an approximation, they are emitted according to an
angular ordering, behaviour known as \it coherence \rm
\cite{coerenza}. In processes involving light quarks only,
coherence is violated at subleading level.
\\
Instead in processes where also heavy quarks are present coherence
is already violated at the next-to-leading level: this means that
soft emissions at large angle are possible even at the
next-to-leading order.
\\
In general soft emissions can be treated according to the eikonal
approximation, by considering the emission of a gluon from the
hard on-shell partons which take part to the Born process. In such
approximation the amplitude factorizes in the Born amplitude
multiplied by the eikonal current:
\begin{equation}
{J}^{\mu }\left( k\right) = -g_S\sum_{i=1}^{n}\frac{v_{i}^{\mu }}{
v_{i}\cdot k}{T}_{i}.
\end{equation}
$v_{i}$ are the 4-velocities of the hard partons: $v_{i}^{2}=0$
for massless, while $v_{i}^{2}=1$ for massive partons. In the
massless case the 4-velocity normalization is given by fixing a
non trivial kinematical invariant. $T_{i}$ are linear operators
acting in the Fock space for the color of the hard partons:
\begin{equation}
{T}_{i}|c_{1}\ldots c_{i}\ldots c_{n}\rangle =\left( {T}_{i}
\right) _{c_{i}^{\prime }c_{i}}|c_{1}\ldots c_{i}^{\prime }\ldots
c_{n}\rangle ,
\end{equation}
where$\left( {T}_{i}\right) _{c^{\prime }c}$ is the generator of
the parton $i$. This space is the direct product of the single
parton spaces:
\begin{equation}
|c_{1}\ldots c_{i}\ldots c_{n}\rangle =|c_{1}\rangle \otimes
|c_{2}\rangle \otimes \cdots \otimes |c_{n}\rangle .
\end{equation}
The color conservation in the QCD interactions may be written in
such a scheme as:
\begin{equation}
\sum_{i=1}^{n}{T}_{i}=0.
\end{equation}
where the sum is over initial/final parton colors.\\ In general
the notation is analogous to the one used for the angular momentum
in quantum mechanics.
\\
The matrices $T_i$ for different practical cases takes the look
\begin{eqnarray}
{\rm quark} &\rightarrow & (T_k)_{ij}=t^a_{ij} \nonumber \\
{\rm antiquark}&\rightarrow &(T_k)_{ij}=-t^a_{ji}\nonumber \\
{\rm gluon}&\rightarrow &(T_k)_{ij}=i f_{bac},
\end{eqnarray}
where $f_{abc}$ are structure constant of $SU(3)$ and $t^a$
matrices of the fundamental representation.
\\
Let us note that, in the soft case, factorization involves the
momenta and the colors of all the hard partons. An important
property of the eikonal current is that it is conserved:
\begin{equation}
k_{\mu }{J}^{\mu }\left( k\right) =\sum_{i=1}^{n}{T}_{i}=0.
\end{equation}
The sum over polarizations may then be substituted with the
diagonal part:
\begin{equation}
S_{\mu \nu }\left( k\right) \rightarrow -g_{\mu \nu }.
\end{equation}
The rate is proportional to the square of the eikonal current
\begin{equation}
-g_{\mu \nu }{J}^{\mu }\left( k\right) {J}^{\nu }\left( k\right)
=-g^2_S\sum_{i<j}^{1,n}\frac{2v_{i}\cdot v_{j}}{v_{i}\cdot
k\,v_{j}\cdot k}{T}_{i}\cdot
{T}_{j}-g^2_S\sum_{i=1}^{n}\frac{v_{i}^{2}}{\left( v_{i}\cdot
k\right) ^{2}\,}{T}_{i}^{2}.  \label{icqu}
\end{equation}
The second sum in the r.h.s. in the equation (\ref{icqu}) gets
contributions only from massive partons and therefore disappears
in the well-know processes as DIS, DY, photon fragmentation, etc.,
which involve only massless partons. If heavy quarks are involved
this term does not disappear, giving rise to logarithms of soft
nature.
\\
In the decay of a heavy flavour into a light flavour, as we will
consider, the eikonal current may be written as:
\begin{equation}
{J}^{\mu }\left( k\right) =\frac{v^{\mu }}{v\cdot k}{T}_{Q}-
\frac{n^{\mu }}{n\cdot k}{T}_{q}.
\end{equation}
We have that $v^{2}=1,\,n^{2}=0.$ We normalize the vector $n$ on
the light cone imposing $n\cdot v=1,$ in such a way that $n_{0}=1$
in the rest frame of the heavy flavour.
\\
In the case of our interest, with just two hard colored partons
(the heavy and the light quark) and with an hard vertex conserving
the color of the quark line, the color conservation reads
$T_{Q}=T_{q}.$ Since
\begin{equation}
{T}_{Q}^{2}={T}_{q}^{2}={T}_{Q}\cdot {T}_{q}=C_{F}
\end{equation}
the contribution for a soft gluon emission is:
\begin{eqnarray}
f_{S}\left( u\right) &=&g^{2}C_{F}\int {\frac{d^{3}l}{(2\pi
)^{3}2\omega }} \left[ -{\frac{v^{2}}{v\cdot
l^{2}}}+{\frac{2n\cdot v}{v\cdot l\,\,\,n\cdot l }}\right] \left\{
\delta \left[ u-o\left( \omega ,t\right) \right] -\delta \left[
u\right] \right\}  \nonumber \\ &=&\int_{0}^{1}d\omega
\int_{0}^{1}dt\left[ \frac{\alpha _{S}C_{F}}{\pi }{
\frac{1}{\omega \,t}}-\frac{\alpha _{S}C_{F}}{\pi
}{\frac{1}{\omega }} \right] \left\{ \delta \left[ u-o\left(
\omega ,t\right) \right] -\delta \left[ u\right] \right\} ,
\end{eqnarray}
being $v\cdot l=\omega $ and $n\cdot k=2\omega t$. We define the
rescaled gluon momentum analogously to (\ref{Eunit}):
\begin{equation}
l^{\mu }\equiv \frac{k^{\mu }}{E_{g}^{\max }}.
\end{equation}
By comparing with the general expression (\ref{generale}),we
obtain:
\begin{equation}
A_{1} =\frac{C_{F}}{\pi },\ \ \ \ \ \ \ S_{1}
=S_{1}(t)=-\frac{\alpha _{S}C_{F}}{\pi }.
\end{equation}
As already discussed in the general case, $S_{1}\neq 0$ represent
a new contribution with respect to processes involving light
quarks only (DIS, Drell-Yan, \dots). \\ Let us note that the soft
function $S_{1}(t)$ actually does not depend on the polar angle
$t$: this fact is related to the choice of the reference frame,
since the heavy quark at rest emits soft gluons isotropically;
this is a violation of coherence at the next-to-leading order
instead of at the NNL one, like in DIS or DY.
\\
By using the soft factorization one can verify the value of
$A_{1}$ as obtained before, by using the collinear factorization.
This last term appears indeed in both schemes having the soft and
the collinear enhancement.

\section{Resummation of Large Logarithms}\label{secresummation}
As long as $\alpha_S \log^2 |X-X_0| \ll 1$ the perturbative
approach is reliable, however, when this term becomes comparable
to 1 problems arise: logarithmically enhanced terms have
comparable sizes to every order of perturbation theory
\begin{equation}\label{doppilog}
\alpha_S^n \log^{2n} |X-X_0| \sim O(1).
\end{equation}
Obviously in this situation the perturbative approach is not
meaningful: one has to resum at least terms such as
(\ref{doppilog}), finding an improved perturbative expansion with
a larger domain of applicability.
\\
The theory of resummation of large infrared logarithms has been
developed in the last twenty years for QCD processes
\cite{pp,kodairatrentadue,catanitrentadue,cttw,curcigreco}: the
resummation of terms like in (\ref{doppilog}) represents the
leading or double logarithmic order. A more refined approximation
is preferable to reduce the dependance on the renormalization
scale\footnote{A strong dependance on the renormalization scale is
in general a signal that neglected higher order terms are large
and important.} and it consists in the resummation of single
logarithmic terms of the form $\alpha^n_S \log^n |X-X_0|$. This is
the next-to-leading approximation we will require for our
calculations. In this way the region of applicability of the
improved perturbative expansion is enlarged to $\alpha_S \log
|X-X_0| \lesssim 1$, a much larger region than the one where the
fixed order calculation is valid.
\\
The resummation of large logarithms has been widely studied and
applied to QCD processes for those observables which exponentiate.
They have particular properties: their matrix elements can be
factorized by expressing the emission of $n$ infrared (soft or
collinear) gluons as the product of $n$ single gluon emissions;
moreover the phase space and, in particular, the kinematical
constraint have to factorize in the same way.
\\
Under such conditions the perturbative expansion in the infrared
region give rise to an exponential series, which resums logarithms
with the required accuracy.
\\
The result of the resummation of large logarithms is accomplished
by the formula \cite{cttw}
\begin{equation}\label{master}
D(X)=K(\alpha_S)\Sigma(\alpha_S;X)+R(X;\alpha_S).
\end{equation}
The functions involved in the formula will be the object of the
calculations we will perform in Chapters \ref{resumtmd} and
\ref{fixedorder} for the process we are going to consider. Here
let us just sum up their meaning and their role:
\begin{itemize}
\item
$\Sigma(X;\alpha_S)$ resums large logarithms in exponentiated
form. It has a perturbative expansion in the form
\begin{equation}
\log \Sigma(X;\alpha_S)=\sum_{n=1}^{\infty}\sum_{m=1}^{n+1} G_{nm}
\alpha_S^n L^m=L g_1(\alpha_S L)+g_2(\alpha_S L)+\alpha_S
g_3(\alpha_S L) + \dots
\end{equation}
where $L=\log X$. The functions $g_i$ have the form
\begin{equation}
g_i(z)=\sum_{n=0}^{\infty} g_{i,n}z^n,
\end{equation}
so that the calculation of $g_1$ resums leading logarithms, that
is terms in the form $\alpha_S^n L^{n+1}$, $g_2$ resums
next-to-leading contributions as $\alpha_S^n L^n$, $g_3$
next-to-next-to-leading and so on. It is worth noting that, even
if the fixed order calculation shows at most two logarithms for
each power of $\alpha_S$, in the exponentiated formula they are
rearranged in such a way that there are only at most $(n+1)$
logarithms for $n$ power of $\alpha_S$. $\Sigma(X;\alpha_S)$ is a
universal function, that is does not depend on the specific
process, but only on general properties of the theory.
\item
$K(\alpha_S)$, the \it coefficient function\rm, is a short
distance process dependent function. It takes into account
constant terms, arising to every order of the perturbative
expansion, which do not exponentiate. It can be calculated order
by order in perturbation theory:
\begin{equation}\label{coefexp}
K(\alpha_S)=1+\kappa_1 \alpha_S + \kappa_2
\alpha_S^2+O(\alpha_S^3)
\end{equation}
\item
$R(X;\alpha_S)$, the \it remainder function\rm, is a process
dependent function which takes into account hard contributions
without any logarithmic enhancement. It depends on the specific
process, the distribution, the kinematics and so on. It can be
calculated in perturbation theory:
\begin{equation}\label{remexp}
R(X;\alpha_S)=r_1(X)\alpha_S+r_2(X)\alpha_S^2+O(\alpha_S^3)
\end{equation}
The property
\begin{equation}
R(X;\alpha_S)\rightarrow 0 \ {\rm for} \ X\rightarrow X_0
\end{equation}
holds for the remainder function.
\end{itemize}
For our calculations we will attain next-to-leading accuracy. In
order to achieve this level of approximation several terms need to
be calculated:
\begin{itemize}
\item
the functions $g_1$ and $g_2$ to resum logarithms with
next-to-leading accuracy. In order to reach this goal, running
coupling effects should be taken into account, as discussed in
section (\ref{running});
\item
the value of $\kappa_1$ in the coefficient function has to be
calculated, because the combination
\begin{equation}
\alpha_S\kappa_1\cdot e^{Lg_1(\alpha_S L)}\sim \kappa_1 \alpha_S^2
L^2
\end{equation}
produces next-to-leading contributions.
\end{itemize}
Moreover, the calculation of the first term of the remainder
function can be relevant to describe hard contributions: the
remainder function is negligible near the border of the phase
space when logarithms become large, but can be relevant for hard
emission, in the opposite limit of the phase space.
\\
Finally let us notice that not every observable exponentiates:
however for our specific case the conditions for the
exponentiation hold and the resummation can be performed. In
particular these requirements are the factorization of matrix
elements for soft and collinear emissions and the factorization of
the phase space. The formers are treated in the next section,
because they involve general properties of QCD, the latter,
instead, strongly depends on the specific observable one is
considering and therefore will be treated in Chapter
\ref{introtmd}.

\subsection{Resummation of Collinear Emissions}
The exponentiation of collinear emissions is quite direct to
demonstrate, because it derives from general properties of the
theory and it passes through the solution of Altarelli-Parisi
evolution equations, already introduced in section
(\ref{evolution}).
\\
Evolution equations can be written in compact form as in
(\ref{dglap})
$$
t\frac{\partial}{\partial t}f_i(x,t)=\sum_j\int_x^1 \frac{dz}{z}
\frac{\alpha_S}{2\pi}P_{ij}(z)f_j(x/z,t).
$$
They are difficult to solve because are integro-differential
equations and can be reduced to simple first order equations by
introducing the Mellin transform\footnote{For a detailed
discussion see for example \cite{peskin,webber}.}:
\begin{equation}
f_N\equiv \int_0^1 dz \ z^{N-1} \ f(z)
\end{equation}
In this way the convolution integral becomes a simple product
\begin{equation}\label{dglapn}
\frac{df_{i,N}(\mu^2)}{d\log\mu^2}=\sum_{j=q_i,\overline{q}_i,g}
\gamma_{ij,N}(\alpha_S(\mu^2))f_{j,N}(\mu^2),
\end{equation}
where $\gamma_{ij,N}$ are the moments of Altarelli-Parisi kernels,
usually referred as \it anomalous dimensions \rm
\begin{equation}
\gamma_{ij,N}(\alpha_S(\mu^2))=\int_0^1 \ dx \ x^{N-1} \
P_{ij}(\alpha_S(\mu^2)).
\end{equation}
Now, the only residual complexity is that the differential
equations (\ref{dglapn}) are coupled: the simplest case to
consider is the flavour non-singlet distribution, defined as
\begin{equation}
f_{NS}=\sum_i(f_{q_i}-f_{\overline{q}_i}).
\end{equation}
In this case (\ref{dglapn}) becomes
\begin{equation}\label{singlet}
\frac{df_{NS,N}(\mu^2)}{d\log\mu^2}=
\gamma_{qq,N}(\alpha_S(\mu^2))f_{NS,N}(\mu^2).
\end{equation}
The lowest order approximation for the anomalous dimension
$\gamma_{qq,N}$ reads:
\begin{equation}
\gamma_{qq,N}^{(0)}=C_F\left[-\frac{1}{2}+\frac{1}{N(N+1)}-2\sum_{k=2}^N
\frac{1}{k}\right].
\end{equation}
For sake of completeness let us recall the lowest order anomalous
dimensions for the other kernels:
\begin{eqnarray}
\gamma_{qg,N}^{(0)}&=&T_R
\left[\frac{2+N+N^2}{N(N+1)(N+2)}\right],\nonumber\\
\gamma_{gg,N}^{(0)}&=&2C_A\left[
-\frac{1}{12}+\frac{1}{N(N-1)}+\frac{1}{(N+1)(N+2)}-\sum_{k=2}^N
\frac{1}{k}\right]-\frac{2}{3}n_f T_R,\nonumber\\
\gamma_{gq,N}^{(0)}&=& C_F \left[\frac{2+N+N^2}{N(N^2-1)}\right].
\end{eqnarray}
The result of equation (\ref{singlet}) is
\begin{equation}
f_{NS,N}(\mu^2)=f_{NS,N}(\mu_0^2)\left(\frac{\alpha_S(\mu^2_0)}{\alpha_S(\mu^2)}\right)^{d_{qq,N}^{(0)}},
\end{equation}
where $d_{qq,N}^{(0)}=\frac{\gamma^{(0)}_{qq,N}}{2\pi\beta_0}$.
\\
Moreover
\begin{equation}\label{exprisom}
\left(\frac{\alpha_S(\mu^2_0)}{\alpha_S(\mu^2)}\right)^{d_{qq,N}^{(0)}}=\exp\left\{\int_{\mu_0}^\mu
\frac{dq^2}{q^2} \
\frac{\alpha_S{(\mu_0)}}{1+\beta_0\alpha_S(\mu_0)\log{\frac{q^2}{\mu^2_0}}}
\frac{\gamma^{(0)}_{qq,N}}{2\pi\beta_0}\right\}
\end{equation}
and if one neglect the running of the coupling
\begin{equation}
\alpha_S(q^2)\rightarrow \alpha_S(\mu^2),
\end{equation}
in turns out that (\ref{exprisom}) is the sum of the exponential
series
\begin{equation}
\sum_{i=0}^\infty
\frac{1}{k!}(\gamma^{(0)}_{qq,N}\alpha_S(\mu^2)\log{\frac{\mu^2}{\mu_0^2}})^k.
\end{equation}
The insertion of the correct behaviour for the running coupling
introduce additional logarithmic corrections.
\\
One can go back to the distribution $f_{NS}$ in the space of
configurations by using an inverse Mellin transform
\begin{equation}
f_{NS}(x)=\frac{1}{2\pi i}\int_C dN \ x^{-N} f_{NS,N}.
\end{equation}
the integration is in general complicated and can be performed
numerically.
\\
It is worth noting that $d_{qq,1}=0$, which corresponds to the
momentum conservation, and $d_{qq,N}<0$ for $N\geq 2$ which states
that the non-singlet distribution function decreases at large $x$.

\subsection{Resummation of Soft Emissions}\label{softresum}
The main problem related to the resummation of soft gluons in QCD
is related to the color algebra and in general to the non abelian
structure of the theory.
\\
Let us at first consider what happens in QED for multiple photon
emission. In general the matrix elements for the emission of $n$
soft photons is the Born matrix elements times $n$ eikonal
currents, defined in this case as
\begin{equation}
J^\mu(k)=\sum_i \ e \ \frac{p^\mu_i}{p_i\cdot k},
\end{equation}
where the summation runs over the $i$ hard emitters involved in
the process as external legs, so that
\begin{equation}
{\cal{M}_n}(p_i,k_1,\dots,k_n)={\cal{M}_0}\prod_j^n \
J^\mu(k_j)\cdot \epsilon^\ast(k_j,\lambda_j).
\end{equation}
The rate of the process turns out to be
\begin{equation}\label{manygamma}
\sigma=\sigma_0 \ \frac{1}{n!} \ \left[ \prod_j^n \
\frac{d^3k_j}{2(2\pi)^3\omega_j} J^\mu(k_j) d_{\mu\nu}
J^\nu(k_j)\right],
\end{equation}
the factor $\frac{1}{n!}$ is introduced to take into account the
bosonic nature of the photons and their indistinguishability. The
tensor $d_{\mu\nu}$ represents the sum over polarization of the
outgoing photons: as pointed out in section (\ref{eikonal}) the
eikonal current is conserved, so that the polarization tensor can
be reduced to
\begin{equation}
d_{\mu\nu}\rightarrow -g_{\mu\nu}.
\end{equation}
The factors in the squared bracket of (\ref{manygamma}) are
integrated over the same region of the phase space, so that the
rate for $n$ soft photon emission can be finally written as
\begin{equation}
\sigma_n=\sigma_0 \ \frac{1}{n!}\left[ \
-\frac{d^3k}{2(2\pi)^3\omega} J^\mu(k) J_\mu(k)\ \right]^n.
\end{equation}
This shows that in QED multiple photon emissions are easily
resummed into an exponential series, because the emission of each
photon is independent from other emissions.
\\
In QCD some differences arise, due to the non abelian nature of
the theory and in particular from the color algebra: in fact when
a gluon is emitted it changes the color of the parent parton and
consequently has an influence on the following emissions.
\\
Color matrices, defined in (\ref{eikonal}), are included in the
eikonal current, which reads
\begin{equation}
J^\mu(k)=\sum_i \ g_S \ T_i \frac{p^\mu_i}{p_i\cdot k}.
\end{equation}
Its square turns out to be
\begin{equation}\label{eikonalqcd}
J^\mu(k)J_\mu(k)=4\pi\alpha_S \ \left[\sum_{i\not= j} T_i\cdot T_j
\frac{p_i\cdot p_j}{(p_i\cdot k)(p_j\cdot k)}+\sum_i T_i^2
\frac{m^2_i}{(p_i\cdot k)}\right].
\end{equation}
Two kind of terms are present in the previous equation: those
where the factor $T_i^2$ appear are easy to factorize because this
term is proportional to the unity matrix and can be treated as a
c-number, so that the factorization of these terms is complete.
\\
This argument does not apply to the first terms in the right hand
side of (\ref{eikonalqcd}) because the terms $T_i\cdot T_j$ are in
general matrices and one should evaluate their matrix elements
between physical states:
\begin{equation}
<{\cal{M}}_0(p_i)|T_i\cdot T_j|{\cal{M}}_0(p_i)>.
\end{equation}
The problem of the exponentiation of soft emissions in QCD has not
therefore a general solution: fortunately in some cases, among
them the one we are interested in, this problem can be trivially
solved. In particular if in the process the number of hard partons
involved is $m=2$ or $m=3$ the conservation of color allows to
reduce the scalar product between color matrices to combination of
square of them, which are c-numbers.
\\
In fact if $m=2$ the conservation of color reads
\begin{equation}
T_1+T_2=0 \ \ \ \Rightarrow \ \ \ T_1=-T_2
\end{equation}
which implies
\begin{equation}
T_1\cdot T_2=T_1^2=T_2^2=C_F
\end{equation}
where the last equality is valid if the emitters are quarks as in
the case of our interest.
\\
For $m=3$ the conservation of color simply implies
\begin{equation}
T_1+T_2=-T_3,
\end{equation}
which allows to reduce the scalar products as
\begin{equation}
T_i\cdot T_j=\frac{1}{2}\left[ T_k^2-T_i^2-T_j^2 \right].
\end{equation}
For $m>3$ the conservation of color implies a number of relation
among the color matrices which is not sufficient to reduce all the
scalar product to c-numbers: this happens since the number of
independent scalar product is $\frac{m(m-1)}{2}$ while the
relations to reduce them are just $m$. For $m>3$ the number of
relations is less than the number of scalar products.
\\
Fortunately this complication do not appear in our calculation,
because the process we are considering has just two hard partons.

\large
\chapter{The radiative decay $B\rightarrow
X_s+\gamma$}\label{capbsgamma}

\section{Phenomenology of the {\it beauty} quark}
Both from a theoretical and experimental point of view, $B$-meson
decays have been widely studied in the last years, above all after
the $B$-factories (Belle and BaBar) and CLEO began to give their
results.
\\
The great interest that $b$ physics arouse is due to the peculiar
role of this quark in the Standard Model: it is the heaviest quark
to hadronize, since the $top$ quark decays before it can give rise
to hadrons, and its physics is a very sensitive test of the
Standard Model. It provides for example a probe to discover new
physics, as well as a way to measure CKM matrix elements and CP
violation parameters.
\\
In what follows we will consider decays of mesons containing a \it
beauty \rm quark and another light quark\footnote{Quarkonium
systems will not be considered.}, such as
$B=\overline{B}^0,B^{+},B_S,\dots$, that is processes such as
\begin{equation}
B\rightarrow X+W_B
\end{equation}
where $X$ represents the hadronic final states and $W_B$ a (real
or virtual) intermediate electroweak boson ($\gamma$, $W^\pm$).
\\
The mass of the $b$ quark is a fundamental quantity of the
Standard Model because it enters many processes useful to measure
the parameters of the theory, such as, for example, CKM matrix
elements.
\\
In spite of this importance it is still a poorly known parameter:
quarks are confined into hadrons and they never appear as free
particles, so a direct measure of their masses, as done for
leptons, is not possible. The determination of quark masses is
indirect and quantitative results depend on the theoretical
framework chosen to define them.
\\
Recent results \cite{pdg} give for the $beauty$ running mass in
the $\overline{MS}$ scheme
\begin{equation}\label{massarunning}
m_{b,\overline{MS}}=4.0-4.5 \ \rm{GeV}
\end{equation}
which corresponds to the pole mass
\begin{equation}\label{massapolo}
m_{b,POLE}=4.6-5.1 \ \rm{GeV}
\end{equation}
A complete discussion about results and theoretical problems
related to the $beauty$ quark mass can be found in \cite{massab}.
\\
Since the mass of the $b$ quark is much larger than the scale
where hadronization occurs, $\Lambda_{QCD}\sim 200$ MeV,
\begin{equation}\label{massablambda}
m_b\gg \Lambda_{QCD}
\end{equation}
the application of perturbative QCD techniques seems to be viable.
The heavy quark expansion technique (see for example \cite{hqet})
allows to control non perturbative effects in inclusive decays and
implies that the rate for hadronic decays $\Gamma(B\rightarrow
X+W_B)$ can be well approximated by the partonic rate
$\Gamma(b\rightarrow q+W_B)$.
\\
In this picture the spectator quark in the meson is supposed not
to be involved in the inclusive process. Obviously in exclusive
decays, where different decay channels are distinguished, the
spectator quark has an important role, but in what follows we will
be devoted to inclusive processes only.
\\
The more relevant decay of the $b$ quark is the transition
\begin{equation}
b\rightarrow c+W^*
\end{equation}
(where the symbol $*$ denotes a virtual particle) but other decays
with very small branching ratio ($\sim 10^{-4}-10^{-5}$) can occur
and can dominate near the border of the phase space, where the
dominant decay is forbidden for kinematical reasons.

\section{Rare $b$ decays}
Decays with very small branching ratio are referred as \it{rare
decays} \rm and are interesting because they are very sensitive to
poorly known parameters of the theory, for example small CKM
elements or test the correctness of the dynamical structure of the
Standard Model.
\\
An important rare decay of the $b$ quark is the transition
\begin{equation}
b\rightarrow u+ W^*
\end{equation}
where the semileptonic decay
\begin{equation}\label{semileptonic}
b\rightarrow u+ W^*\rightarrow u+l+\overline{\nu}_l
\end{equation}
seems to be the cleanest process to measure the CKM element
$|V_{ub}|$.
\\
Beside there are the so called radiative decays, rare decays with
the emission of a photon, such as
\begin{equation}\label{bsgamma}
b\rightarrow s+\gamma
\end{equation}
\begin{equation}\label{bdgamma}
b\rightarrow d+\gamma
\end{equation}
For example (\ref{bdgamma}) is interesting because it is sensible
to CP violation. Several theoretical problems are involved in its
study, making it a challenging issue for particle physics. Its
branching ratio is very small \cite{revgreub}
\begin{equation}
6.0 \ 10^{-6}\leq BR(B\rightarrow X_d\gamma)\leq2.6 \ 10^{-5}
\end{equation}
\begin{figure}[h]
\begin{center}
\mbox{\epsfig{file=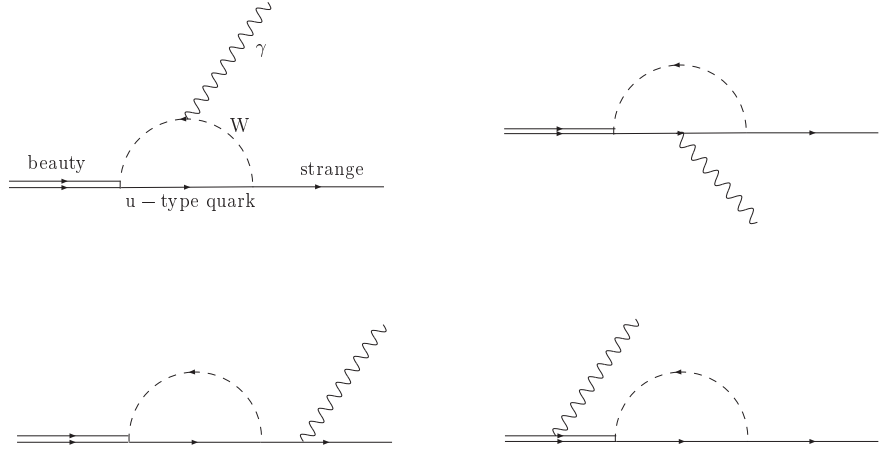, height=6cm}}
\caption{\label{vertice}Vertex for $b\rightarrow s\gamma$ in the
Standard Model.}
\end{center}
\end{figure}
In the rest of this thesis we will deal with the process
(\ref{bsgamma}): the world average for the experimental branching
ratio is \cite{revgreub, jessop}
\begin{equation}\label{branchingratio}
BR(B\rightarrow X_s\gamma)=(3.34\pm 0.38)10^{-4}
\end{equation}
In general, inclusive decay rates of the $b$ quark can be
calculated in perturbation theory in a model independent way.
However exclusive decay rates are easier to measure, though they
cannot be calculated from first principles.
\\
Since the exclusive channel $B\rightarrow K\gamma$ violates
angular momentum conservation, the simplest process to observe is
$B\rightarrow K^\ast\gamma$. The first measurement of the
exclusive rate for the final states $K^\ast(892)\gamma$ has been
performed at CLEO \cite{cleomis}.
\\
To measure inclusive decay rates two independent methods are used
\cite{stone}: the cleanest way is to measure and sum decay rates
for different exclusive channels as $K\gamma n\pi$ where $n\le 4$
and at most one pion is a $\pi^0$. The previous constraint are
necessary to reduce the background but introduce some model
dependence.
\\
The previous method can be integrated with an independent one,
which consist in measuring of the high energy photon, emitted in
the process. The photon spectrum has to be separated from the
background of other processes and this is performed by using a
neural network. This method was used at CLEO to perform the first
measure of the inclusive rate for $b\rightarrow s\gamma$.
\\
Since in the Standard Model flavour changing neutral current do
not occur, a decay such as (\ref{bsgamma}) must be mediated by a
loop even at the lowest level.
\\
As shown in the figure (\ref{vertice}), the particles in the loop
are a virtual W and a virtual \it u-type \rm quark, that is \it
up, charm \rm or \it top\rm. The different contribution of the
three quarks in the loop is mainly due to the ratio of CKM
elements: since $V_{ub}$ is much smaller than $V_{cb}$ and
$V_{tb}$ the $u$ quark contribution in the loop can be safely
neglected. The $t$ quark is favored by its larger mass, but $c$
quark contribution should not be left out. It turns out to amount
about $30\%$.
\\
The photon can be radiated by each charged line, as shown in the
figure.
\\
Since this decay is loop mediated it is a sensitive probe of new
physics: for example the $W$ boson may be substituted by a charged
Higgs, in an extension of the Standard Model with two scalar
doublets \cite{gsw2} or by a supersimmetric particle. If the mass
of new physics particles is not larger by many order of magnitude
than Standard Model particles an effect should be detectable.
However it should be pointed out that efforts to find signals of
new physics in past years have not given results and, at the
moment, the branching ratio for this process agrees with the
prediction of the theory.
\\
Recent calculations for the inclusive branching ratio gives the
prediction \cite{branchteorico,branchteorico2}
\begin{equation}
BR(B\rightarrow X_s\gamma)=(3.60 \pm 0.30) \ 10^{-4}
\end{equation}
Another useful application of $b\rightarrow s\gamma$ is the
measure of CKM elements: in particular this process may provide a
way to parametrize non perturbative effects in non perturbative
functions that can be used as an input in (\ref{semileptonic}),
allowing a precise measure of $|V_{ub}|$ \cite{leibovich}.
\\
The reason why we are interested in this process is that
$b\rightarrow s\gamma$ shows a non trivial interplay between
electroweak and strong interactions: the process at quark level is
mediated by electroweak interactions, but coloured particles are
responsible of gluon emissions which can be treated by well known
tools of QCD (resummation, evolution equations, \dots).
\\
Taking into account QCD corrections different scale appear in the
process: $m_t\sim m_W$ is the mass of the heavy (virtual)
particles in the loop, $m_b$, the $b$ quark mass, is the hard
scale of the process, $\Lambda_{QCD}$ is the scale where non
perturbative strong effects appear and they can be ordered as
\begin{equation}
m_t \sim m_W \gg m_b \gg \Lambda_{QCD}
\end{equation}
The interplay between these scales makes the process difficult to
handle: they can be disentangled by building an effective theory
where the heavy particles are integrated out and an effective low
energy hamiltonian is considered, like in Fermi hamiltonian for
$\beta$ decays.
\\
In the next section such an hamiltonian will be briefly described.

\section{Effective Hamiltonian for $b\rightarrow s\gamma$}
QCD corrections to the vertex in figure (\ref{vertice}) heavily
change the lowest order result\footnote{They enhance the rate of a
factor 5: in practice this is a peculiar case where perturbative
corrections are more relevant than the lowest order \cite{bmmp}.}
and need to be handled carefully: due to the presence of different
energy scales large logarithms of the form
\begin{equation}\label{logs}
\alpha^n_S \ \log^p(m^2_W/m^2_b)  \ \ \ \rm with \ \ p\le n
\end{equation}
make their appearance at higher orders of perturbation theory.
These logarithms are significant because $m_W\gg m_b$ and the term
$\frac{\alpha_S}{\pi}\log m^2_W/m^2_b$ is too large to be a good
expansion parameter. Terms such as (\ref{logs}) can be resummed
with Renormalization Group Equations to take them under control.
\\
This can be done by building an effective hamiltonian of the
form\footnote{Small CKM matrix elements are neglected.}
\cite{gsw2,gsw}
\begin{equation}\label{hamiltoniana}
{\cal H}_{eff}(x) = -{4G_F \over \sqrt{2}} \ V^*_{ts}V_{tb} \
\sum_{j=1}^8 C_j(\mu_b) \ \hat{\cal O}_j(x;\mu_b).
\end{equation}
This way long-distance effects, both perturbative and
non-perturbative, are factorized in the matrix elements of the
operators $\hat{\cal O}_j$, while the short-distance effects are
contained in the coefficient functions $C_j(\mu_b)$, calculable in
perturbation theory.
\\
In order to build the effective hamiltonian, the heavy particles
in the loop, the $W$ boson and the $top$ quark, have to be
integrated out, so that they do not appear as physical degrees of
freedom. The integration of the two heavy particles at the same
time corresponds to neglect strong coupling running effects
between $m_t$ and $m_W$.
\\
Then the large logarithms are included in the coefficient
functions $C_j(\mu)$ by scaling down the coefficients to
$\mu=O(m_b)$. Logarithms can be resummed in the coefficients
$C_j(\mu_b)$ with the required accuracy: for example the leading
approximation involves the resummation of terms with $n=p$, the
next-to-leading one terms with $p=n-1$ and so on. This is
performed in \cite{gsw2,bkp, cmm,completenlo,altri} in
$\overline{MS}$ scheme.
\\
At the end, an hamiltonian describing a 5 quark theory is
obtained.
\\
A suitable basis for the operators $\hat{\cal O}_j$ is given by
six four-quark operators,the current-current operators $\hat{\cal
O}_1$, $\hat{\cal O}_2$ and the so called penguin operators
$\hat{\cal O}_3-\hat{\cal O}_6$, and by the magnetic penguin
operators (electromagnetic) $\hat{\cal O}_7$ and (chromomagnetic)
$\hat{\cal O}_8$ \cite{gsw2, gsw}:
\begin{eqnarray}\label{base2}
\hat{\cal O}_1 &=& (\overline{c}_{L,\beta} \gamma_\mu
b_{L,\alpha})(\overline{s}_{L,\alpha} \gamma_\mu
c_{L,\beta})\nonumber\\
\hat{\cal O}_2 &=& (\overline{c}_{L,\alpha} \gamma_\mu
b_{L,\alpha})(\overline{s}_{L,\beta} \gamma_\mu
c_{L,\beta})\nonumber\\
\hat{\cal O}_3 &=& (\overline{s}_{L,\alpha} \gamma_\mu
b_{L,\alpha})(\sum_q \overline{q}_{L,\beta} \gamma_\mu
q_{L,\beta})\nonumber\\
\hat{\cal O}_4 &=& (\overline{s}_{L,\alpha} \gamma_\mu
b_{L,\beta})(\sum_q \overline{q}_{L,\beta} \gamma_\mu
q_{L,\alpha})\nonumber\\
\hat{\cal O}_5 &=& (\overline{s}_{L,\alpha} \gamma_\mu
b_{L,\alpha})(\sum_q \overline{q}_{R,\beta} \gamma_\mu
q_{R,\beta})\nonumber\\
\hat{\cal O}_6 &=& (\overline{s}_{L,\alpha} \gamma_\mu
b_{L,\beta})(\sum_q \overline{q}_{R,\beta} \gamma_\mu
q_{R,\alpha})\nonumber\\
\hat{\cal O}_7 &=& {e\over 16\pi^2}m_{b,\overline{MS}}(\mu_b)
\overline{s}_{L,\alpha}\sigma^{\mu\nu}b_{R,\alpha}F_{\mu\nu}\nonumber\\
\hat{\cal O}_8 &=& {g\over 16\pi^2}m_{b,\overline{MS}}(\mu_b)
\overline{s}_{L,\alpha}\sigma^{\mu\nu}T^a_{\alpha\beta}b_{R,\alpha}G^a_{\mu\nu},
\end{eqnarray}
where $m_{b,\overline{MS}}(\mu_b)$ is the $b$ mass in the
$\overline{MS}$ scheme, evaluated at $\mu_b$ and $q=u,d,s,c$ or
$b$. In the operators $\hat{\cal O}_7$ and $\hat{\cal O}_8$ a term
proportional to $m_s$ is present, but we have omitted it because
$m_s\ll m_b$ and the approximation $m_s\rightarrow 0$ is
justified.
\\
For several reasons a different choice for the basis can be
performed and we will follow the definition given in \cite{cmm}.
This basis reads
\begin{eqnarray}\label{base}
\hat{\cal O}_1 &=& (\overline{s}_{L} \gamma_\mu T^a
c_{L})(\overline{c}_{L} \gamma^\mu T^a
b_{L})\nonumber\\
\hat{\cal O}_2 &=& (\overline{s}_{L} \gamma_\mu
c_{L})(\overline{c}_{L} \gamma^\mu
b_{L})\nonumber\\
\hat{\cal O}_3 &=& (\overline{s}_{L} \gamma_\mu b_{L})(\sum_q
\overline{q} \gamma^\mu
q)\nonumber\\
\hat{\cal O}_4 &=& (\overline{s}_{L} \gamma_\mu T^a b_{L})(\sum_q
\overline{q} \gamma^\mu T^a
q_{L})\nonumber\\
\hat{\cal O}_5 &=& (\overline{s}_{L}
\gamma_{\mu_1}\gamma_{\mu_2}\gamma_{\mu_3} b_{L})(\sum_q
\overline{q}\gamma^{\mu_1}\gamma^{\mu_2}\gamma^{\mu_3}
q)\nonumber\\
\hat{\cal O}_6 &=& (\overline{s}_{L}
\gamma_{\mu_1}\gamma_{\mu_2}\gamma_{\mu_3} T^a b_{L})(\sum_q
\overline{q} \gamma^{\mu_1}\gamma^{\mu_2}\gamma^{\mu_3}T^a
q)\nonumber\\
\hat{\cal O}_7 &=& {e\over 16\pi^2}m_{b,\overline{MS}}(\mu_b)
\overline{s}_{L,\alpha}\sigma^{\mu\nu}b_{R,\alpha}F_{\mu\nu}\nonumber\\
\hat{\cal O}_8 &=& {g\over 16\pi^2}m_{b,\overline{MS}}(\mu_b)
\overline{s}_{L,\alpha}\sigma^{\mu\nu}T^a_{\alpha\beta}b_{R,\alpha}G^a_{\mu\nu}.\nonumber\\
\end{eqnarray}
It is important to underline that (\ref{base2}) and (\ref{base})
are equivalent from a physical point of view and can be turned one
into other with proper transformations.
\\
The dimension of these operators is six: higher-dimension
operators have coefficients suppressed by inverse powers of the
masses of the integrated particles ($t$ and $W$) and do not
contribute in first approximation.

\section{Lowest Order Amplitude for $b\rightarrow s\gamma$}
Now to calculate the inclusive rate for $b\rightarrow s\gamma$ one
has to evaluate the matrix elements of the operators of the basis
(\ref{base}) between external states.
\\
Let us consider at first the lowest level\footnote{In this case
the expression \it Born \rm amplitude to indicate the lowest order
amplitude could be misleading because, as we saw, the process is
loop mediated even in the lowest approximation.}. The matrix
element for a generic operator $\hat{\cal O}_i$ is
\begin{equation}\label{elmatrice}
<s\gamma|\hat{\cal O}_i|b>.
\end{equation}
A straightforward calculation shows that the operators $\hat{\cal
O}_1$-$\hat{\cal O}_6$ have vanishing matrix elements. Moreover
the operator $\hat{\cal O}_8$ at the lowest level describes the
process $b\rightarrow s~g$, so that it enters only in radiative
corrections.
\\
At the lowest level the only operator giving contribution is
$\hat{\cal O}_7$, whose matrix element reads
\begin{equation}\label{elmato7}
<s\gamma|\hat{\cal O}_7|b>=m_b
\frac{e}{8\pi^2}\overline{u}(P)\epsilon\!\!/q\!\!\!/
\frac{1+\gamma_5}{2}u(p)
\end{equation}
where $\epsilon$ is the photon polarization, $q^\mu$ the photon
momentum, $P^\mu$ and $p^\mu$ are respectively the heavy and light
quark momenta.
\\
The inclusive rate is
\begin{equation}\label{rate}
\Gamma_0= \ |V^*_{ts}V_{tb}|^2 \ \frac{4G_F}{m_b} \int_\Phi \
|<s\gamma|\hat{\cal O}_7|b>|^2 \  d\Phi
\end{equation}
Let us perform the calculation in the heavy quark rest frame,
where $$P^\mu=(m_b,\vec{0}),$$ for a massless \it strange \rm
quark.
\\
The inclusive rate turns out to be
\begin{equation}\label{born}
\Gamma _0\simeq \frac{\alpha _{em}}{\pi
}\frac{G_{F}^2\,m_{b}^{3}m_{b, \overline{MS}}^{2}\left(
m_{b}\right) \,|V_{tb}V_{ts}^{\ast }|^{2}}{32\pi
^{3}}C_{7}^{2}\left( \mu_{b}\right),
\end{equation}
As we pointed out, the parameter $m_b$, appearing in physical
results, should be carefully considered: the lowest order rate
contains both $m_b=m_{b,POLE}$, coming from the phase space, and,
in principle, $m_{b,\overline{MS}}(\mu)$, coming from the
effective (renormalized) operator \cite{ali2}. However, as long as
QCD corrections are not concerned, the renormalization point is
not important and we can safely choose $\mu=m_b$. As we will see,
when strong interaction are calculated, the definition of the $b$
mass in the $\overline{MS}$ scheme must be coherent with the point
where the coefficient functions are evaluated.

\section{QCD Radiative Corrections for $b\rightarrow s\gamma$}
QCD corrections to the process we are considering involve the
calculation of Feynman diagrams for real and virtual gluon
emissions.
\\
Once the effective hamiltonian is built, one can calculate the
n-th order QCD correction to the rate in perturbation theory, by
evaluating the contributions of gluon emissions, that is processes
such as
\begin{equation}
b\rightarrow s\gamma \ g_1 \dots g_n.
\end{equation}
This can be accomplished by evaluating the coefficient functions
$C_j(\mu)$ at the order $\alpha_S^n$ and the matrix elements
\begin{equation}
<s\gamma g_1 \dots g_n|\hat{O}_j|b>
\end{equation}
for the effective operators. To obtain sensible physical results,
these terms must be coherently evaluated in the same
renormalization scheme, for example $\overline{MS}$, as usually
done in literature. In particular $\alpha_S$ corrections to the
inclusive rate of $b\rightarrow s\gamma$ are calculated in
\cite{pott} for gluon brehmsstrahlung and in \cite{greub,greub2}
for virtual diagrams.
\\
Not all the operators of the basis (\ref{base}) give a
contribution as relevant as the others to the inclusive rate: it
turns out that $\hat{\cal O}_7$ is responsible about for 85\% of
the total rate and the remaining contributions are mainly due to
the operators $\hat{\cal O}_2$ and $\hat{\cal O}_8$. In practice
the other operators can be neglected because they are associated
to very small coefficient functions or give vanishing amplitudes.
One has just to pay some attention to the operators $\hat{\cal
O}_5$ and $\hat{\cal O}_6$, giving non negligible terms, which can
be taken into account with a redefinition of the coefficients
functions of $C_7$ and $C_8$ \cite{pott,greub,greub2}
\begin{eqnarray}\label{ceffective}
\tilde{C}_{1\dots 6}&=&C_{1\dots 6}\nonumber\\
\tilde{C}_7&=&C_7-\frac{1}{3}C_5-C_6\nonumber\\
\tilde{C}_8&=&C_8+C_5.
\end{eqnarray}
In this way one can mainly consider the operator $\hat{\cal O}_7$
and smaller terms given by $\hat{\cal O}_2$ and $\hat{\cal O}_8$.
\\
The evolution of the coefficient functions $C_i^{eff}(\mu)$ is
governed by the renormalization group equation
\begin{equation}
\mu\frac{d}{d\mu}C_i^{eff}(\mu)=\gamma_{ij}^{eff}(\alpha_S)C_j^{eff}
\end{equation}
where $\gamma_{ij}^{eff}$ is the anomalous dimension matrix
related to the effective coefficient functions defined in
(\ref{ceffective}). In recent years, efforts to achieve a
next-to-leading accuracy has been done, to reduce the
renormalization scale dependence of the amplitude, which was found
to be the largest source of uncertainty \cite{bmmp}. This request
needs the knowledge of the (effective) coefficient functions up to
$O(\alpha_S)$
\begin{equation}
C_i^{eff}=C^{(0)eff}_i(\mu)+\frac{\alpha_S(\mu)}{4\pi}C^{(1)eff}_i(\mu)+O(\alpha_S)
\end{equation}
and the anomalous dimension matrix up to $O(\alpha_S^2)$
\begin{equation}
\gamma_{ij}^{eff}=\frac{\alpha_S}{4\pi}\gamma^{(0)eff}_{ij}+\left(\frac{\alpha_S(\mu)}{4\pi}\right)^2\gamma^{(1)eff}_{ij}+O(\alpha^3_S)
\end{equation}
This huge work has been performed by different groups in the last
years and final results are listed in \cite{completenlo,cmm} and
in references therein.
\\
Further details about the QCD corrections are discussed in the
chapters \ref{resumtmd} and \ref{fixedorder}, where explicit
calculations are shown. Here let us finally recall the result for
the contribution of the leading operator $\hat{\cal O}_7$ to the
order $\alpha_S$ as calculated in \cite{pott,greub,greub2}:
\begin{eqnarray}\label{rateoneloop}
\Gamma_7&=&\Gamma_0\tilde{C}_7^2(\mu)\left[1+\frac{\alpha_S}{\pi}\left(\frac{16}{9}-\frac{4}{9}\pi^2+\frac{16}{3}\log\frac{m_b}{\mu}\right)\right]=\nonumber\\
\Gamma_7&=&\Gamma_0\tilde{C}_7^2(\mu)\left[1+\frac{\alpha_S}{\pi}\left(\frac{16}{9}-\frac{4}{9}\pi^2+\frac{4}{3}\log\frac{m_b}{\mu}\right)\right]\left(\frac{m^2_{b,\overline{MS}}(\mu)}{m^2_{b,\overline{MS}}(m_b)}\right).\nonumber\\
\end{eqnarray}
The two expressions in (\ref{rateoneloop}) reduce one another
because of the relation
\begin{equation}
m_{b,\overline{MS}}(\mu)=m_{b,\overline{MS}}(m_b)\left(1+\frac{2\alpha_S(m_b)}{\pi}\log\frac{m_b}{\mu}\right).
\end{equation}
and can be chosen depending which mass we want in the lowest order
amplitude, $m_{b,\overline{MS}}(\mu)$ or
$m_{b,\overline{MS}}(m_b)$.

\section{Photon Spectrum}
The spectrum of the photon is the first distribution calculated
and measured for $b\rightarrow s\gamma$.
\\
The kinematics of this process is very simple: at the lowest level
it is a two body decay, so that the photon energy is fixed by the
kinematics: in the approximation of a massless \it strange \rm
quark  the spectrum is discrete and monochromatic and the photon
energy is:
\begin{equation}
E_\gamma=\frac{m_b}{2}.
\end{equation}
This approximation is obviously too na\"{i}ve, because it does not
take into account two fundamental effects:
\begin{itemize}
\item
gluon emissions radically change the spectrum, making the process
a multi-body decay: the spectrum becomes continuous
\begin{equation}
0\le E_\gamma\le E_\gamma^{MAX}=\frac{m_b}{2}
\end{equation}
QCD corrections have to calculated in perturbation theory. As
discussed in Chapter \ref{basic} an improved perturbative
expansion is needed to have reliable results in the whole phase
space;
\item
non perturbative effects play a role: for small exchanged momenta,
non perturbative effects make their appearance, related to final
states effects (\it hadronization\rm ) and initial state effects
(\it Fermi motion\rm). Obviously they cannot be calculated in
perturbation theory and need to be factorized in some
fragmentation or structure function. They will be discussed in
Chapter \ref{compare}. They spread the spectrum around the
endpoint.
\end{itemize}
The photon spectrum was first calculated in \cite{ali} for a
massive \it strange \rm quark. A comparison with the data present
at the time is performed in \cite{ali2}, where a model to deal
with the Fermi motion effects is introduced. The comparison, used
to extract the ratio $\left| \frac{V_{ts}}{V_{cb}}\right|$, agrees
with the prediction of the Standard Model, based on the unitarity
triangle. Latest experimental results from CLEO collaboration are
shown in \cite{cleo}.
\begin{figure}[h]
\begin{center}
\mbox{\epsfig{file=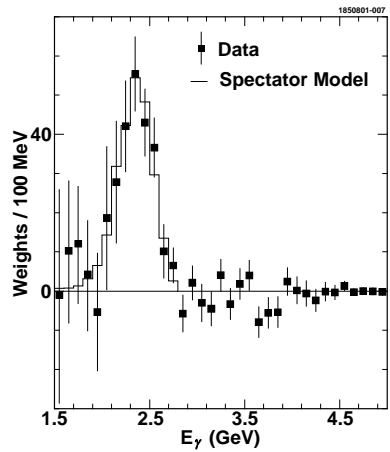, height=6cm}}
\caption{\label{spectrum}Photon spectrum in $b\rightarrow s\gamma$
measured at CLEO \cite{cleo}.}
\end{center}
\end{figure}
Other details about the photon spectrum will be discussed in
section (\ref{threshold}).
\\
Let us only remark that this is a distribution where large
logarithms appear near the border of the phase space, related to
emission of infrared gluons. This happens in the region near the
endpoint of the photon spectrum, for $E_\gamma\rightarrow m_b/2$.
These large logarithms spoil the perturbative expansion in that
region and the usual approach to handle them is their resummation
according to the theory described in section
(\ref{secresummation}).
\\
In section (\ref{resumtmd}) this will be shown and the limits of
this approach will be described.
\\
To conclude let us underline that the photon spectrum has been the
first distribution studied for the process we are dealing with:
the photon spectrum is sensitive to longitudinal degrees of
freedom, along the photon direction. On the contrary the
transverse momentum distribution of the strange quark with respect
to the photon direction is sensitive to transverse degrees of
freedom, so that it can give complementary information about the
dynamics of the process.
\\
In Chapter \ref{compare} the photon spectrum and the transverse
momentum distribution will be compared: being sensitive to
different degrees of freedom they can give complementary
informations about several features of $b$ physics, such as non
perturbative effects, reliability of the perturbative expansion
and so on.

\part{Transverse Momentum Distribution in $b\rightarrow s\gamma$}
\chapter{Preliminary Observations}\label{introtmd}

\section{Motivations}
In the next chapters the calculations concerning the transverse
momentum distribution in the rare decay $b\rightarrow s\gamma$
will be discussed and described in great detail. This topic was
introduced in \cite{noi1} and the calculation was completed in
\cite{noi2}: here we want to underline the motivations which
induce us to analyze this argument and its relevance in the
outline of $b$ physics.
\\
In the last years experiments dedicated to the study of $b$
physics have been set up: in particular BaBar at Fermilab and
Belle at KEK began to carry out a wide program of measurements of
parameters involved in the physics of this quark. The rich
phenomenology of the $b$ physics opens many possibilities both
experimentally and theoretically: the final aim remains a precise
test of the Standard Model, with the determination of every
parameter involved in the predictions of the theory and a
compelling comparison with experimental data. This can show the
level of accuracy of the theory.
\\
Since the discovery of the $b$ quark the importance of QCD
corrections in processes involving it was evident: the tree level
process is always mediated by an electroweak current, but
emissions of gluon by initial or final quark change radically the
theoretical prediction. The most manifest case is just related to
$b\rightarrow s\gamma$ which is one of the few processes where
radiative corrections have a larger size than the lowest order.
\\
Perturbative QCD was widely applied to the $b$ physics in past
years and in particular also to its decays: the $b$ quark is the
lightest quark where this kind of approach can be considered
viable. Its mass, as discussed in the chapter \ref{capbsgamma}, is
larger than the typical scale of hadronization, see eq.
(\ref{massablambda}), and the strong coupling constant $\alpha_S$
can be considered a good parameter for the perturbative expansion,
being reasonably smaller than one \cite{pdg}:
\begin{equation}
\alpha_S(m_b)\cong 0.22.
\end{equation}
For lighter quark these considerations are not valid: even the
quark $charm$ is too light for a reasonable application of pQCD
without heavy corrections.
\\
This has driven to consider perturbative QCD as an essential tool
in $b$ decays: in particular also more refined tools of the theory
have been applied to this kind of processes, for example the
resummation of large logarithms for semi-inclusive observables has
become a standard technique.
\\
In the last twenty years QCD has had amazing confirmations of its
predictions in experiments of high energy physics, for example at
LEP: however in these cases the scale of energy where the
processes occurred was much higher with respect to the $b$ decays.
Typically these experiments where performed at scales larger than
$Z^0$ mass,
\begin{equation}
M_{Z^0}\cong 91 {\rm GeV},
\end{equation}
where the strong coupling constant is about an half the coupling
constant at the $b$ mass energy:
\begin{equation}
\alpha_S(M_{Z^0})=0.12.
\end{equation}
Moreover in $b$ physics theoretical prediction are complicated by
the presence of parameters in principle unknown and that cannot be
extracted by the theory: in particular the CKM matrix elements and
the quark mass which, as discussed in Chapter \ref{capbsgamma},
are poorly known. These parameters are affected by uncertainties
which can confuse QCD predictions. They are usually eliminated by
considering ratio of widths of different processes, which cancel
the dependance on the quark mass and in several cases also on CKM
matrix elements. A rather good theoretical quantity is
represented, for instance, by  the semileptonic branching ratio:
\begin{equation}\label{slbf}
B_{SL}=\frac{\Gamma_{SL}}{\Gamma_{TOT}},
\end{equation}
which turns out to be marginally in agreement with present data
\cite{altpet}.
\\
These reasons drove us to calculate with high accuracy
(next-to-leading level resummation + fixed order hard
contributions) a distribution in a decay of the $b$ quark.
\\
The second question to answer is why we decided to study this
particular decay, $b\rightarrow s\gamma$: as stated in chapter
\ref{capbsgamma} this is a rare decay, has a small branching
ratio, presents many experimental difficulties for the
measurements and therefore seems not to be the most immediate
choice. However the reasons we have studied it is that has
interesting and suitable features: being at the lowest level a two
body decay its kinematics is very simple, so that one can face the
problem of radiative corrections and resummation without
additional and unimportant complications. Moreover the results are
in many cases very general and can be extended to other processes
such as the semileptonic decay $b\rightarrow u+e+\overline{\nu}_e$
or the phenomenologically more important $b\rightarrow
c+e+\overline{\nu}_e$, where the mass of the final quark is a big
kinematical complication.
\\
The tool of resummation was mainly applied, in past years, to
processes involving light quarks only, while in this case the
presence of an heavy quark in the initial states gives additional
and new effects.
\\
In our case this scheme is justified by the fact that the double
logarithm appearing to order $\alpha_S$ can become rather large
(with respect to 1 coming from the tree level):
\begin{equation}
-\frac{\alpha_S C_F}{4\pi} \log^2\frac{p_t^2}{m_b^2} \sim -0.7
\end{equation}
if we push the transverse momentum to such small values as
$p_t~\sim~\Lambda_{QCD}~=~300$ MeV. The single logarithm can also
become rather large, having a large numerical coefficient:
\begin{equation}
-\frac{5\alpha_S C_F}{4\pi} \log\frac{p_t^2}{m_b^2} \sim 0.6.
\end{equation}
The purpose of resumming classes of such terms therefore seems
quite justified. If we consider running coupling effects, i.e. if
the (frozen) coupling evaluated at the hard scale $Q~=~m_B~=~5.2$
GeV is replaced by the coupling evaluated at the gluon transverse
momentum,
\begin{equation}
\alpha_S(m_b) \rightarrow
\alpha_S(p_t)=0.45~~~~~~{\rm{for}}~~~~~p_t~=~1~\rm{GeV},
\end{equation}
the logarithmic terms have sizes of order:
\begin{equation}
-\frac{\alpha_S(p_t) C_F}{4\pi} \log^2\frac{p_t^2}{m_b^2} \sim
-0.5
\end{equation}
and
\begin{equation}
-\frac{5\alpha_S C_F}{4\pi} \log\frac{p_t^2}{m_b^2} \sim 0.8.
\end{equation}
The main difference with respect to resummation in $Z^0$ decays is
a hard scale smaller by over an order of magnitude, i.e. a
coupling larger by a factor 2 and infrared logarithms smaller by a
factor 3.
\\
Finally this decay, even if rare, has been widely studied
experimentally because it seemed to be a privileged process for
the discovery of new physics: matter of fact all the discrepancies
between experimental data and theoretical predictions could be
explained and up to now no signals of new physics have been found.
However the large effort to measure this process have produced a
huge amount of experimental data, which will be possibly compared
with our predictions.
\\
The last question to make clear is why we decide to study the
transverse momentum distribution: the main reason is that for this
process the photon spectrum has been calculated, measured and the
role of non perturbative contributions has been discussed, with
the introduction of a structure function, called \it shape
function \rm \cite{generale}. The transverse momentum is sensitive
to a complementary kinematics with respect to the photon spectrum,
so that it can give additional informations about the process.
\\
In the following chapters the calculation of the transverse
momentum distribution will be discussed: according to the theory
of resummation we decided to achieve a next-to-leading accuracy in
the calculation, for the reasons discussed in section
(\ref{secresummation}).
\\
This level of accuracy requires the calculation of the coefficient
controlling double and single logarithmic effects and the
resummation of logarithms in the function $g_1$ and $g_2$
introduced in section (\ref{secresummation}). This part, the
next-to-leading-logarithmic approximation (NLL), is described in
Chapter \ref{resumtmd}.
\\
In order to achieve a complete next-to-leading order accuracy
(NLO), hard terms to the order $\alpha_S$ have been calculated in
Chapter \ref{fixedorder}, which shows the computation of real and
virtual Feynman diagrams to the order $\alpha_S$.
\\
The resummation of the transverse momentum show different results
with respect to the photon spectrum and a comparison between these
two complementary distributions is performed in chapter
\ref{compare}.

\section{Kinematics}\label{tmdkinematics}
The specific process we are considering is
\begin{equation}\label{processoadronico}
B\rightarrow h_s+X+\gamma,
\end{equation}
where $h_s$ is a strange hadron, typically a $K^\ast$, and $X$
every other hadronic state. Without any generality loss we can
consider the B-meson rest frame and define as reference direction
the direction of flight of the photon (axis $+z$).
\\
Let us define the transverse momentum $p_t$ of the strange hadron
with respect of this direction: it is in practice the projection
of the momentum of $h_s$ on the plane $x-y$.
\begin{figure}[h]
\begin{center}
\mbox{\epsfig{file=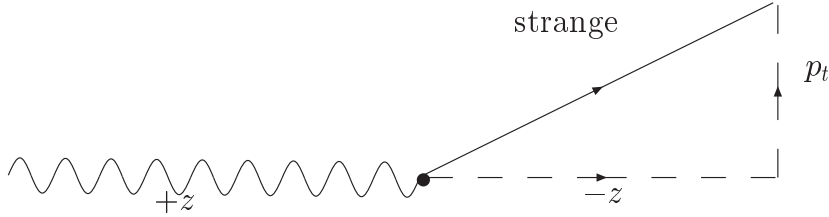, width=12cm}}
\caption{Kinematics of the transverse momentum distribution.}
\end{center}
\end{figure}
At the lowest order of the partonic level the process is
\begin{equation}
b\rightarrow s+\gamma,
\end{equation}
so that, being a two body decay, the photon and the strange quark
are emitted along the same direction and the transverse momentum,
as defined above, vanishes.
\\
However it should be reminded that the physical process that can
be detected and measured is the hadronic one
(\ref{processoadronico}) and radiative corrections change
radically the spectrum at the partonic level, so that different
sources of transverse momentum can be identified:
\begin{itemize}
\item  {\it Multiple gluon emissions}: the emission of gluons by the
initial or the final quark makes the process a multi-body decay
\begin{equation}
b\rightarrow s\gamma g_1\dots g_n
\end{equation}
and spreads the transverse momentum spectrum. This effect can be
calculated order by order in perturbation theory and produce a
transverse momentum $p_p$ for the strange quark because of
momentum conservation.
\begin{figure}[h]
\begin{center}
\mbox{\epsfig{file=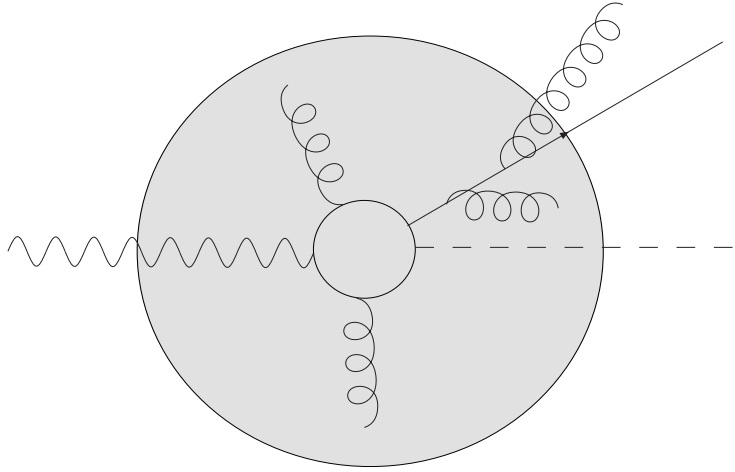, width=10cm}} \caption{Multiple
gluon emission from the initial or the final quark in the heavy
quark-meson rest frame.}
\end{center}
\end{figure}
\item {\it Initial state effects}: a  first non-perturbative effect is the well-known Fermi
motion \cite{generale}. Classically, this is a small vibratory
motion of the $b$ quark in the $B$ meson, caused by the momentum
transfer with the valence quark. In the quantum theory, the
interaction with the light quark produces also some virtuality of
the heavy quark. Let us work in the $B$ rest frame. We can
parametrize the heavy quark momentum as:
\begin{equation}
p_b = m_{B} v+p_{f}.
\end{equation}
$v^{\mu }$ is the velocity of the heavy quark,
\begin{equation}
v^{\mu } =(1;0,0,0)
\end{equation}
and $p_{f}$ is the residual momentum associated to the Fermi
motion, such that
\begin{equation}
|p_{f}^{\mu }|\sim \Lambda.
\end{equation}
where $\Lambda $ is the QCD scale. We expect that Fermi motion
produces a distribution of intrinsic transverse momenta of the $b$
quark of order
\begin{equation}
|p_{f}| \sim \Lambda .  \label{fermimom}
\end{equation}
Roughly speaking, we may say that transverse momenta distributions
in the region
\begin{equation}
|p_t|\sim \Lambda   \label{nonpert}
\end{equation}
are sensitive to the oscillations of the heavy quark in the
transverse plane (x , y).
\begin{figure}[h]
\begin{center}
\mbox{\epsfig{file=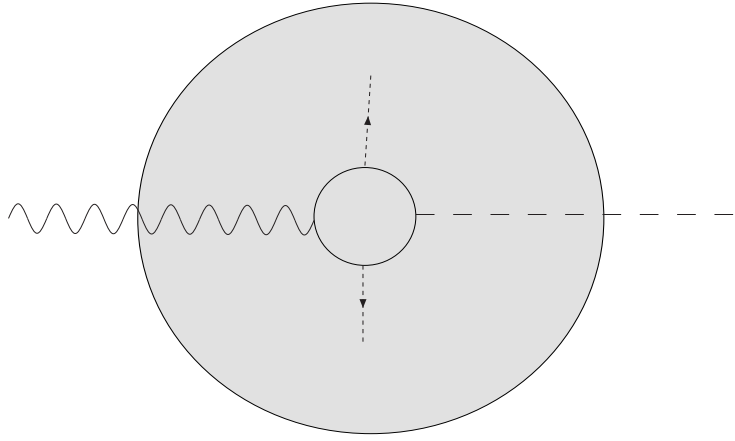, width=10cm}} \caption{Sketch of
Fermi Motion effects on the transverse momentum.}
\end{center}
\end{figure}
\item {\it Final state effects}: a second non-perturbative effect is the fragmentation of the
strange quark into the hadron $h_{s}.$ According to the idea of
local parton-hadron duality, hadronization is expected to modify
the partonic distribution by a quantity of the order of the QCD
scale:
\begin{equation} \label{hadrmom}
|p_{h}|\sim \Lambda .
\end{equation}
\begin{figure}[h]
\begin{center}
\mbox{\epsfig{file=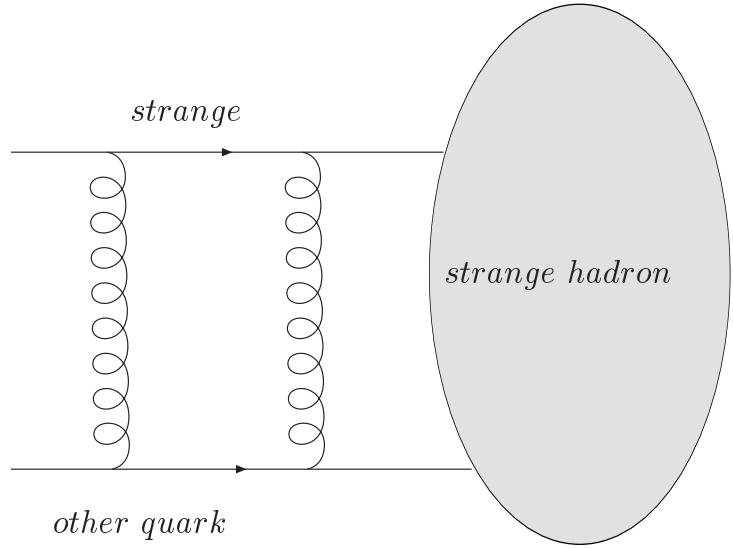, width=10cm}} \caption{Sketch of
the effects of the hadronization on the transverse momentum.}
\end{center}
\end{figure}
\end{itemize}
The total transverse momentum can be written as the sum of the
perturbative and non-perturbative components:
\begin{equation}
p_{t} = p_p + p_{f} +p_{h}.
\end{equation}
and the distribution is controlled by perturbative effects only as
long as $|p_t| \rm \gg \Lambda_{QCD}$.
\\
Let us note that initial and final state effects involve momenta
of the same order and in principle they cannot be separated by any
energy scale.
\subsection{Kinematics for Single Gluon
Emission}\label{seckinungluone} 
\begin{figure}[h]
\begin{center}
\mbox{\epsfig{file=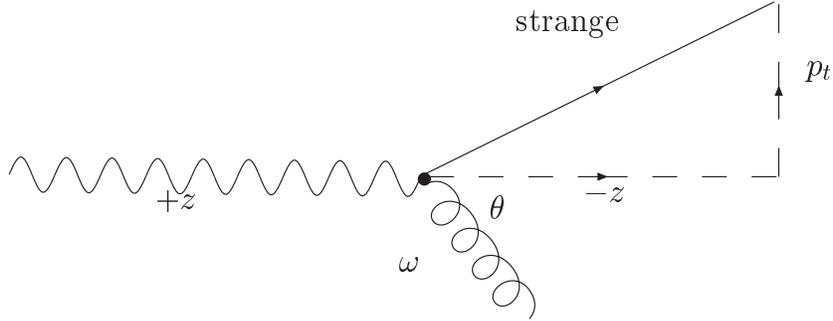, width=12cm}}
\caption{\label{figonegluon}Kinematics for one gluon emissions.}
\end{center}
\end{figure}

Let us consider the kinematics at
the partonic level for one gluon emission, described in figure
(\ref{figonegluon}).
\\
The mass of the strange quark can be neglected.
\\
Let us define the fraction of energy of the gluon as
\begin{equation}
\omega=\frac{2E_g}{Q}=\frac{2E_g}{m_b},
\end{equation}
being the quark mass $m_b$ the hard scale of the process.
\\
Let us define as angular variable
\begin{equation}
t=\frac{1-\cos\theta}{2},
\end{equation}
where $\theta$ is defined as the angle between the opposite of the
photon direction (axis -z) and the strange quark direction, as
shown in the figure.
\\
$\omega$ and $t$ are unitary variables and the Born kinematics
(vanishing transverse momentum) is achieved for $\omega\rightarrow
0$ or $t\rightarrow 0$.
The transverse momentum $p_t$ (since now this symbols will denote
the modulus of the transverse momentum) of the strange quark can
be expressed in terms of these two unitary variables: since
$\vec{p}_t=-\vec{k}_t$, being $\vec{k}_t$ the transverse momentum
of the gluon, because of momentum conservation, it can be obtained
from geometrical considerations that
\begin{equation}
p^2_t=k^2_t=m^2_b\left( \omega^2 t(1-t) \right).
\end{equation}
Let us define the adimensional variable
\begin{equation}
x=\frac{p^2_t}{m^2_b},
\end{equation}
so that the kinematical constraint (see section (\ref{seclogs}))
turns out to be
\begin{equation}\label{vincolox}
\delta\left[x-o(\omega,t)\right]=\delta\left[x-\omega^2
t(1-t)\right].
\end{equation}
This kinematical constraint is infrared safe, according to the
definition given in section (\ref{seclogs}), since it vanishes in
the soft and in the collinear limit:
\begin{eqnarray}
o(\omega,t)\rightarrow 0  \ \ \ &{\rm for}& \ \ \
\omega\rightarrow 0\nonumber\\
o(\omega,t)\rightarrow 0  \ \ \ &{\rm for}& \ \ \ t\rightarrow 0.
\end{eqnarray}
It turns out from kinematics that the limits on $x$ are
\begin{equation}
0\leq x \leq \frac{1}{4}.
\end{equation}

\chapter{Resummation of Transverse Momentum Distribution}\label{resumtmd}
This Chapter is dedicated to the resummation with next-to-leading
order accuracy of the large logarithms appearing in the transverse
momentum distribution near the border of the phase space.
\\
The first step is the calculation of the coefficients controlling
the double and single logarithms.
\\
Then they are resummed to every order by introducing the impact
parameter space \cite{pp}, according to the theory outlined in
section (\ref{secresummation}).

\section{Calculation of the Coefficients $A_1$ and $B_1$}
In this section the distribution of transverse momentum of the
\it{strange} \rm quark with respect to the photon direction is
calculated with logarithmic accuracy for one gluon emission. The
partonic process is
\begin{equation}\label{onegluon}
b \rightarrow s + \gamma + g
\end{equation}
This computation involves the coefficients $A_1$ and $B_1$ which
control the leading and next-to-leading contributions in the
resummed distribution.
\\
As defined in section (\ref{seckinungluone}) we consider the
variable :
\begin{equation}
x\equiv {\frac{p_t^2}{Q^2}}\hskip 1cm ,\hskip 1cm Q=m_b
\end{equation}
The distribution of interest is therefore:
\begin{equation}\label{distribuzione}
d\left( x\right) =\frac{1}{\Gamma _{0}}\frac{d\Gamma }{dx}.
\end{equation}
The kinematical constraint to impose on the radiative gluon is
given in (\ref{vincolox})
$$ o(\omega,t)=\omega^2 t(1-t),$$
so that $p_t$ distribution  is infrared safe.
\\
The aim of this section is the determination of the logarithmic
structure at one loop: in order to extract the logarithms arising
at this stage the kinematical constraint (\ref{vincolox}) can be
expressed with its lowest order approximation
\begin{equation}\label{vincoloapprox}
\tilde{o}\left( \omega ,t\right) =\omega ^{2}t.
\end{equation}
Using the decomposition of the matrix element discussed in section
(\ref{seclogs}) (see eq. (\ref{decomposizionematrel})) we can
write to the order $\alpha_S$:
\begin{equation}
d(x)=\delta (x)+\alpha _{S}\int_{0}^{1}d\omega
\int_{0}^{1}dt\left[ {\frac{ A_{1}}{\omega
t}}+{\frac{S_{1}(t)}{\omega }}+{\frac{C_{1}(\omega )}{t}} \right]
\left[ \delta (x-\omega ^{2}t)-\delta (x)\right] . \label{distrib}
\end{equation}
The first $\delta$-function describes the real emissions, while
the second one describes the virtual emissions. The function $F_1$
introduced in (\ref{decomposizionematrel}) can be neglected
because it does not produce any logarithmic enhancement: it will
be taken into account in Chapter (\ref{fixedorder}).
\\
By integrating over the gluon energy and polar angle one obtains
the following contributions \footnote{Let us note that the
contribution in eq.(\ref{double}) is called \it double logarithmic
\rm because it becomes proportional to $\log^2 x$ in the
integrated rate $\int_0^y dx \ {1 \over \Gamma_0} \ {d\Gamma \over
dx}$ . For the same reason eq.(\ref{singlesoft}) and
(\ref{singlecoll}) are called \it single logarithms\rm.}:
\begin{itemize}
\item  leading double-logarithmic term: \newline
\begin{equation}\label{double}
\alpha _{S}A_{1}\int_{0}^{1}\frac{d\omega }{\omega
}\int_{0}^{1}\frac{dt }{t}[\delta (x-\omega ^{2}t)-\delta
(x)]=-{\frac{\alpha _{S}A_{1}}{2}} \left( {\frac{\log
x}{x}}\right) _{+};
\end{equation}
\item  next-to-leading soft term:
\begin{equation}\label{singlesoft}
\alpha _{S}\int_{0}^{1}dt\int_{0}^{1}d\omega
{\frac{S_{1}(t)}{\omega }} [\delta (x-\omega ^{2}t)-\delta
(x)]={\frac{\alpha _{S}S_{1}}{2}}\left( { \frac{1}{x}}\right)
_{+}+\mathcal{O}(1);
\end{equation}
\item  next-to-leading collinear term:
\begin{equation}\label{singlecoll}
\alpha _{S}\int_{0}^{1}d\omega \int_{0}^{1}dt{\frac{C_{1}(\omega
)}{t}} [\delta (x-\omega ^{2}t)-\delta (x)]=\alpha
_{S}\,C_{1}\left( {\frac{1}{x}} \right) _{+}+\mathcal{O}\left(
\frac{1}{\sqrt{x}}\right) .
\end{equation}
\end{itemize}
Let us underline that the collinear term has a different rest with
respect to the soft one: this rest is divergent for $x\rightarrow
0$ even if it does not give rise logarithmic
enhancement\footnote{Let us underline that every contribution of
the rest to D(x), eq. (\ref{distrint}) is finite.}.
\\
By summing the contributions, the one soft-gluon emission
distribution reads:
\begin{equation}\label{onegluonemission}
f(x)=\delta (x)-\frac{\alpha _{S}A_{1}}{2}\left( {\frac{\log
x}{x}}\right) _{+}+\alpha _{S}B_{1}\left( \frac{1}{x}\right) _{+}
\end{equation}
where
\begin{equation}\label{coefficiente}
B_{1}=C_{1}+{\frac{S_{1}}{2}}=-{\frac{5}{4}}{\frac{C_{F}}{\pi }}
\end{equation}
In usual hard processes, such as DIS and DY, $S_1=0$ so that
$B_{1}=C_1$. In our case, single logarithmic effects are more
pronounced because this coefficient is almost factor 2 larger.

\section{Exponentiation in the Impact Parameter Space}\label{multiplesoft}
In this section we are going to deal with the resummation of
multiple soft gluon emission: this is performed by factorizing QCD
amplitudes for multiple soft gluon emission and introducing the
impact parameter space which factorizes the kinematical constraint
for transverse momentum
\cite{pp,kodairatrentadue,cattrendem,russi,curcigreco}. After that
one gets the exponentiation of one gluon distribution.
\\
Before dealing with the general case, let us treat in some detail
the simplest non-trivial case, namely double emission.
\begin{equation}
b\rightarrow s+\gamma +g_{1} + g_{2}
\end{equation}
According to transverse momentum conservation we have
\begin{equation}
p_t=-{k}_{t1}-k_{t2}.
\end{equation}
Multi-gluon amplitudes factorize in the infrared limit at leading
level, so that
\begin{equation}
\frac{1}{\Gamma _{B}}\frac{d^{2}\Gamma _{2}}{d{k}_{t1}d{k}_{t2}}%
\left( {k}_{t1},k_{t2}\right) \simeq \frac{1}{2}\frac{1}{%
\Gamma _{B}}\frac{d\Gamma _{1}\left( {k}_{t1}\right) }{d{k}_{t1}}%
\frac{1}{\Gamma _{B}}\frac{d\Gamma _{1}\left( {k}_{t2}\right) }{d%
{k}_{t2}}.
\end{equation}
where $k_{1}$ and $k_{2}$ are gluons momenta, while $\Gamma_{i}$
is the width of one gluon emission process.
\\
The distribution therefore reads:
\begin{eqnarray}\label{everyorder}
\frac{1}{\Gamma _{0}}\frac{d\Gamma }{dp_t}\left( p_t\right)
&=&\delta \left( p_t\right) +\int dp_{t1}\delta \left(
p_t+k_{t1}\right) \frac{1}{\Gamma_{0}}\frac{d\Gamma_{1}\left(
k_{t1}\right) }{d{k}_{t1}}+  \\
&&+\frac{1}{2}\int d{k}_{t1}k_{t2}\delta \left( p_t+%
{k}_{t1}+{k}_{t2}\right) \frac{1}{\Gamma_{0}}\frac{d\Gamma
_{1}\left( {k}_{t1}\right) }{d{k}_{t1}}\frac{1}{\Gamma _{0}}%
\frac{d\Gamma _{1}\left( {k}_{t2}\right) }{d{k}_{t2}}+\cdots
.\quad
\end{eqnarray}
The quark transverse momentum  distribution is inclusive with
respect to gluon radiation: we do not detect the transverse
momenta of individual gluons, but only their sum, in an indirect
way, by detecting the transverse momentum of the light quark. To
solve the $k_t$-constraint, we go to the distribution in impact
parameter space $b$ by taking a two-dimensional Fourier transform,
\begin{equation}
{1 \over \Gamma_0}\frac{d\widetilde{\Gamma }}{d{b}}\left(
{b}\right) \equiv \int_{-\infty}^{+ \infty} \int_{-\infty}^{+
\infty}d{p_t}\exp \left[ i{p_t}\cdot {b}\right] \,{1 \over
\Gamma_0}\frac{d\Gamma \left( {p_t}\right) }{d{p_t}}.
\label{bparam}
\end{equation}
Inserting eq. (\ref{everyorder}) in the r.h.s. of eq.
(\ref{bparam}), one obtains
\begin{eqnarray}
\frac{1}{\Gamma_{0}}\frac{d\widetilde{\Gamma }}{d{b}}\left( {b%
}\right)  &=&1+\frac{1}{\Gamma_0}\frac{d\widetilde{\Gamma }_{1}}{d{b}}\left( {b}%
\right) +\frac{1}{2}\left[ \frac{1}{\Gamma_0}\frac{d\widetilde{\Gamma }_{1}}{d{b}}%
\left( {b}\right) \right] ^{2}.
\end{eqnarray}
Let us now consider the general case. The transverse momentum
distribution can be written as
\begin{equation}
\frac{1}{\Gamma _{0}}\frac{d\Gamma }{d{p}_t}\left( {p}_t\right)
=\delta \left( {p}_t\right) +\sum_{n=1}^{\infty }\frac{1}{\Gamma _{0}}%
\frac{d\Gamma ^{\left( n\right) }}{d{p}_t}\left( {p}_t\right) ,
\label{basicmult}
\end{equation}
where the distribution due to $n$ real gluons is given by
\begin{equation}
\frac{d\Gamma ^{\left( n\right) }}{d{p}_t}\left( {p}_t\right)
=\int_{-\infty}^{+\infty}\prod_{l=1}^{n}d{k}_{tl}\frac{d^{n}\Gamma \left( {k%
}_{t1},{k}_{t2},\cdots {k}_{tn}\right) }{d{k}_{t1}d{k%
}_{t2}\cdots d{k}_{tn}}\delta \left( {p}_t+{k}_{t1}+{k%
}_{t2}+\cdots +{k}_{tn}\right) .  \label{distribn}
\end{equation}
In the infrared limit, the matrix elements for multiple emission
factorize into the single emission ones, so that
\begin{equation}
\frac{1}{\Gamma _{0}}\frac{d^{n}\Gamma _{n}\left( {k}_{t1},{k}%
_{t2},\cdots {k}_{tn}\right) }{d{k}_{t1}d{k}_{t2}\cdots d%
{k}_{tn}}\simeq \frac{1}{n!}\prod_{l=1}^{n}\frac{1}{\Gamma _{0}}\frac{%
d\Gamma _{1}}{d{k}_{tl}}\left( {k}_{tl}\right) . \label{IRfactor}
\end{equation}
The factorization of the amplitudes represented by eq.\thinspace
(\ref {IRfactor}) is a dynamical property of QCD. By substituting
eq.(\ref {distribn}) into (\ref{IRfactor}) and (\ref{basicmult})
into (\ref{distribn}), we obtain:
\begin{equation}
\frac{1}{\Gamma _{0}}\frac{d\Gamma \left( p_t\right) }{dp_t}%
=\delta \left( p_t\right) \,+\,\sum_{n=1}^{\infty }\,\frac{1}{n!}%
\,\int_{0}^{Q}\,\prod_{l=1}^{n}d^{2}k_{tl}\,\frac{1}{\Gamma _{0}}%
\frac{d\Gamma _{1}\left( k_{tl}\right) }{dk_{tl}}\,\delta \left(
p_t+k_{t1}+k_{t2}+\cdots +k_{tn}\right) .
\end{equation}
Substituting eq.\thinspace (\ref{IRfactor}) in eq.\thinspace
(\ref{bparam}), we obtain the exponentiation of the effective
one-gluon distribution in impact parameter space:
\begin{eqnarray}
\frac{1}{\Gamma_{0}}\frac{d\widetilde{\Gamma }}{d{b}}\left( {b%
}\right)  &=&1+\frac{1}{\Gamma_0}\frac{d\widetilde{\Gamma }_{1}}{d{b}}\left( {b}%
\right) +\frac{1}{2}\left[\frac{1}{\Gamma_0} \frac{d\widetilde{\Gamma }_{1}}{d{b}}%
\left( {b}\right) \right]^{2} +\cdots \nonumber \\
&\stackrel{{\rm i.r. \ limit}}{=}&\exp \left[
\frac{d\widetilde{\Gamma }_{1}}{d{b}}\left( {b}\right) \right] .
\end{eqnarray}
The original distribution in momentum space is recovered by an
inverse transform,
\begin{equation}\label{inverse}
\frac{d\Gamma \left( p_t\right) }{dp_t}=\int \frac{db%
}{4\pi ^{2}}\exp \left[ -ip_t\cdot b\right] \frac{d%
\widetilde{\Gamma }}{db}\left( b\right) .
\end{equation}
Let us stress that the exponentiation of the one-gluon
distribution is a consequence of two different facts:
\begin{enumerate}
\item  Factorization of QCD amplitudes in the infrared limit;
\item  Factorization of the kinematical constraint going to impact parameter
space.
\end{enumerate}

\section{Higher orders}\label{prescrizioni}
In order to achieve a next-to-leading logarithmic resummation
higher order effects have to be taken into account. This can be
performed by considering the two following prescriptions during
the calculation:
\begin{enumerate}
\item  the bare coupling have to be substituted with a running coupling at two loop
level evaluated at transverse momentum squared of the gluon
\cite{abcmv}:
\begin{equation}
\alpha_{S}\rightarrow \alpha_{S}(k_t^{2}={p_t}^2)=
\alpha_{S}(Q^{2}x), \label{copsost}
\end{equation}
where the running coupling constant is evaluated with
next-to-leading accuracy, according to equation (\ref{alfarun}),
\begin{equation}
\alpha^{NLO}_{S}(Q^{2}x)={\frac{1}{\beta _{0}}}{\frac{1}{\log ({s\,}x)}}-{\frac{%
\beta _{1}}{\beta _{0}^{3}}}{\frac{\log [\log ({s\,}x)]}{\log ^{2}({s\,}x)}+O%
}\left( \beta _{1}^{2}\right) {.}  \label{arun}
\end{equation}
$s$ is defined as:
\begin{equation}
s\equiv \frac{Q^{2}}{\Lambda ^{2}}.
\end{equation}
The first two coefficients of the $\beta $-function, $\beta_{0}$
and $\beta_{1},$ are independent of the renormalization scheme
(they are universal) and have the values:
\begin{equation}
\beta _{0}={\frac{33-2n_{f}}{12\pi },\qquad }\beta _{1}={\frac{153-19n_{f}}{%
24\pi ^{2}},}
\end{equation}
where $n_{f}=3$ is the number of active flavours. The coupling
depends on the transverse momentum only, therefore the integration
over energy and angle in equation (\ref{distrib}) does not involve
any running of the coupling; it is sufficient to make the
substitution (\ref{copsost}) in (\ref{distrib}).
\item  the two loop contribution in the double logarithmic term need to be included:
\begin{equation}
A_{1}\alpha _{S}\rightarrow A_{1}\alpha _{S}+A_{2}\alpha _{S}^{2}.
\label{mettoA2}
\end{equation}
where the universal coefficient $A_2$ is \cite{kodairatrentadue}
\begin{equation}
A_2 = {C_F\over 2\pi^2} [ C_A ({67\over18} - {\pi^2\over6}) -
{5\over9} \ n_f].
\end{equation}
According to what stated in the first point one can simply
substitute (\ref {mettoA2}) in (\ref{distrib}).
\end{enumerate}
In general the  resummation of transverse momentum logarithms
involves the following two functions, both possessing power series
expansion in $\alpha_{S}$:
\begin{eqnarray}
A\left( \alpha _{S}\right) &=&\sum_{n=1}^{\infty }A_{n}\,\alpha_{S}^{n}=A_{1}\alpha _{S}+A_{2}\alpha _{S}^{2}+\cdots ,   \\
B\left( \alpha _{S}\right) &=&\sum_{n=1}^{\infty
}{B}_{n}\,\alpha_{S}^{n}={B}_{1}\alpha _{S}+\cdots ,
\end{eqnarray}

\section{Resummation of the Transverse Momentum
Distribution}\label{transvresum} In this section we will show the
results about the resummation of transverse momentum distribution:
the results shown here are based on \cite{noi1}.
\\
In order to perform the resummation with next-to-leading accuracy
we must follow the prescription in section (\ref{prescrizioni}):
by substituting (\ref{copsost}) and (\ref{mettoA2}) equation
(\ref{distrib}) reads
\begin{eqnarray}\label{distribcor}
d(x)&=&\delta (x)+\alpha^{NLO}_{S}(Q^{2}x)\int_{0}^{1}d\omega
\int_{0}^{1}dt\left[
{\frac{A_{1}+\alpha^{NLO}_S(Q^{2}x)A_2}{\omega
t}}\right.+\nonumber\\&+&{\frac{S_{1}(t)}{\omega}}+\left.{\frac{C_{1}(\omega
)}{t}} \right] \left[ \delta (x-\omega ^{2}t)-\delta (x)\right]
\end{eqnarray}
After the integration over the phase space we obtain the effective
distribution of a single gluon in the transverse momentum space,
accurate to the next-to-leading level:
\begin{equation}\label{effectiveungluone}
d_R(x)=-{\frac{A_{1}}{2\beta _{0}}}{\frac{\log x}{x\log
sx}}+{\frac{A_{1}\beta
_{1}}{2\beta _{0}^{3}}}{\frac{\log x\log \log sx}{x\log ^{2}sx}}+{\frac{%
B_{1}}{\beta _{0}}}{\frac{1}{x\log sx}}-{\frac{A_{2}}{2\beta
_{0}^{2}}}{\frac{\log x}{x\log ^{2}sx}.}
\end{equation}
The inclusion of virtual diagrams introduces "+" distributions.
\\
The resummation of transverse momentum is performed in the impact
parameter space \cite{pp,kodairatrentadue}.
\\
In the impact parameter space the general formula (\ref{master})
becomes
\begin{equation}\label{masterb}
\frac{1}{\Gamma _{0}}\frac{d\tilde{\Gamma}}{db}=K\left(
\alpha_{S}\right) \,\Sigma\left( b;\alpha _{S}\right) +R\,\left(
b;\alpha_{S}\right)
\end{equation}
Let us define:
\begin{equation}\label{impardis}
\overline{d}(b)\equiv \frac{1}{\Gamma_{0}}\frac{d\widetilde{\Gamma }}{d%
b}\left( b\right) .
\end{equation}
Let us recall two important mathematical properties of
(\ref{impardis}): it is real and it depends only on $|b|$.
\\
Now we use the relation
\begin{equation}
\frac{1}{\Gamma_{0}}\frac{d\Gamma \left( p_t\right) }{dp_t}=%
\frac{1}{\pi }\frac{1}{\Gamma_{0}}\frac{d\Gamma }{{dp_t}^{2}}=\frac{1}{\pi Q^{2}%
}\frac{1}{\Gamma_{0}}d\left( x\right)
\end{equation}
and substitute it in the equation (\ref{bparam}):
\begin{equation}\label{impardis2}
\overline{d}(b)=\int_{0}^{1}dx\,d(x)\int_{0}^{2\pi }\frac{d\phi
}{2\pi }\ \exp \left[ iQb\sqrt{x}\cos \phi \right] \,,
\end{equation}
By combining equations (\ref{frv}) and (\ref{impardis2}), and
integrating over the azimuthal angle, one obtains
\begin{equation}
\overline{d}(b)=1+\int_{0}^{1}dx\,\,d_R(x)\,[J_{0}(Qb\sqrt{x})-1].
\label{esatta}
\end{equation}
$J_{0}(s)$ is the Bessel function of zero order defined by the
integral
\begin{equation}
J_{0}(s)\equiv \int_{0}^{2\pi }\frac{d\phi }{2\pi }\exp \left[ is\,\cos \phi %
\right] ,
\end{equation}
from which follows that $J_{0}(0)=1,$ so that the infrared singularity in $%
x=0$ is screened.
\\
As shown in section (\ref{multiplesoft}), by iterating for the
multiple gluon emission, the single emission term exponentiate and
we obtain
\begin{equation}
\Sigma(b)=\exp \left\{ \int_{0}^{1}dx\,\,d_{R}(x)\,[J_{0}(Qb\sqrt{x}%
)-1]\right\} .
\end{equation}
where the meaning of the function $\Sigma(b)$, is described in
section (\ref{secresummation}).
\\
In order to integrate over the $x$ variable, the transverse
momentum, we use the following approximation of the Bessel
function \cite{kodairatrentadue}, such an approximation remains
valid up to the next-to-leading level:
\begin{equation}
J_{0}(s)-1\simeq -\Theta (s-b_{0})  \label{japprox}
\end{equation}
where $b_{0}=2e^{-\gamma _{E}}\approx 1.12.$ We have therefore:
\begin{equation}\label{brisomm}
\Sigma(b)\simeq \exp \left[ -\int_{b_{0}^{2}/\left(
Q^{2}b^{2}\right) }^{1}\,d_{R}\left( x\right) \,dx\right] .
\end{equation}
Where $\theta \left( b/b_{0}-1/Q\right) $ in front of the integral
is understood. The impact parameter distribution coincides with
the cumulative distribution (\ref{distrint}) in the transverse
momentum space. It is important to notice that the step
approximation in equation (\ref{japprox}) cuts the small
transverse momenta region, since implies
\begin{equation}
k_t^{2}>\frac{b_0^2}{b^{2}}.
\end{equation}
\begin{figure}[t]
\begin{center}\label{graficob}
\psfig{bbllx=90pt, bblly=335pt, bburx=650pt, bbury=740pt,
file=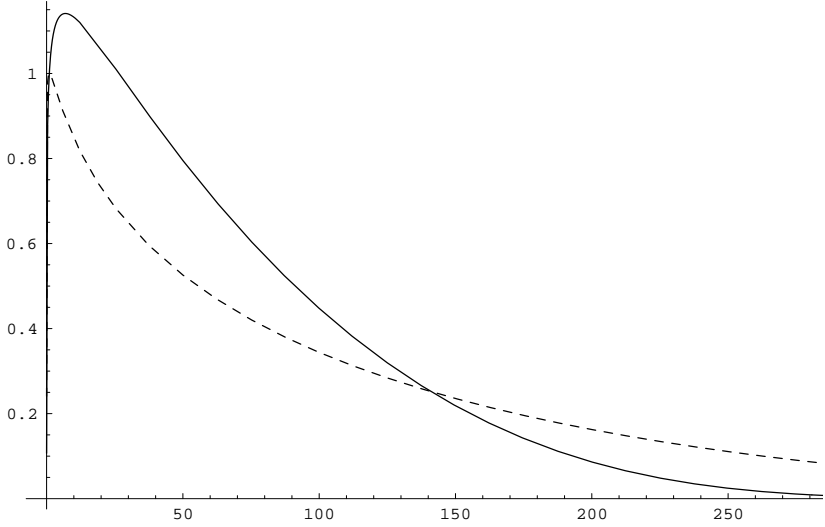, height=9cm, width=13cm} \vspace{-1.5cm}
\caption{Plot of the function $f(b)$  in the variable $y=Q^2 b^2
/b_0^2$ $(\alpha_S=0.22)$ Solid line: NLO; dotted line: LO.}
\end{center}
\end{figure}
\noindent The approximation (\ref{japprox}) is therefore
equivalent to a prescription for the non perturbative effects
related to the infrared pole in the coupling. To order
$\alpha_{S},$ one has:
\begin{equation}
\Sigma(b)=\exp \left[ \int_{b_{0}^{2}/\left( Q^{2}b^{2}\right)
}^{1}dx\left( \frac{1}{2}A_{1}\alpha _{S}{\frac{\log x}{x}}-B%
_{1}\alpha _{S}{\frac{1}{x}}\right) \right]
\end{equation}
By integrating over $x$ one obtains:
\begin{equation}
\Sigma(b)=\exp \left[ -{A_{1} \over 4}\alpha _{S}\log ^{2}\left( \frac{Q^{2}b^{2}%
}{b_{0}^{2}}\right) -B_{1}\alpha_{S}\log \left( \frac{Q^{2}b^{2}}{%
b_{0}^{2}}\right) \right] .
\end{equation}
In order to obtain the resumed distribution in impact parameter to
every order let us define $d_R$ as in eq.(\ref{effectiveungluone})
and substitute it in the eq.(\ref{brisomm}). Then let us integrate
over $x$ and write the hard scale logarithm as a function of two
loop coupling, by using the equation
\begin{equation}
\log s={\frac{1}{\beta _{0}\alpha _{S}}}+{\frac{\beta _{1}}{\beta _{0}^{2}}}%
\log \left( \beta_{0}\alpha_{S}\right) .
\end{equation}
The impact parameter resummed distribution may be written as the
exponential of a series of functions, as discussed in section
(\ref{secresummation}):
\begin{equation}
\Sigma(b)=\exp \left[
L\,g_{1}(\beta_{0}\alpha_{S}L)+g_{2}(\beta_{0}\alpha_{S}L)+\alpha_{S}\,g_{3}(\beta
_{0}\alpha _{S}L)+\cdots \right]
\end{equation}
where
\begin{equation}
L\equiv \log \frac{Q^{2}b^{2}}{b_{0}^{2}}
\end{equation}
and $\alpha_{S}\equiv \alpha_{S}(Q^{2})$.
\\
We obtain the following expressions for the $g_{1}(\omega )$ and
$g_{2}(\omega )$ functions:
\begin{eqnarray}\label{g1g2}
g_{1}(\omega ) &=&{\frac{A_{1}}{2\beta _{0}}}\frac{1}{\omega
}\left[ \omega +\log (1-\omega )\right] ,  \label{lo} \\
g_{2}(\omega ) &=&{-\frac{A_{2}}{2\beta _{0}^{2}}}\left[ {\frac{\omega }{%
1-\omega }}+\log (1-\omega )\right]+ \nonumber \\ &+&{\frac{A_{1}\beta _{1}}{2\beta _{0}^{3}}%
}\left[ {\frac{\omega }{1-\omega }}+{\frac{\log (1-\omega )}{1-\omega }}-{%
\frac{1}{2}}\log ^{2}(1-\omega )\right]-{\frac{B_{1}}{\beta
_{1}}}\log (1-\omega ). \label{nlo}
\end{eqnarray}
The expansion to order $\alpha^2_S$ of the exponent reads:
\begin{equation}\label{expansion}
\log \Sigma(b;\alpha_S) = -{1\over 4}A_1\alpha_S L^2 - B_1
\alpha_S L - {1 \over 6}A_1 \beta_0 \alpha_S^2 L^3 - {1\over 4}
A_2 \alpha_S^2 L^2 - {1 \over 2} B_1 \beta_0 \alpha_S^2 L^2
\end{equation}
Let us note that a single constant $B_1$ controls the
single-logarithmic effects in any order. The physical reason is
that a soft gluon and a collinear one with the same transverse
momenta are emitted with the same effective coupling
$\alpha_S(k^2_\perp)$.
\\
The function $\Sigma(b,\alpha_S)$ is plotted in fig.
(\ref{graficob}).
\\
It resum the logarithmic enhancements and becomes singular when
\begin{equation}
\omega \rightarrow 1^{-}  \label{unomeno}
\end{equation}
Since
\begin{equation}
\omega =\beta _{0}\alpha _{S}L\approx \frac{\log Q^{2}b^{2}}{\log
Q^{2}/\Lambda ^{2}},
\end{equation}
the singularity occurs when the transverse strange momentum
becomes as small as the hadronic scale,
\begin{equation}
p_{\perp } \approx \frac{1}{b}\approx \Lambda .
\end{equation}
The singularity (\ref{unomeno}) is produced by the infrared pole
(Landau pole) in the running coupling and signals an intrinsic
limitation of resummed perturbation theory, in agreement with
previous qualitative analysis. Let us note that the function
$g_{2}$ has basically a pole singularity in the limit
(\ref{unomeno}), while $g_{1}$ has only a softer, logarithmic,
singularity.

\chapter[Transverse Momentum vs. Threshold Comparison]{Comparison between Transverse Momentum Distribution
and Threshold Distribution}\label{compare}

\section{The Threshold Distribution}\label{threshold}
For the rare decay $b\rightarrow s\gamma$ the first distribution
to be studied was the threshold distribution, that is the spectrum
of the jet invariant mass and it is strictly related by the
kinematics to the photon spectrum , so that in practice they are
the same quantity.
\\
This distribution is particularly interesting both experimentally
and theoretically near the endpoint of the spectrum, namely where
\begin{eqnarray}
E_\gamma&=&\frac{m_b}{2}\nonumber\\
m^2_X&\rightarrow&0.
\end{eqnarray}
The theoretical importance lies on the appearance of interesting
effects of logarithmic enhancement near the endpoint, which drove
to apply techniques of resummation in $b$ physics. Moreover in the
endpoint region non perturbative effects become relevant and the
interplay between resummed perturbative distribution and non
perturbative bound state effects turns out to be crucial to
understand.
\\
From the experimental point of view physicists studying rare
decays such as  $b\rightarrow s\gamma$ or $b\rightarrow
ue^-\overline{\nu}_e$ have to face the problem of the large
background coming from the transition $b\rightarrow c$, which is
overwhelming in almost the whole phase space. Since the mass of
the quark $charm$ cannot be neglected in the process
\begin{equation}
b\rightarrow ce^- \overline{\nu}_e
\end{equation}
it turns out that the endpoint of the electron spectrum is reached
for
\begin{equation}
E^{b\rightarrow c}_e=\frac{m^2_b-m^2_c}{2m_b},
\end{equation}
so that in the endpoint region of the electron spectrum for
$b\rightarrow u$
\begin{equation}
E^{b\rightarrow u}_e=\frac{m_b}{2}
\end{equation}
the dominant transition $b\rightarrow c$ is forbidden and its
background eliminated.
\\
An analogous problem appear for $b\rightarrow s\gamma$, where the
annoying background is due to radiative photon emissions from the
tower of transitions $b\rightarrow c \rightarrow s$ and one has to
discriminate from strange hadrons coming from the process of
interest and the ones coming from charm quark decays. However in
the endpoint, for
\begin{equation}
E_\gamma>\frac{m^2_b-m^2_c}{2m_b},
\end{equation}
again the dominant transition is forbidden and the spectrum of
$b\rightarrow s\gamma$ can be cleanly measured.
\\
Results about fixed order calculations \cite{ali} and resummed
calculations \cite{ugo} has been performed for this distribution
in past years and non perturbative effects has been included by
defining a structure function, called \it shape function
 \rm\cite{generale}, based on the heavy quark effective theory
\cite{hqet}.
\\
In this chapter these results will be compared with the analogous
ones found for the transverse momentum distribution, showing
peculiar differences between these two quantities.

\section{Threshold distribution at Order $\alpha_S$}
Let us consider the energy spectrum of the photon in the rare
decay $b\rightarrow s\gamma$ near the endpoint:
\begin{equation}
f(z)=\frac{1}{\Gamma _{0}}\frac{d\Gamma }{dz}
\end{equation}
with
\begin{equation}
z\equiv \frac{2E_{\gamma }}{m_{B}} =1-\frac{m_{X}^{2}}{
m_{B}^{2}}.
\end{equation}
The photon spectrum then coincides with the mass distribution of
the jet. The kinematic constraint is:
\begin{equation}
o\left( \omega ,t\right) =\omega t.
\end{equation}
Let us note that it is infrared safe. Explicitly the distribution
reads
\begin{equation}
f(z)=\delta (1-z)+\alpha_{S}\int_{0}^{1} d\omega \int_{0}^{1} dt \left[ {%
\frac{A_{1}}{\omega t}}+{\frac{S_{1}(t)} \omega }+{\frac{C_{1}(\omega )%
}{t}}\right] \left[ \delta (1-z-\omega t)-\delta (1-z)\right] .
\end{equation}
By integrating over the energy and the polar angle we obtain:
\begin{equation}
f(z)=\delta (1-z)-\alpha_{S} A_{1}\left( {\frac{\log \left[ 1-z\right] }{1-z}%
}\right) _{+} +B_{1}\alpha_{S}\left( {\ \frac{1}{1-z}}\right) _{+}
,
\end{equation}
where the finite corrections have been neglected and we have
defined
\begin{equation}
B_{1}=S_{1}+C_{1}.
\end{equation}
Note the symmetry between the variables $\omega $ e $t$ implying
that the coefficient of the next-to-leading term is the sum of the
soft and the collinear coefficient and so it is for $\omega
\leftrightarrow t$ symmetries. Large infrared contributions are
included by replacing the bare coupling with the running coupling
constant evaluated at the transverse momentum squared
\begin{equation}
\alpha_{S}\rightarrow \alpha_{S}k_t^{2},
\end{equation}
where
\begin{equation}
k_t^{2}\simeq Q^{2}\omega^{2}t.
\end{equation}
Here the symmetry between soft and collinear contributions starts
to be broken.

\section{Resummation of the Threshold Distribution}
In this case the resummation can be performed in Mellin space:
\begin{equation}  \label{mellin}
{\frac{1}{\Gamma_0}} \ \Gamma_N = \int_0^1 \ dx\ x^{N-1}
{\frac{1}{\Gamma_0}} \ {\frac{d\Gamma }{dx}}
\end{equation}
The general formula (\ref{master}) for the N-th moment of the rate
can be written as \cite{catanitrentadue,cttw},
\begin{equation}
\frac{1}{\Gamma_{0}} \ \Gamma _{N}=\mathcal{C}\left(
\alpha_{S}\right) \,f_{N}\left( \alpha _{S}\right) +R_{N}\left(
\alpha_{S}\right) ,
\end{equation}
where now the large logarithm contains the $N$-variable:
\begin{equation}
\qquad \qquad \qquad \qquad L\equiv \log \frac{N}{N_{0}}\qquad
\qquad \left( \mathrm{threshold\,\,case}\right) ,
\end{equation}
with $N_{0}\equiv \exp \left[ -\gamma_{E}\right] .$ The functions
$g_{i}$ in the exponent are different with respect to the ones in
the $p_{\perp }$ case and the leading and next-to-leading ones
read \cite{ugo,americani}:
\begin{eqnarray}
g_{1}\left( \lambda \right) &=&-{\frac{A_{1}}{2\beta _{0}}}\frac{1}{\lambda }%
\left[ (1-2\lambda )\log (1-2\lambda )-2(1-\lambda )\log
(1-\lambda )\right]
,  \nonumber\\
g_{2}\left( \lambda \right) &=&\frac{\beta _{0}A_{2}-\beta
_{1}A_{1}}{2\beta _{0}^{3}}\left[ \log (1-2\lambda )-2\log
(1-\lambda )\right] -\frac{\beta _{1}A_{1}}{4\beta _{0}^{3}}\left[
\log ^{2}(1-2\lambda )-2\log
^{2}(1-\lambda )\right] + \nonumber \\
&&+\frac{S_{1}}{2\beta _{0}}\log (1-2\lambda )+\frac{C_{1}}{\beta
_{0}}\log (1-\lambda ).  \label{g1eg2}
\end{eqnarray}
The expansion to order $\alpha _{S}^{2}$ of the exponent reads:
\begin{equation}
\log f_{N}=-\frac{1}{2}A_{1}\alpha _{S}{L}^{2}-\alpha _{S}\left(
S_{1}+C_{1}\right) \,{L}-\frac{1}{2}A_{1}\beta _{0}\alpha _{S}^{2}L^{3}-%
\frac{1}{2}A_{2}\alpha _{S}^{2}L^{2}-\left(
S_{1}+\frac{1}{2}C_{1}\right) \beta _{0}\,\alpha
_{S}^{2}\,{L}^{2}.
\end{equation}
Let us comment on the above results. The single-logarithmic
effects at one loop are controlled by the constant
\begin{equation}
S_{1}+C_{1}= -\frac{7}{4}\frac{C_{F}}{\pi },
\end{equation}
i.e. by the sum of the soft and the collinear coefficients, which
is different from the $p_{\perp }$-case (cf. eqs.
(\ref{coefficiente}) and (\ref {expansion})). At two-loop they are
instead controlled by a different constant,
\begin{equation}
S_{1}+\frac{1}{2}C_{1}=-\frac{11}{8}\frac{C_{F}}{\pi }.
\end{equation}
The soft and the collinear terms begin to differentiate at this
order and the soft one has a two times larger coefficient.
Contrarily to the $p_{\perp }$ case, two different constants are
needed to describe the single logarithmic effects. The dynamical
difference between soft and collinear terms is that, for a fixed
jet mass, the transverse momentum of a soft gluon is substantially
smaller that of a collinear gluon \cite {abcmv,ugo,ks}.

\section{Singularities of the Threshold Distribution}
The functions $g_{1}$ and $g_{2}$ in (\ref{g1eg2}) (and therefore
also the resummed distribution) have two different singularities
\cite{ugo,montp,lep3proc}:
\begin{description}
\item[$i)$]  the first one occurs when
\begin{equation}
\frac{1}{2}\,=\,\lambda \,\approx \,\frac{\log Q^{2}/m_X^{2}}{\log
Q^{2}/\Lambda ^{2}},  \label{approx}
\end{equation}
or, equivalently, when
\begin{equation}
m_X^{2}\,\approx \,\Lambda \,Q,  \label{onehalf}
\end{equation}
where $m$ is the mass of the final hadronic jet $s+\widehat{X}.$
In the last member of (\ref{approx}), we have used the
approximation $N\approx Q^{2}/m_X^{2}$ \cite{mangano}. The
singularity (\ref{approx}) signals the occurrence of
non-perturbative effects in region (\ref{onehalf}), to be
identified with the well-known Fermi motion \cite{nostri}; it is
related to soft-gluon effects,  i.e. to the terms proportional to
$A_{1},\,S_{1}$ and $A_{2}$, and not to collinear ones, the term
proportional to $C_{1}.$ Fermi-motion effects are therefore
controlled by soft and not by collinear dynamics. This fact allows
a factorization of Fermi-motion effects by means of a function
taking into account soft dynamics only, the well-known shape
function\footnote{The shape function is also called structure
function of the heavy flavours.} \cite{generale}. In this region,
initial bound state effects become relevant while final-state
binding effects can be neglected \cite{ugo,congiulia2}.
\item[$ii)$]  the second singularity occurs at
\begin{equation}
\lambda =1,
\end{equation}
or
\begin{equation}
m_X^{2}\approx \Lambda^{2}
\end{equation}
and is related to final-state hadronization effects. Both soft and
collinear terms are singular in this region and there are
non-perturbative effects related to initial as well as final
bound-state dynamics.
\end{description}

\section{The Fermi Motion and the Shape Function}
The main idea underlying the concept of \it shape function \rm
\cite{generale} is the factorization of initial bound state
effects in a non perturbative function, based on the Heavy Quark
Effective Theory \cite{hqet}, which can be extracted from a
process, such as $b\rightarrow s\gamma$ to be considered in other
processes such as $b\rightarrow ue^- \overline{\nu}_e$.
\\
This approach apply to the decay of an heavy hadron $B$,
containing a $b$ quark, into an inclusive hadronic state $X$,
having a large energy and a small invariant mass; the kinematics
of interest is therefore \cite{congiulia2}
\begin{eqnarray}
\frac{m^2_X}{E_X}&\sim& O(\Lambda_{QCD})\nonumber\\
E_X&\gg& \Lambda_{QCD}.
\end{eqnarray}
Being the heavy quark mass very large with respect to the typical
scale of QCD \cite{hqet}, the Heavy Quark Effective Theory (HQET)
can be applied
\begin{equation}
m_b\geq E_X \gg \Lambda_{QCD}.
\end{equation}
The invariant mass of the final hadronic scale is small compared
to its energy, but is large enough to justify a perturbative
approach: however non perturbative effects related to the initial
state appear and the shape function is introduced to factorize
them.
\\
At the partonic level the process turns out to be
\begin{equation}
b\rightarrow \hat{X}+\dots
\end{equation}
where $\hat{X}$ differ from $X$ because it does not contain the
valence quark contained in $B$.
\\
The heavy quark and heavy hadrons frames are the same in first
approximation: they differ because the heavy quark has a vibratory
motion inside the hadron, due to the exchange of soft gluons with
the valence quark.
\\
Therefore the heavy quark momentum can be written as
\begin{equation}
P^\mu=m_B v^\mu+k^\mu,
\end{equation}
where $m_B$ is the mass of the decaying hadron, $v^\mu$ is the
velocity of the heavy hadron, taken without loss of generality in
its rest frame
\begin{equation}
v^\mu=(1,0,0,0).
\end{equation}
$k^\mu$ is the residual momentum, taking into account the exchange
of gluons between the heavy quark and the valence quark: according
to the HQET
\begin{equation}
k^\mu\sim O(\Lambda_{QCD}).
\end{equation}
The distribution of residual momentum $k^\mu$ is of non
perturbative origin and represents in practice the Fermi
Motion\footnote{The term Fermi Motion is taken from nuclear
physics, where a similar phenomenon occurs.} of the heavy quark
inside the heavy hadron.
\\
In practice the Fermi Motion implies a change in the momentum
available for the final state, so that the partonic invariant mass
and the hadronic one differ and in particular
\begin{equation}\label{shapekin}
m^2_{\hat{X}}=m^2_X+2E_Xk_+\frac{m^2_X}{2E_X}k_+k^2,
\end{equation}
being $k_+=k_0-k_z$.
\\
Dropping terms of order $O(\Lambda^2_{QCD})$ the approximation
obtained is
\begin{equation}
m^2_{\hat{X}}\simeq 2E_X k_+.
\end{equation}
This result has two main consequence: the invariant mass at
partonic level is affected by non perturbative effects and should
be calculated by a convolution with them; moreover the good
variable describing Fermi Motion seems to be the component $k_+$.
\\
Applying this formalism to the photon spectrum, for example, the
distribution of interest is
\begin{equation}
\Phi(E_\gamma)=\frac{1}{\Gamma_0}\int_0^E \ dE_\gamma \
\frac{d\Gamma}{dE_\gamma}
\end{equation}
which can be calculated as
\begin{equation}\label{shapedist}
\Phi(E_\gamma)=\int \ dk_+ \ \Phi^{pert}(E_\gamma-k_+) \ f(k_+)
\end{equation}
where $f(k_+)$ is the shape function, describing the probability
that the heavy quark has a "+" component of the momentum $k_+$.
\\
According to the HQET this quantity is formally expressed as
\begin{equation}
f(k_+)=<B|h^\dagger_v\delta(k_+-iD_+)h_v|B>,
\end{equation}
where $D_+=D_0-D_z$ is a component of the covariant derivative,
$B$ is the $B$ meson state and $h_v$ is the field describing an
heavy quark with velocity $v$ in HQET.
\\
An alternative way to express this quantity, avoiding $\delta$
distributions, is to consider
\begin{equation}
F(k_+)=<B|h^\dagger_v\frac{1}{k_+-iD_++i\epsilon}h_v|B>
\end{equation}
and the obtain the shape function by mean of the optic theorem
\begin{equation}
f(k_+)=-\frac{1}{\pi}{\rm Im} F(k_+).
\end{equation}
An alternative physical interpretation of the shape function is
that it represent the probability that the decaying quark has an
effective mass $m_\ast$ at disintegration time.
\\
In fact the momentum available for the final state is
\begin{equation}
Q^\mu=m_B v^\mu-q^\mu+k^\mu\simeq m_\ast v^\mu - q^\mu,
\end{equation}
up to terms of order $O(\Lambda^2_{QCD})$, so that one introduces
the effective mass
\begin{equation}
m_\ast=m_B+k_+.
\end{equation}
In practice the physical interpretation is that, in order to take
into account Fermi Motion, one can consider a decaying quark
on-shell, but with an effective $m_\ast$ instead of an off-shell
quark with a virtuality $k_+$. In this way (\ref{shapedist})
becomes
\begin{equation}
\Phi(E_\gamma)=\int_{-\infty}^{m_B} \ \Phi(E_\gamma,m_\ast) \
f(m_\ast) \ dm_\ast.
\end{equation}
A final remark needs to be inserted: this approach is viable to
describe effects due to soft gluon emissions. Hard collinear
gluons cannot be described in this pattern, as it can be inferred
by (\ref{shapekin}). For an hard collinear gluon none of the terms
in (\ref{shapekin}) can be neglected and the following
approximation are not possible.
\\
In conclusion the shape function approach is based on the HQET and
allows to factorize non perturbative effects due to the emission
of soft gluons and it is consistent in the kinematical region
$m^2_X=O(E_X \Lambda_{QCD})$. It can be extracted by experimental
data or calculated by lattice QCD \cite{shapelattice}. From a
physical point of view these effects are related to the Fermi
Motion of the initial heavy quark.

\section{Comparison between Transverse Momentum and Threshold Distributions}
It is interesting the comparison between the results of section
(\ref{transvresum}) for the transverse momentum distribution and
this chapter for the threshold distribution \cite{noi1}.
\begin{itemize}
\item
In the threshold distribution two different coefficients $S_1$ and
$C_1$ control the next-to-leading terms at every order, because
soft and collinear logarithms differentiate.
\\
On the contrary in the transverse momentum distribution (in the
next-to-leading logarithmic approximation) just a coefficient,
indicated by $B_1$, is needed to describe single logarithmic
effects. In practice soft and collinear terms act in the same way,
with the same dynamics.
\\
From a physical point of view the reason that for a given
invariant mass (or photon energy) soft gluons and hard collinear
gluons have a different transverse momentum, that is a different
effective coupling\footnote{Let us recall that the effective
coupling is $\alpha_S(k^2_t)$.}. On the contrary in the transverse
momentum distribution soft and collinear single logarithms cannot
be differentiated because, tautologically, having the same
transverse momentum they couple in the same way.
\item
The different structure of the singularities is the most
interesting comparison.
\\
In the threshold case two singularities are present: a first
singularity, closer to the origin\footnote{Formally the region of
the spectrum from $E_\gamma=0$ to $E_\gamma=m_b/2$ corresponds to
the evolution from $\lambda=0$ to $\lambda=1$ in $\lambda$ space.}
is found in $\lambda=\frac{1}{2}$. The soft nature is underlined
by the fact that only the soft coefficient $S_1$ controls this
singularities (apart double logarithmic coefficient $A_1$ and
$A_2$), while the collinear coefficient $C_1$ does not appear.
This merely soft nature suggests that it is related to initial
bound state emissions, namely the Fermi Motion, and can be removed
by mean of the shape function.
\\
A second singularity is found closer to the endpoint spectrum, in
$\lambda=1$: this singularity has not a soft nature only, it is
related to hadronization effects and cannot be faced with a shape
function approach. However, in the threshold distribution this
singularity appears \it after \rm the other one, which is already
factorized by the shape function and therefore is less important.
\\
In the $p_t$ spectrum just a singularity appears, for $\omega=1$,
which is of the same nature of the one discussed above for
$\lambda=1$. This singularity is not related to Fermi Motion
effects only: this reflects the considerations discussed in
section (\ref{tmdkinematics}), where with na\"{i}ve kinematical
arguments we concluded that hadronization and Fermi Motion effects
cannot be separated in the transverse momentum case.
\item
the previous point has an immediate kinematical explanation: while
in the threshold distribution Fermi Motion effects can be
separated by a scale from hadronization effects (and this allows
their factorization) in the $p_t$ case this does not occur.
\\
In fact the final hadronic state has a squared invariant mass
\cite{congiulia2}
\begin{equation}
m^2_X=p^2_X=(P-q)^2,
\end{equation}
being $P$ the momentum of the heavy quark and $q$ the momentum of
the probe, in our case the photon. Fermi Motion effects implies a
change of order $\Lambda$ in the momentum of the heavy quark, so
that the squared invariant mass turns out to be
\begin{equation}
m^{\prime 2}_X=(P\pm\Lambda-q)^2\approx m^2_X\pm
2E_X\Lambda+O(\lambda^2),
\end{equation}
that is a variation of order $\Lambda$ in the momentum of the
initial quark produces a larger effects in the invariant mass of
the hadronic state
\begin{equation}
\delta m^2_X\approx 2E_X\Lambda,
\end{equation}
being $E_X\cong \frac{m_b}{2}$.
\\
Hadronization effects implies a variation of order $\Lambda$ in
the momentum of the final hadronic state, but this leads just to a
change of order $\Lambda^2$ in the invariant mass
\begin{equation}
\delta m^2_X\approx \Lambda^2.
\end{equation}
It is evident that non perturbative effects of these two different
types act at different scales.
\\
In the $p_t$ case this does not happen: a variation of order
$\Lambda$ in the initial or in the final state does produce a same
effect, roughly speaking of the order $\Lambda$ as argued in
section (\ref{tmdkinematics}). There is not a scale separating
these two effects in this case.
\item
previous considerations imply that, since in the $p_t$ case a
singularity in the resummed formula, having soft nature only, is
not present, a shape function approach, introduced to factorize
Fermi Motion effects, is not viable. One could try a different
approach based on the effective theory introduced in \cite{UC} or
the collinear effective theory developed in \cite{separati}, but
this topic has not yet been developed.
\item
let us note that, in general, the singularities of the functions $%
g_{i}\left( \omega \right) $ are more severe than those of the functions $%
g_{i}\left( \lambda \right) $ for $\lambda \rightarrow 1/2$ or
$1.$ For example, $g_{1}\left( \omega \right) $ has a logarithmic singularity, while $%
g_{1}\left( \lambda \right) $ has an additional pre-factor
$1-2\lambda $ or $ 1-\lambda $ which softens the singularity.
\end{itemize}
Owing to the different singularity structure, the $p_{\perp
}$-distribution is complementary to the threshold one and gives
independent information about non-perturbative physics.

\chapter{Complete $O(\alpha_S)$ Calculation}\label{fixedorder}
This chapter is devoted to the explicit calculation of one loop
Feynman diagrams (both real and virtual), describing the
transverse momentum distribution in $b\rightarrow s\gamma$. The
results shown here are based on \cite{noi2}.
\\
The calculation of Feynman diagrams to the order $\alpha_S$ is
needed in the resummation program to extract the first term $k_1$
in the perturbative expansion of the coefficient function
$K(\alpha_S)$, introduced in equation (\ref{coefexp}).
\\
This constant is necessary to achieve the next-to-leading accuracy
which we have required, as discussed in section
(\ref{secresummation}).
\\
While the coefficients controlling the logarithmic enhancement can
be computed by general properties of QCD, such as the evolution of
the light quark through the Altarelli-Parisi kernel or the eikonal
approximation, $k_1$ is process dependent and the explicit
evaluation of Feynman diagrams is needed.
\\
After the calculation of the diagrams a matching procedure must be
performed to avoid double counting in the logarithmic terms: in
practice one has to subtract the logarithmic terms already
included and resummed in the function $\Sigma(x;\alpha_S)$, by
expanding the latter to the order $\alpha_S$.
\\
Moreover the calculation of real diagrams can allow the
calculation of the first term $r_1(x)$ in the expansion of the
reminder function $R(\alpha_S;x)$ (see equation (\ref{remexp})):
this term is not strictly required for next-to-leading
resummation, but can become relevant in the hard region of the
phase space. Let us underline that, due to computational
difficulties, remainder functions are usually neglected, since
they are not very relevant for the resummation program, and they
are known for few processes.
\\
The calculation is performed inserting the operator $\hat{O}_7$ of
the basis in equation (\ref{base}) in the vertex and calculating
corrections due to (real or virtual) gluon emissions: this is the
leading approximation because this is the only operator having a
logarithmic enhancement. Precisely the other operators shows
infrared divergences that can be reabsorbed in the matrix element
of $\hat{O}_7$, by writing an effective coefficient $C^{eff}_7$
according to \cite{cmm}. This implies that the contribution to the
coefficient function, for those operators, comes only from virtual
diagrams, which are kinematics independent and therefore can be
taken from literature: this is done in section (\ref{finalcoef}).

\section{Complete Calculation of Real Diagrams}
\begin{figure}[h]
\center\mbox{\epsfig{file=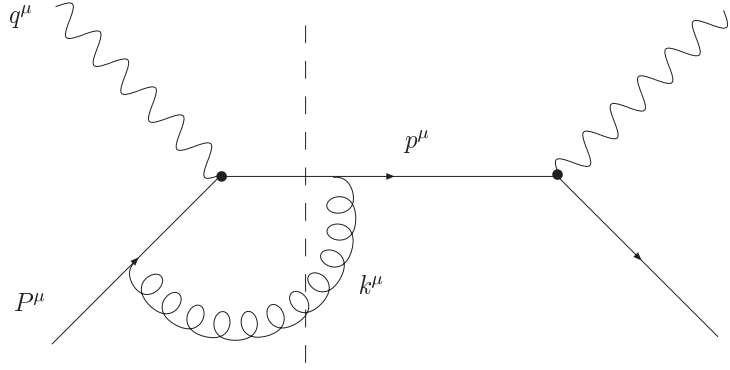, width=10cm}}
\caption{\label{diagvertice} Interference diagram: product of
diagrams describing the emission from the initial and the final
quark.}
\end{figure}

\begin{figure}[h]
\center\mbox{\epsfig{file=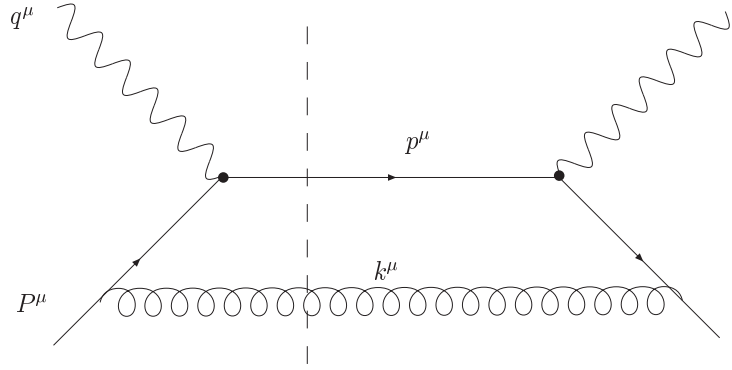, width=10cm}}
\caption{\label{diagladder} Ladder diagram: square of the diagram
describing the emission of the gluon from the initial quark.}
\end{figure}

\begin{figure}[h]
\center\mbox{\epsfig{file=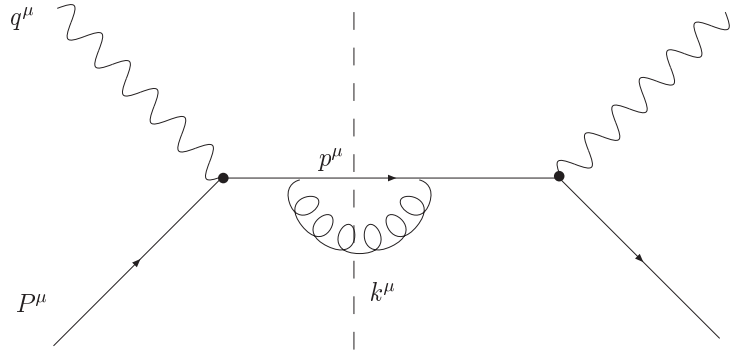, width=10cm}}
\caption{\label{diagautoenergia} Interference diagram.}
\end{figure}

Let us start with the calculation of gluon bremsstrahlung from the
initial or the final quark.
\\
The three amplitudes coming from the product of these two diagrams
can be depicted with cut diagrams, according to Cutkowski rules
and the optical theorem \cite{peskin}. They are shown in figures
(\ref{diagvertice},\ref{diagladder},\ref{diagautoenergia}).
The calculation of the diagram in figure (\ref{diagvertice})
involves the calculation of the trace:
\begin{equation}
2\frac{{\rm Tr}\left[
(\hat{P}+m_b)\gamma^\mu\hat{q}(1+\gamma_5)(\hat{p}+\hat{k})\gamma^\rho\hat{p}\gamma_\mu\hat{q}(1+\gamma_5)(\hat{P}-\hat{k}+m_b)\gamma_\rho\right]}{(P\cdot
k)(p\cdot k)},
\end{equation}
where the operator $\hat{O}_7$ was inserted in the vertex,
according to the equation (\ref{elmato7}).
\\
The attribution of momenta is given in the figures.
\\
Let us recall that the mass of the strange quark is neglected, so
that $p^2=q^2=k^2=0$ and $P^2=m^2_b$.
\\
The calculation of the trace is performed in dimensional
regularization, in order to regularize infrared divergencies, in
$n=4+\epsilon$ dimensions, with $\epsilon>0$.
\\
The trace can be calculated according to the rules written in
appendix (\ref{feynrules}): here $\hat{n}=\gamma^\mu n_\mu$.
The ladder diagram in figure (\ref{diagladder}) produces the
contribution
\begin{equation}
\frac{{\rm
Tr}\left[(\hat{P}+m_b)\gamma^\rho(\hat{P}-\hat{k}+m_b)\gamma^\mu\hat{q}(1+\gamma_5)\hat{p}\gamma_\mu\hat{q}(1+\gamma_5)(\hat{P}-\hat{k}+m_b)\gamma_\rho\right]}{(P\cdot
k)^2}.
\end{equation}
Finally the diagram (\ref{diagautoenergia}) involves the trace
\begin{equation}
\frac{{\rm Tr}\left[
(\hat{P}+m_b)\gamma^\mu\hat{q}(1+\gamma_5)(\hat{p}+\hat{k})\gamma^\rho\hat{p}\gamma_\rho(\hat{p}+\hat{k})\gamma^\mu\hat{q}
(1+\gamma_5)\right]}{(p\cdot k)^2}.
\end{equation}
After the calculation of the traces\footnote{Tedious details and
results about this topic are not important.} we can insert the
kinematics of section (\ref{seckinungluone}), by defining:
\begin{eqnarray}
\omega&=&\frac{2E_\gamma}{m_b}\nonumber\\
t&=&\frac{1-\cos\theta}{2}\nonumber\\
y&=&\frac{2E_s}{m_b}.
\end{eqnarray}
The conservation of the quadrimomentum implies that
\begin{equation}
y=\frac{1-\omega}{1-\omega(1-t)},
\end{equation}
so that the matrix element, to be integrated over the phase space,
can be expressed in terms of just two variables, $\omega$ and $t$.
\\
Summing the real diagrams the contribution to the matrix element
in dimensional regularization turns out to be:
\begin{eqnarray}
M(\omega,t)&=&\\&=&-\frac{1}{\Gamma(1+\epsilon/2)}\frac{1}{4(1-\omega(1-t))^3(\epsilon+2)}\left[
(1-t)^{\frac{\epsilon+2}{2}}(1-\omega)^3
\left(\frac{1-\omega}{(1-\omega(1-t))}\right)^\epsilon\right.\nonumber\\
&&\left(-8+8(1-t)\omega-4\omega^2(1+t)-4\epsilon
(1-(1-t)\omega+\omega^2(1+t))-\epsilon^2\omega^2
(1-t)\right)\left.\right]\nonumber
\end{eqnarray}
The quantity of interest is (see equation (\ref{vincolox}))
\begin{equation}
\frac{1}{\Gamma_0}\frac{d\Gamma}{dx}=\frac{1}{2m_b}\frac{M(\omega,t)}{\omega^{1-\epsilon}t^{1-\epsilon/2}}d\Phi(\omega,t)\left[x-\omega^2
t(1-t)\right].
\end{equation}
$d\Phi(\omega,t)$ is the phase space described in appendix
\ref{phasespace}.
\\
The decomposition in (\ref{decomposizionematrel}) allows to
underline terms giving rise to double logarithmic enhancement,
(collinear or soft) logarithmic enhancement and finite terms.
\begin{eqnarray}
A_1&=&\frac{1}{\Gamma(1+\epsilon/2)};\\
C_1(\omega)&=&\frac{\omega^\epsilon (\omega(\epsilon+2)-4)}{4\Gamma(1+\epsilon/2)};\nonumber\\
S_1(t)&=&\frac{1-(1-t)^{1+\epsilon/2}}{\Gamma(1+\epsilon/2)};\nonumber\\
F_1(\omega,t)&=&\frac{\omega t(2\omega^2
t^3-\omega(\omega^3-2\omega^2+8\omega-6)t^2)}{2(1-\omega(1-t))}+\nonumber\\&+&\frac{\omega
t(4\omega^4-10\omega^3+15\omega^2-13\omega^2+4)t-(1-\omega)^2(3\omega^2-2\omega+2))}{2(1-\omega(1-t))}\nonumber.
\end{eqnarray}
The function $F_1$, since it does not give rise to infrared
divergencies, can be calculated for $\epsilon\rightarrow 0$.
\\
We are interested in the cumulative distribution, which takes from
real emissions a contribution of the form:
\begin{equation}
D_R(x)=\int_0^xdx^\prime \int_0^1 d\omega \int_0^1 dt \ {1 \over
\Gamma_0} \ {d\Gamma \over dx^\prime}(\omega,t;\epsilon) \
\delta[x^\prime-\omega^2t(1-t)].
\end{equation}
\begin{figure}[h]
\center\mbox{\epsfig{file=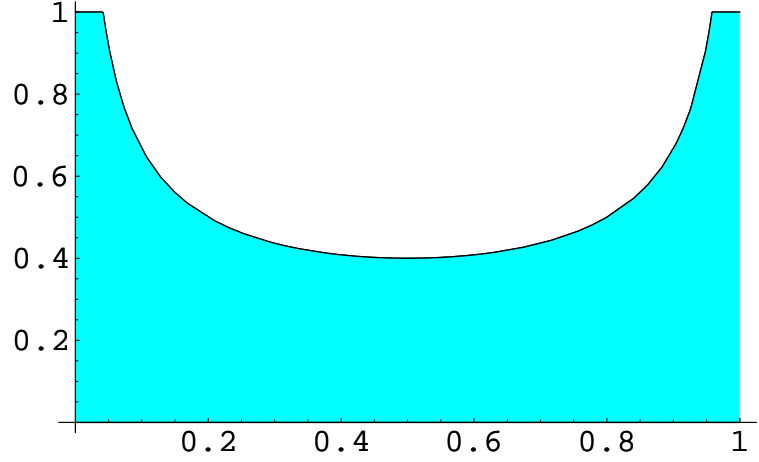, width=10cm}}
\caption{\label{spaziofasi1}Region of the phase space to
integrate: vertical axes $\rightarrow\omega$, horizontal axes
$\rightarrow t$.}
\end{figure}
After the integration over the phase space we expect four kinds of
terms:
\begin{itemize}
\item
Poles in the regulator $\epsilon$: they parametrize the infrared
singularities and cancel in the sum with virtual diagrams because
the distribution we are dealing with is infrared-safe;
\item
Logarithmic terms diverging for $x\rightarrow 0$: with the
matching procedure they are subtracted and inserted in the
function $\Sigma(x;\alpha_S)$;
\item
Constant terms: they enter the coefficient function $K(\alpha_S)$;
\item
Remainder functions: terms that vanish in the limit $x\rightarrow
0$.
\end{itemize}
\begin{figure}
\begin{minipage}{15cm}
\parbox{7cm}{
\center\mbox{\epsfig{file=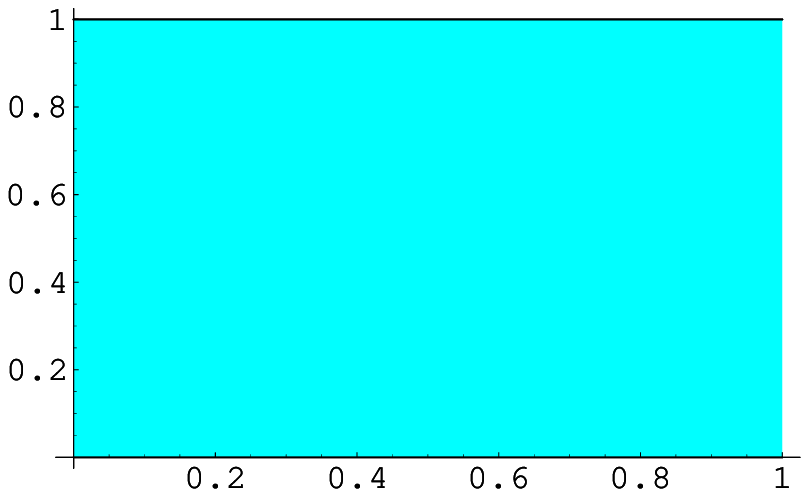, width=6cm}}
\caption{\label{spaziofasi2} Whole phase space.} }
\parbox{7cm}{
\center\mbox{\epsfig{file=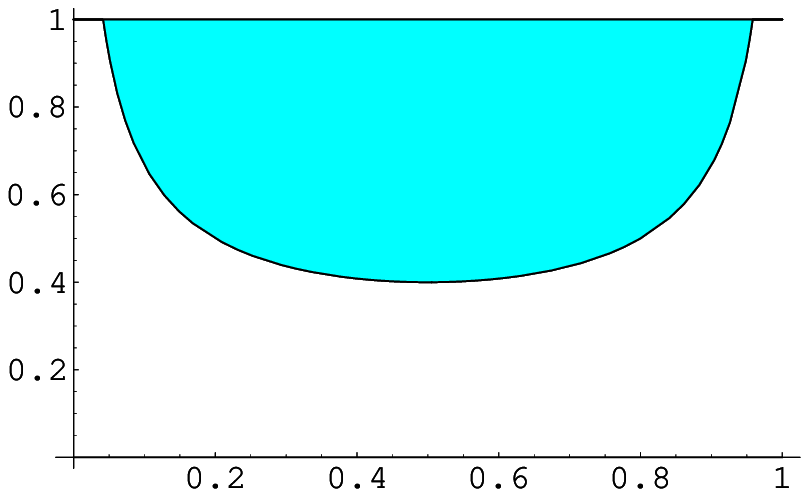, width=6cm}}
\caption{\label{spaziofasi3} Complementary region.}
\end{minipage}}
\end{figure}
Integrating over $x^\prime$ we have:
\begin{equation}\label{cumulativax}
D_R(x)= \int_0^1 d\omega \int_0^1 dt \ {1 \over \Gamma_0} \
{d\Gamma \over dx^\prime}(\omega,t;\epsilon) \
\theta[x-\omega^2t(1-t)].
\end{equation}
The remaining integrations are non-trivial because of the
simultaneous presence of the kinematical constraint (see figure
(\ref{spaziofasi1}) and by the dimensional regularization
parameter $\epsilon$. By using the identity
\begin{equation}
\theta[x-\omega^2t(1-t)]~=~1-\theta[\omega^2t(1-t)-x],
\end{equation}
we separate these two effects and rewrite the distribution
$D_R(x)$ as a difference between an integral over the whole phase
space (fig. \ref{spaziofasi2}) and a integral over the
complementary region (fig. \ref{spaziofasi3}):
\begin{equation}
D_R(x)=\int_0^1 d\omega \int_0^1 dt{1 \over \Gamma_0} \ {d\Gamma
\over dx}(\omega,t;\epsilon) \ - \ \int_0^1 d\omega \int_0^1 dt {1
\over \Gamma_0} \ {d\Gamma \over dx}(\omega,t;0) \
\theta[\omega^2t(1-t)-x]~+~O(\epsilon).
\end{equation}
In this way we have separated the two main sources of difficulty
in the computation of equation (\ref{cumulativax}).
\\
The first integral must be evaluated for $\epsilon\neq 0$ because
it contains poles in $\epsilon$, but is done over a very simple
domain, independent of $x$: the integration can be performed
without the introduction of special techniques with an advanced
program of symbolic calculation. The result can be expanded in
powers of $\epsilon$ and at this stage double and single poles in
the regulator and constant arise.
\\
The second integral does not contain any pole in $\epsilon$ and
therefore one can take the limit $\epsilon\rightarrow 0$ in the
integrand: it strongly depends on the kinematical constraint,
which makes the analytical computation very complicated.
\\
At first let us integrate over the variable $\omega$, since the
analytical dependence on it in the constraint (\ref{vincolox})
seems to be simpler. The limits of integration are
\begin{equation}
\frac{\sqrt{x}}{\sqrt{t(1-t)}}<\omega<1.
\end{equation}
Then, after the change of variable $y=\frac{1}{t}-1$, the
integration can be performed over $y$, with the limits of
integration
\begin{equation}\label{limity}
\tau < y <\frac{1}{\tau},
\end{equation}
being
\begin{equation}\label{tau}
\tau=\frac{1-\sqrt{1-4x}}{1+\sqrt{1-4x}}.
\end{equation}
The analytic integration is now very complicated and seems to
resist to the attack of usual techniques: a powerful tool that can
overcome this difficulty is the use of harmonic polylogarithms
\cite{remiddi}. The general properties of this approach are
outlined in appendix \ref{apphpl}.
\\
The first step is the introduction of a proper basis of function:
the one we singled out is
\begin{eqnarray}\label{baseint}
g[0;y]&\equiv& {1 \over y} \nonumber \\ g[-1;y]&\equiv& {1 \over
y+1} \nonumber \\ g[-2;y]&\equiv& {1 \over {\sqrt{y}(1+y)}}\nonumber \\
g[-3;y]&\equiv& -{\sqrt{x} \over
{2({1-\sqrt{x}\sqrt{y})}\sqrt{y}}}.
\end{eqnarray}
According to the theory described in appendix \ref{apphpl}, the
harmonic polylogarithms (HPL) of weight 1 are defined as:
\begin{eqnarray}
J[a;y]&\equiv&\int_0^y dy^\prime \ g(a;y^\prime)  \ \ \ \rm{for}\;   a \neq 0 \nonumber\\
J[0;y]&\equiv& \log y.
\end{eqnarray}
In terms of usual functions, they read:
\begin{eqnarray}
J[-1;y]&\equiv& \log (1+y)\nonumber\\
J[-2;y]&\equiv& 2\arctan \sqrt y\nonumber\\
J[-3;y]&\equiv& \log (1-\sqrt{x} \sqrt{y})
\end{eqnarray}
The HPL's of weight 2 are defined for $(u,v)\not= (0,0)$ as
\begin{equation}
J[u,v;y]\equiv \int_0^y dy^\prime \ g[u;y^\prime]\int_0^{y^\prime}
 dy^{\prime\prime} g[v;y^{\prime\prime}]
\end{equation}
and $J[0,0;y]=1/2\log^2y$. HPL's of higher weight may be defined
in an analogous way. They will not be used here.
\\
The integration is not trivial: it is performed through chains of
integration by parts of every terms, using the property
\begin{equation}
\frac{d}{dx}J(\vec{m}_w;x)=g(a;x)J(\vec{m}_{w-1};x)
\end{equation}
and algebraic manipulations. For example
\begin{equation}
\int \ y^n\ J(a,w;y)dy=\frac{y^{n+1}}{n+1} J(a,w;y)-\int \
\frac{y^{n+1}}{n+1} \ g(a;y) \ J(w;y)dy
\end{equation}
The remaining integral can be simplified by using partial
fractioning and following integrations by parts.
\\
After this process every function appearing in the result is
expressed in terms of algebraic functions or harmonic
polylogarithms: in particular every transcendental function is
expressed as a combination of hpl's.
\\
Once the undefined integral is calculated, one has to substitute
the limits of equation \ref{limity} and the final result is,
summing the two integrals:
\begin{equation}\label{completoreale}
D_R(x)= C_F {\alpha_S \over \pi}\ \left({m^2_b \over
4\pi\mu^2}\right)^{\epsilon/2}\ {1\over \Gamma(1+\epsilon/2)}\
\left[{2\over \epsilon^2} \ - {5 \over 2\epsilon}\ -{1 \over 4} \
\log^2 x - \ {5 \over 4} \ \log x + \ {1 \over 4} \
+r_1(x)\right],
\end{equation}
where $r_1(x)$ is a function vanishing for $x\rightarrow 0$.
\\
From the matching procedure $r_1(x)$ can be recognized as the
first term in the expansion of the remainder funtion\footnote{Let
us remember that virtual diagrams do not depend on the kinematics,
so that they cannot contribute to the remainder function.} (see
section (\ref{finalcoef})) and turns out to be \normalsize
\begin{eqnarray}\label{taudef}
&r_1(\tau)&=\frac{(\tau-1)(49\tau^8+468\tau^7+1797\tau^6+3642\tau^5+4450\tau^4+3642\tau^3+1797\tau^2+468\tau+49)}{12(\tau+1)^5(\tau^2+3\tau+1)^2}
\nonumber\\
&+&
\frac{-5-61\tau-317\tau^2-912\tau^3-1622\tau^4-1934\tau^5-1622\tau^6-912\tau^7-317\tau^8-61\tau^9-5\tau^{10}}{4(\tau+1)^6(\tau^2+3\tau+1)^2}\log\tau
\nonumber\\
&-&J[0,-3,\tau]+J[0,-3,1/\tau]-2J[0,-1,\tau]+J[-1,0,\tau]+J[-1,-3,\tau]-J[-1,-3,1/\tau]
\nonumber\\
&-&2\sqrt \tau
\arctan(\sqrt\tau)\frac{(\tau+1)(2\tau^2+7\tau+2)}{(\tau^2+3\tau+1)^2}+\frac{\pi}{2}\sqrt
\tau\frac{(\tau+1)(2\tau^2+7\tau+2)}{(\tau^2+3\tau+1)^2}+\frac{49}{12}
\nonumber\\
&+&\frac{5}{4}\log\tau-\frac{5}{2}\log(\tau+1)+\log^2(\tau+1).
\end{eqnarray}\large
Let us notice that $\tau$ behaves as $x$ for small values of the
transverse momentum
\begin{equation}
\tau(x)=x+O(x^2)
\end{equation}
and it is a unitary variable
\begin{eqnarray}
\tau\rightarrow 0 \ \ \ &\rm{for}& \ \ \ x\rightarrow 0\\
\tau\rightarrow 1 \ \ \ &\rm{for}& \ \ \ x\rightarrow 1/4.\\
\end{eqnarray}
The relation (\ref{tau}) may be inverted as
\begin{equation}
x=\frac{\tau}{(\tau+1)^2}.
\end{equation}
One can easily check that $r(\tau)$ vanishes for $\tau\sim
x\rightarrow 0$, by using the properties
\begin{equation}
J[0,-1,0]=J[0,-3,0]=J[-1,0,0]=J[-1,-3,0]=0
\end{equation}
\begin{equation}
\lim_{\tau\to 0}J[0,-3,1/\tau]=\lim_{\tau\to
0}J[-1,-3,1/\tau]=-\frac{\pi^2}{3}.
\end{equation}
Finally let us observe that the extraction of the remainder
function is based on a matching procedure: in order to avoid
double counting we have subtracted the logarithmic terms already
included in the resummation of logarithms, namely the equation
(\ref{onegluonemission}) after the integration over $x$,
\begin{equation}
-\frac{\alpha_{S}A_{1}}{4}\log^2 x+\alpha_{S}B_{1}\log x.
\end{equation}
The constant appearing in (\ref{completoreale}) has to be summed
to the virtual contribution to get the coefficient function.

\section{Complete Calculation of Virtual Diagrams}
\begin{figure}[h]
\begin{minipage}{15cm}
\parbox{7cm}{
\mbox{\epsfig{file=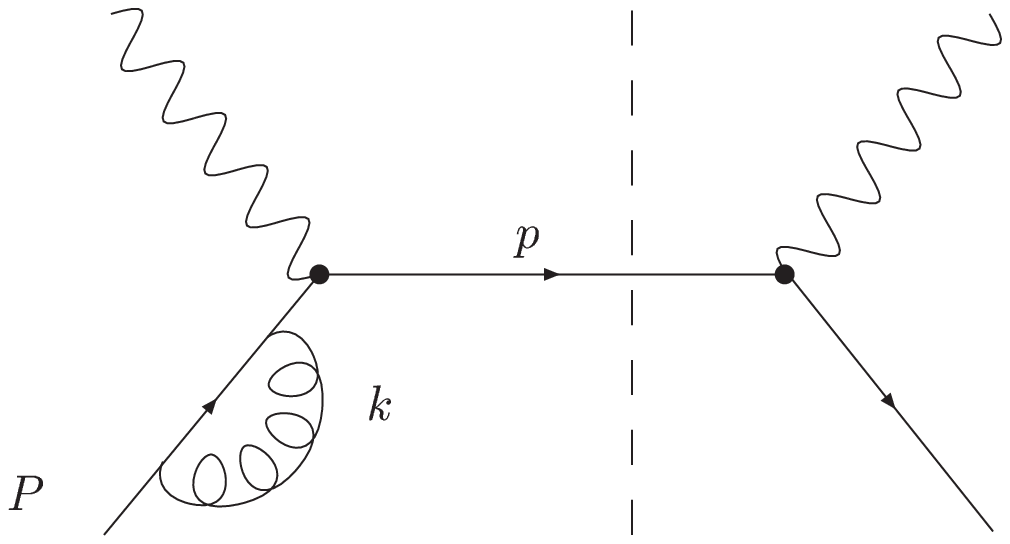, width=7cm}}
\caption{\label{self1}Heavy quark self-energy diagram.}}
\parbox{7cm}{
\mbox{\epsfig{file=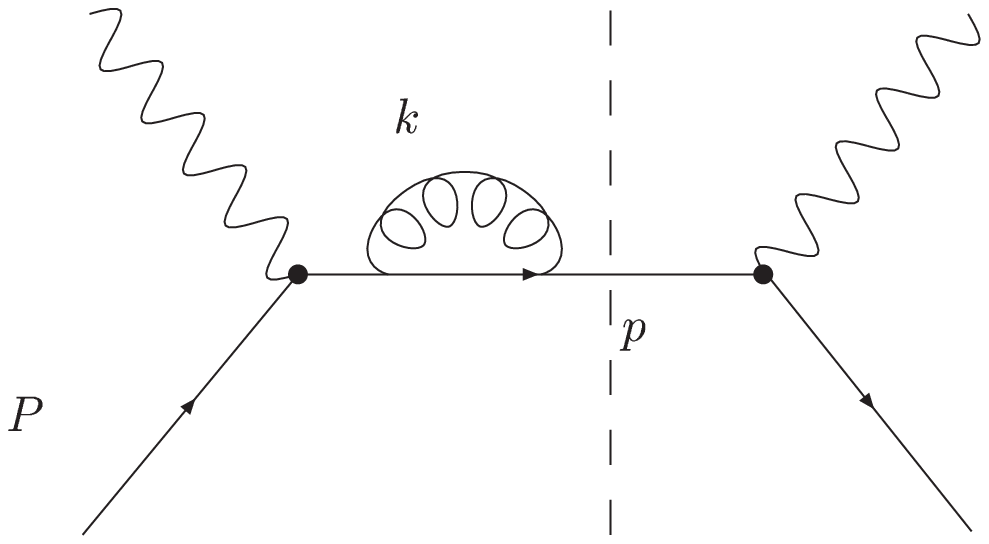, width=7cm}}
\caption{\label{self2}Light quark self-energy diagram.}}
\end{minipage}
\end{figure}

\begin{figure}[h]
\begin{center}
\mbox{\epsfig{file=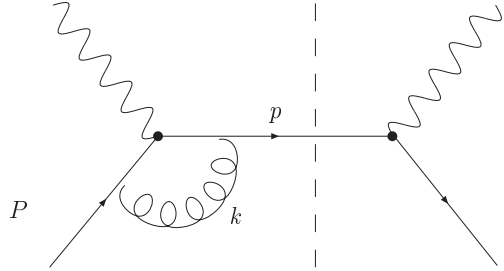, width=7cm}}
\caption{\label{vert}Vertex correction diagram.}
\end{center}
\end{figure}

Virtual corrections to $b\rightarrow s\gamma$ have been calculated
in \cite{greub,greub2} for a massive strange quark; we present
here the computation in the massless case.
\\
The diagrams consist of self-energy corrections to the heavy and
light lines (see fig. (\ref{self1}) and (\ref{self2})) and of
vertex corrections to the operator $\hat{\mathcal{O}}_7$ (see
figure  (\ref{vert})); we compute them in the $\overline{MS}$
scheme so as to be consistent with the (known) coefficient
functions $C_i$. The computation can be done with standard Feynman
parameter technique \cite{peskin} or by a reduction using the
integration by part identities \cite{chettak}.
In order to show both techniques, let us start by calculating the
self-energies with Feynman parameters. The diagrams are calculated
in $n=4-\epsilon$ dimensions, to regularize ultraviolet
divergencies, and off-shell to regularize infrared divergencies,
$p^2\not= m^2$. The renormalization provide for the cancellation
of ultraviolet divergencies by counterterms in a proper scheme,
decided once for all. All the calculations can be consistent with
the chosen renormalization scheme.
\\
The calculation of the diagram in figure (\ref{self1}) gives
\begin{equation}
\Sigma=4\pi\alpha_S\mu^\epsilon\int \ \frac{d^nk}{(2\pi)^n} \
\frac{1}{k^2}\frac{1}{(P-k)^2-m_b^2}
\left[\gamma^\mu(\hat{P}-\hat{k}+m_b)\gamma_\mu\right]-(Z_0-1)m_b-(Z_2-1)\hat{p}.
\end{equation}
$\mu$ is the renormalization scale.
\\
The diagram can be written as
\begin{equation}
\Sigma=A(P^2)+B(P^2)\hat{P}
\end{equation}
where $A(P^2)$ and $B(P^2)$ can be easily calculated, by
introducing Feynman parameters as in appendix \ref{feynpar}. They
turn out to be
\begin{eqnarray}
A(P^2)&=&4\pi\alpha_S\int\frac{dk^\prime}{(2\pi)^n}\int_0^1
\frac{nm_b}{\left[k^{\prime
2}+(P^2-m^2_b)x-P^2x^2\right]^2}-(Z_0-1)m_b,\nonumber\\
B(P^2)&=&4\pi\alpha_S\int\frac{dk^\prime}{(2\pi)^n}\int_0^1
\frac{(2-n)(1-x)}{\left[k^{\prime
2}+(P^2-m^2_b)x-P^2x^2\right]^2}-(Z_2-1)m_b.
\end{eqnarray}
After a straightforward calculation one obtains the value of the
counterterms which cancel the ultraviolet singularities of the
diagram, in $\overline{MS}$ scheme they turn out to be
\begin{eqnarray}
(Z_0-1)&=&\frac{\alpha_S}{\pi}\left[\frac{2}{\epsilon}-\gamma+\log
4\pi\right],\nonumber\\
(Z_2-1)&=&\frac{\alpha_S}{\pi}\left[-\frac{1}{2\epsilon}+\frac{\gamma}{4}-\frac{1}{4}\log
4\pi\right].
\end{eqnarray}
Once the ultraviolet divergencies have been cancelled, the
diagrams can be calculated on-shell and the infrared singularities
can be regularized by taking $n=4+\epsilon$.
\\
The contribution given by the diagram to the rate is
\begin{equation}
\Sigma_1=\frac{\partial A}{\partial P^2}|_{P^2=m^2_b}
2m_b+B(m^2_b)+2m^2_b\frac{\partial B}{\partial P^2}|_{P^2=m^2_b},
\end{equation}
which reads
\begin{equation}
\Sigma_1=\frac{\alpha_S}{\pi}\left[\frac{1}{\epsilon}+\frac{\gamma}{2}
+\frac{1}{2}\log\frac{m^2_b}{4\pi\mu^2}-1+\frac{1}{4}\log\frac{m^2_b}{\mu^2}\right].
\end{equation}
For the massless self-energies in figure (\ref{self2}) the case is
more involved. The only term contributing in $\Sigma_1$ is $B(0)$.
We need to calculate just $B(p^2)$ off-shell, subtract the
ultraviolet divergencies with the counterterm and to determine
$B(0)$, which reads
\begin{equation}
B(0)=\Sigma_1=\frac{\alpha_S}{\pi}\left[-\frac{1}{2\epsilon}-\frac{\gamma}{4}+\frac{1}{4}\log
4\pi\right] .
\end{equation}
Let us now briefly describe the evaluation of the vertex
correction within the second method. One has to compute the scalar
integral:
\begin{equation}
\mathcal{V}=\int {d^nk \over (2\pi)^n} \ {N(k^2,P\cdot k, p\cdot
k;\epsilon)\over k^2 [(k-P)^2-m^2](k-p)^2},
\end{equation}
where
\begin{equation}
N(k^2,P\cdot k, p\cdot k;\epsilon) = 32 P\cdot k -32 p\cdot k - 16
m^2_b +\mathcal{O}(\epsilon^2) k^2 + \mathcal{O}(\epsilon^2)
P\cdot k p\cdot k + \mathcal{O}(\epsilon^2)  {(p\cdot k)}^2  .
\end{equation}
$\mathcal{V}$ has at most a double pole in $\epsilon$ coming from
the product of the soft and the collinear singularities. The terms
in the numerator $N$, which vanish in the soft limit
$k_{\mu}\rightarrow 0$, do not give rise to soft singularities and
therefore produce at most a simple pole coming from the collinear
or the ultraviolet region. Therefore the $\mathcal{O}(\epsilon^2)$
terms in $N$ do not contribute in the limit $\epsilon\rightarrow
0$. Moreover, since the only terms in $N$ which can give
ultraviolet divergencies are quadratic in $k$ and can produce at
most a single pole in $\epsilon$, $\mathcal{V}$ can be safely
calculated in $n=4+\epsilon$ dimensions to regularize just
infrared divergencies.
\\
\\
By expressing the scalar products in the numerator as linear
combinations of the denominators as
\begin{eqnarray}
\label{rotate} k\cdot p &=& \frac{1}{2} \left( k^2 - (k-p)^2
\right),
\nonumber \\
k\cdot P &=& \frac{1}{2} \left( k^2 - (k-P)^2 + m_b^2\right),
\end{eqnarray}
we can reduce $\mathcal{V}$ to a superposition of scalar integrals
of the form:
\begin{equation}
{\rm T}[a,b,c]=\int {d^nk \over (2\pi)^n} \ {1 \over [k^2]^a \
[(k-P)^2-m_b^2]^b \ [(k-p)^2]^c}
\end{equation}
with $a,b,c\le 1$. The above amplitudes can be related to each
other by identities of the form \cite{chettak}:
\begin{equation}
\int d^n k \ {\partial \over \partial k^\mu} \  {  v^{\mu}
 \over [k^2]^a [(k-P)^2-m^2]^b[(k-p)^2]^c  } =0
\end{equation}
with $v^{\mu}=k^\mu,p^\mu,P^\mu$. By explicitly evaluating the
derivatives and re-expressing the scalar products using eqs.
(\ref{rotate}), one obtains relations among amplitudes with
shifted indices.
\\
\begin{figure}[h]
\begin{center}
\mbox{\epsfig{file=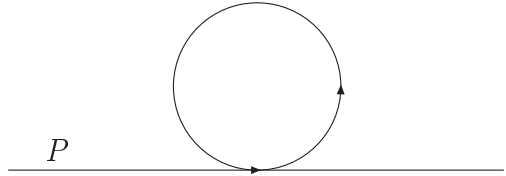, width=7cm}}
\caption{\label{tadpole}Massive tadpole diagram.}
\end{center}
\end{figure}
\\
By solving the above identities, one can reduce all the amplitudes
to the tadpole (figure \ref{tadpole}), and obtains for the
integral\footnote {Such a strong reduction of 3-point function to
a vacuum amplitude is possible because the only scale in the
process is the heavy quark mass $m_b$. Virtual corrections have
indeed the lowest-order kinematics $P^2 = m_b^2,~P\cdot p =
m_b^2/2,~p^2 = q^2 = 0$.}:
\begin{equation}
\mathcal{V} = \left(+{16 \over \epsilon} -8+8\epsilon\right)
{1\over m^2_b} \ {\rm T}[0,1,0],
\end{equation}
where
\begin{equation}
{\rm T}[0,1,0]= C_F{\alpha_S\over 16\pi} \ \left({m^2\over
4\pi\mu^2}\right)^{\epsilon / 2} \ {\Gamma(-\epsilon / 2)\over
1+\epsilon/2} \ m_b^2.
\end{equation}
The contribution associated to the vertex correction reads
\begin{equation}
\mathcal{V}=C_F\frac{\alpha_S}{\pi}\ \left({m^2\over
4\pi\mu^2}\right)^{\epsilon / 2}\
\Gamma\left(1-\frac{\epsilon}{2}\right)\
\left[-\frac{2}{\epsilon^2}+\frac{2}{\epsilon}-2\right].
\end{equation}
Summing self-energies and vertex corrections, and subtracting the
$1/\epsilon$ poles according to the $\overline{MS}$ scheme, one
obtains for their contribution to the rate $D_V$ \footnote{To
factorize $\Gamma_0$ one has to replace
$m_{b,\overline{MS}}(\mu_b)$ by $m_{b,\overline{MS}}(m_b)$ using
the formula $m_{b,\overline{MS}}(\mu_b)=m_{b,\overline{MS}}(m_b) (
1+\frac{3}{2} {C_F\alpha_S\over\pi})$.}:
\begin{equation}
D_V=C_F{\alpha_S\over\pi} \ \left({m^2_b\over
4\pi\mu^2}\right)^{\epsilon/2} \ \Gamma\left(1-{\epsilon\over
2}\right) \ \left[-{2\over \epsilon^2}+{5 \over 2\epsilon} + 4
\log{m_b\over \mu_b} -3\right].
\end{equation}
We have kept the factor in front of the square bracket unexpanded
to simplify the computation of the total rate.
\\
The virtual corrections to the remaining operators
$\hat{\mathcal{O}}_{i\not= 7}$ contain only (simple) ultraviolet
poles in $\epsilon$, which are removed by renormalization; their
contributions to $D_V$ amount only to finite constants and $\log
m_b/\mu_b$.

\section{Final Result: the Coefficient Function and the Remainder Function}\label{finalcoef}
\begin{figure}[h]
\begin{center}
\mbox{\epsfig{file=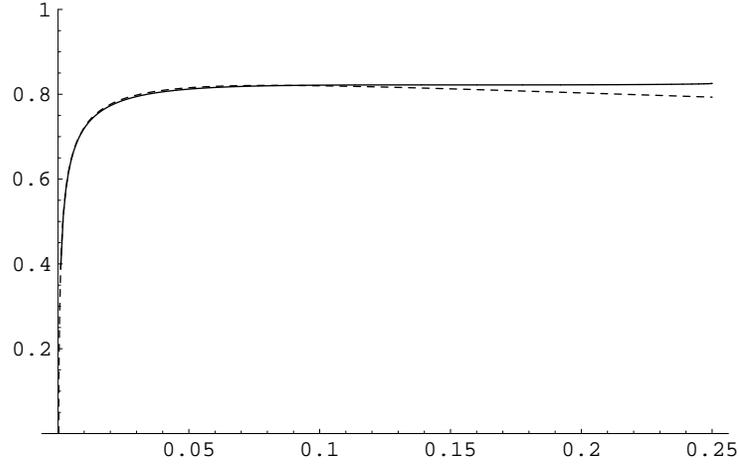, width=10cm}}
\caption{\label{cfr}Comparison between the full fixed order
calculation for $D(x)$ (solid line) and the logarithmic
approximation (dashed line).}
\end{center}
\end{figure}

Summing real and virtual contributions, the transverse momentum
distribution for the decay $b\rightarrow s\gamma$ reads, to
$O(\alpha_S)$:
\begin{equation}\label{final}
D(x)=1~+~C_F{\alpha_S\over\pi}\ \left[-{1 \over 4} \ \log^2 x - \
{5 \over 4} \ \log x + \kappa_1 +\ r_1(x)\right].
\end{equation}
As expected, the result contains a double logarithm and a single
logarithm of $x$, a finite term $\kappa_1$ and a function $r_1(x)$
vanishing in the limit $x\rightarrow 0$.
\\
By expanding the resummed formula (\ref{master}) to order
$\alpha_S$ one obtains:
\begin{eqnarray}\label{final2}
D(x) &=& \left( 1+\frac{C_F\alpha_S}{\pi} \kappa_1 \right)
\left(1-\frac{A_1}{4}\alpha_S \log^2 x + B_1 \alpha_S \log x
\right)+ \frac{C_F\alpha_S}{\pi} r_1(x) \nonumber
\\
&=& 1-\frac{A_1}{4}\alpha_S \log^2 x + B_1 \alpha_S \log
x+\frac{C_F\alpha_S}{\pi} \kappa_1 +\frac{C_F\alpha_S}{\pi}
r_1(x)+O(\alpha_S^2).\nonumber\\
\end{eqnarray}
The matching procedure consists in the identification of the
values of $\kappa_1$ and $r_1(x)$ and the computation allows to we
check the values for $A_1$ and $B_1$, evaluated in \cite{noi1}
using general properties of QCD radiation.
\\
The value of the first term in the expansion of the coefficient
function is
\begin{equation}\label{coefficient}
k_1= - \ {11 \over 4}- \ {\pi^2 \over 12} +~ 4\log{m_b\over \mu},
\end{equation}
The value of the first term in the expansion of the remainder
function is given in equation (\ref{taudef}).
\\
As explained in previous sections, the remaining operators
$\hat{\mathcal{O}}_{i\not= 7}$ contribute to $D(x)$ only by finite
terms $\tilde{r}_i$ and remainder functions. Since the constants
$\tilde{r}_i$ come from virtual diagrams alone, we can quote their
result from \cite{greub,greub2} and present an improved formula
for the coefficient function, in analogy with \cite{acg}:
\begin{equation}\label{coefcompleto}
K(\alpha_S)~=~ 1 + {\alpha_S \over 2\pi} \sum_{i=1}^8
{C^{(0)}_i(\mu_b) \over C^{(0)}_7(\mu_b)}\left(\Re\  \tilde{r}_i +
\gamma^{(0)}_{i7}\log{m_b\over \mu_b}\right)+{\alpha_S \over
2\pi}{C^{(1)}_7(\mu_b) \over
C^{(0)}_7(\mu_b)}+\mathcal{O}(\alpha_S^2)
\end{equation}
where
\begin{eqnarray}
\tilde{r}_{i} &=& r_i ~~~~~ i\not= 7
\nonumber \\
\tilde{r}_7 &=& \frac{8}{3}\left( f-4\log\frac{m_b}{\mu_b}\right)=
-\frac{22}{3}-\frac{2\pi^2}{9}.
\end{eqnarray}
Let us remark that only the coefficients related to the operators
with $i=1,2,7,8$ are relevant, because the others are multiplied
by very small coefficient  functions and can be neglected:
\begin{eqnarray}
r_1&=&-{1 \over 6}r_2\nonumber\\
\Re \ r_2&=& -4.092-12.78(0.29-m_c/m_b)\nonumber\\
r_8&=& {4 \over 27}(33-2\pi^2).
\end{eqnarray}
The analytic expressions for the coefficient functions as well as
a standard numerical evaluation are given in \cite{cmm}. The
anomalous dimension $\gamma^{(0)}_{77}$ is derived from the
coefficient of the logarithmic term in $k_1$. The values of
$\gamma_{i7}^{(0)}$ are \cite{cmm}:
\begin{equation}
\gamma^{(0)}_{i7}=\left(-{208 \over 243}, {416 \over 81}, -{176
\over 81}, -{152 \over 243}, -{6272\over 81}, {4624 \over 243},{32
\over 3},-{32 \over 9}\right).
\end{equation}
\\
The calculation of the coefficient function in
(\ref{coefcompleto}) complete the program of next-to-leading
resummation.
\\
The remainder function is in general negligible for small
transverse momenta, for larger value of transverse momenta a small
change in the fixed order prediction can be noticed, as shown in
figure (\ref{cfr}). The figure shows that for larger values than
$x=0.1$ the remainder function produces a correction of the order
of $O(10-15\%)$ with respect to the only logarithmic contribution.
\\
Obviously this happen where logarithms are already quite small and
this effect can be taken into account near the fully inclusive
region of the spectrum, but can be discarded in the semi-inclusive
one (small $p_t$).
\\
\\

\chapter{Mass Effects from the Final Quark}\label{masseffects}
\large
\section{Motivations}
In this chapter preliminary results are shown, concerning the
structure of infrared logarithms in heavy flavour decays, where
both the decaying quark and the final quark are massive.
\\
Example of these transitions are $top \rightarrow beauty$, $beauty
\rightarrow charm$, $beauty \rightarrow strange$ (with a massive
strange quark).
\\
The main phenomenological difference with respect to transitions
like $beauty \rightarrow up$ or $beauty \rightarrow strange$ (if
we neglect the strange quark mass) is that in this case the
hadronic final jet branches from a massive quark.
\\
In order to be plain, we deal with a specific physical
distribution, the photon spectrum distribution in $b \rightarrow
s\gamma$, but it should be clear that the results apply to other
$Q \rightarrow Q$ transitions with few changes in kinematics.
\\
In particular from a phenomenological point of view it should be
very interesting to apply these results to the transition
$$
b\rightarrow c.
$$
In fact it has a much higher rate than the rare process we are
considering and it is widely investigated from an experimental
point of view. However, being the kinematics more complicated, we
decided to start from a simpler process, anyhow having a great
phenomenological importance.
\\
In this chapter we analyze the main differences with respect to
the massless case, due to soft gluon emissions, which are affected
by the the presence of a mass in the final state.

\section{Kinematics}
Let us consider the photon spectrum in the decay
\begin{equation}\label{procadronico}
B\rightarrow X_s + \gamma
\end{equation}
where $X_S$ is an hadronic jet containing a strange quark.
\\
At the tree level in the partonic process it is
\begin{equation}\label{procpartonico}
b\rightarrow s\gamma.
\end{equation}
Two practical reasons drove us to consider this specific process:
first of all at the tree level it is a two body decay and its
kinematics is very simple, moreover the results for this decay
neglecting the strange quark mass are well known and it can be
interesting to compare the effects due to the presence of the
mass. When we will need such a comparison, the case with the
strange mass neglected will be referred as the \it massless
case\rm.
\\
At the tree level the photon energy is fixed by kinematics
\begin{equation}\label{energiatree}
E_\gamma = {m^2_b - m^2_s \over 2m_b}.
\end{equation}
In the following calculations $m_s\not= 0$, but $$\mu = {m_s^2
\over m^2_b} \ll 1$$ Radiative corrections are given by gluon
emissions
\begin{equation}\label{moltigluoni}
b\rightarrow s\gamma g_1 \dots g_n
\end{equation}
which spread and shift the photon spectrum peak.
\\
Let us define the invariant jet mass
$$M^2_X = (\sum_i p_i)^2$$
where $p_i$ are parton (strange quark + gluons) momenta.
\\
The photon energy is related to the invariant mass by the relation
\begin{equation}\label{energia}
E_\gamma = {m_b^2 - M^2_X \over 2m_b}.
\end{equation}
Let us define a unitary variable $X_\gamma$ as
\begin{equation}\label{variabile}
X_\gamma = E_\gamma \ {2 m_b \over m^2_b- m^2_s}
\end{equation}
so that $X_\gamma \rightarrow 1$ is the elastic limit.
\\
Another useful relation is
\begin{equation}\label{relazione}
(1 - X_\gamma) = {m^2_X \over m^2_b - m^2_s}
\end{equation}
where
\begin{equation}\label{massaridotta}
m^2_X = M^2_X - m^2_s.
\end{equation}
If $m_s=0$
$$
(1-X_\gamma) = {m^2_x \over m_b^2} = {M^2_X \over m^2_b}
$$
and we can recognize the kinematics of the massless case.
\\
The correct hard scale in the process of jet evolution is
\begin{equation}\label{scala}
Q =2E_X
\end{equation}
as discussed in \cite{ugoscala}. $E_X$ is the energy of the
hadronic jet and the factor 2 is introduced to simplify the
calculations.
\\
The jet energy is related to the invariant mass by the relation
\begin{equation}\label{energiajet}
E_X = {m^2_b + M^2_X \over 2m_b}
\end{equation}
Because of (\ref{energia}) $M^2_X$-distribution and photon
spectrum are strictly related: in the following we will calculate
relations for the $M^2_X$-spectrum because it involves just the
evolution of the jet, and then we will go back again to the photon
spectrum using (\ref{energia}).

\section{One Soft Gluon Emission}
Let us now consider $\alpha_S$ corrections and in particular one
soft real gluon emission
\begin{equation}\label{ungluone}
b \rightarrow s\gamma g.
\end{equation}
Let us define
\begin{equation}\label{momentojet}
p^\mu_X = p^\mu_s +p^\mu_g
\end{equation}
so that $p^2_X = M^2_X$.
\\
It is well known that the rate for one real gluon emission is
affected by infrared singularities of soft and collinear nature,
which are cancelled by virtual emissions, for infrared-safe
variables.
\\
Here we are focused on the eikonal approximation, described in
section (\ref{seclogs}). As it was discussed there, this
approximation allows to study the structure of logarithms arising
from soft gluon emissions and it lies on the definition of an
eikonal current
\begin{equation}\label{corrente}
J^\mu =g_S T_b {p^\mu \over p_b\cdot p_g} - g_S T_s {p^\mu \over
p_s \cdot p_g}
\end{equation}
where $p^\mu_b$ is the beauty quark momentum and $T_i$ are  color
generators.
\\
The squared matrix element is given by
\begin{eqnarray}\label{elementomatrice}
-g_{\mu\nu}J^\mu J^\nu &=& 4\pi\alpha_S \left(2 T_b\cdot T_s {p_b
\cdot p_s \over (p_b \cdot p_g)(p_s \cdot p_g)} - T^2_b {m^2_b
\over (p_b\cdot p_g)^2} - T^2_s {m^2_s \over (p_s \cdot
p_g)^2}\right)\nonumber \\ &=& {\cal{M}}_{bs} + {\cal{M}}_{bb} +
{\cal{M}}_{ss}
\end{eqnarray}
The difference with respect to the massless case is that we need
to calculate all the three terms in the previous formula, in
particular the third one is different from zero in this case:
moreover the different kinematics produces corrections to the
other terms.
\\
The color structure is trivial because of the presence of just two
coloured partons in the hard vertex.
\\
The hard vertex conserves the color, so that
$$ T_b = T_s $$
which implies
$$ T^2_b = T^2_s = T_b \cdot T_s = C_F. $$
Because of the relation (\ref{energia}) we can calculate the
invariant mass distribution to get the photon spectrum: the
kinematical constraint can be derived by (\ref{momentojet}) whose
square is
\begin{equation}
m^2_s = M^2_X - 2p_X\cdot p_g = M^2_X - 2E_gE_X\left(1 - \sqrt {1
- {4M^2_X \over Q^2}} \cos \theta\right)
\end{equation}
where $\theta$ is the angle between the strange quark and the
gluon.
\\
Let us now define
\begin{equation}\label{omega}
\omega = {2E_g \over Q} = {E_g \over E_X}
\end{equation}
\begin{equation}\label{zeta}
z =\cos\theta
\end{equation}
the constraint becomes
\begin{equation}\label{vincolo}
{1 \over Q^2} \ \delta\left[{m^2_X \over Q^2} - {1 \over 2}\omega
\left(1-\sqrt {1 - {4M^2_X\over Q^2}}z\right)\right].
\end{equation}
Now we have to integrate the squared matrix element over the phase
space with the kinematical constraint
\begin{equation}\label{Mdistr}
\frac{1}{\Gamma_0} \left.\frac{d\Gamma}{dM^2_X}\right)_{SOFT} =
\int {d^3p_g \over (2\pi)^32E_g} \left(-g_{\mu\nu} J^\mu
J^\nu\right) {1 \over Q^2} \delta\left[{m^2_X \over Q^2} - {1
\over 2}\omega \left(1-\sqrt{1 - \frac{4M^2_X}{Q^2}}\
z\right)\right].
\end{equation}
The kinematical constraint does not introduce particular changes
in the limits of integration of the phase space, which turn out to
be
\begin{eqnarray}
0<&\omega&<1,\nonumber\\
-1<&z&<1.
\end{eqnarray}
Let us consider the three terms in the eikonal current separately:
the dominant contribution, the one giving the leading logs in the
massless case, is
\begin{equation}\label{interferenza}
{\cal{M}}_{bs} = 8\pi\alpha_S \ \left({p_b\cdot p_s \over
(p_b\cdot p_g)(p_s \cdot p_g)}\right).
\end{equation}
The integrations gives
\begin{equation}\label{interfintegr}
{1 \over \Gamma_0} \left.{d\Gamma \over
dM^2_X}\right)_{bs}=\alpha_S {C_F \over \pi} \ {1 \over m^2_X} \
{E_X \over p_X} \log \left[{E_X + p_X \over E_X - p_X}\right]
\end{equation}
where $p_X = \sqrt{E^2_X - M^2_X}$.
\\
Recalling that
\begin{eqnarray}\label{massainv}
M^2_X &=& m^2_b(\mu + (1-\mu)(1-X_\gamma)) \nonumber
\\  &\approx& m^2_b(\mu + (1-X_\gamma))
\end{eqnarray}
we have
\begin{equation}\label{doppiologmassive}
\frac{1}{\Gamma_0} \left.\frac{d\Gamma}{dM^2_X}\right)_{bs}= {1
\over m^2_b - m^2_s} \ \alpha_S {C_F \over \pi} {1 \over 1-
X_\gamma} \log\left[\mu + (1-X_\gamma)\right]
\end{equation}
up to terms with no logarithmic enhancement.
\\
The argument of the logarithm signals that the most interesting
feature in this case is the interplay between the variable $( 1 -
X_\gamma)$ and the parameter $\mu$; in fact we can clearly
discriminate three different regimes:
\begin{itemize}
\item
if $(1 - X_\gamma) \gg \mu $ we fall in the massless case, getting
the usual double logarithmic structure
$$ \log[1-X_\gamma] \over 1-X_\gamma;$$
\item
if $(1 - X_\gamma) \ll \mu $, that is very near the endpoint we
have a quasi-collinear logarithm
$$ \log[\mu] \over 1-X_\gamma.$$
The collinear enhancement disappears because of the mass screen
from the massive final parton: in this case we have a single log
structure;
\item
if $(1 - X_\gamma) \sim \mu $ both the terms contribute and we
have an interplay between the massless and massive case.
\end{itemize}
\begin{figure}[h]
\center\mbox{\epsfig{file=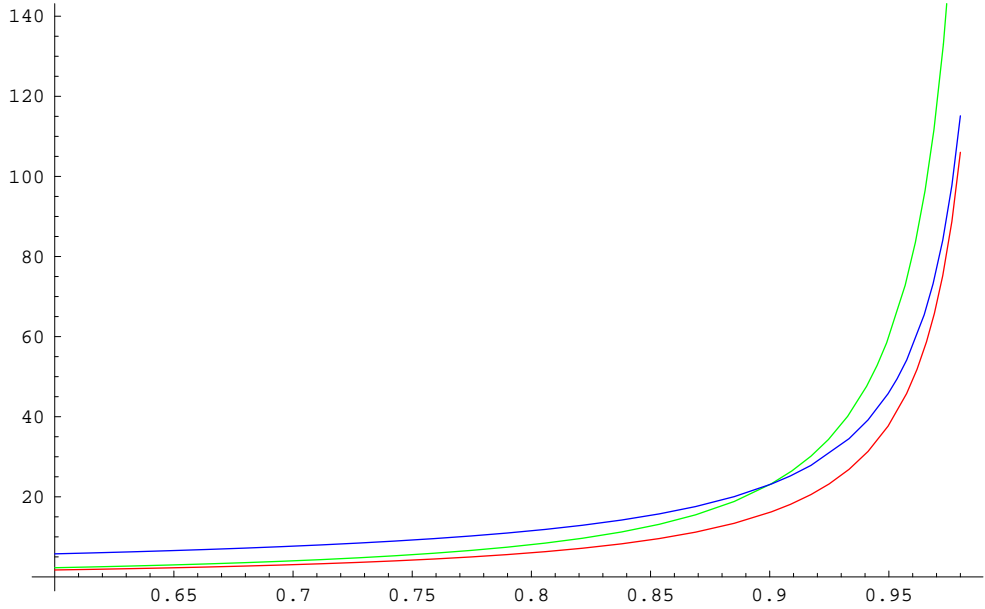, width=13cm}} \caption{The
figure shows the different regime discussed above in the double
logarithmic approximation. The red line represents the full result
obtained in (\ref{doppiologmassive}), compared with the massless
case (green line) and with purely massive case (blue line). The
picture graphically shows the behaviours and the approximations
discussed in the section: far from the endpoint the mass is
negligible and the massless result is reproduced; approaching the
endpoint the mass effect makes its appearance.}
\end{figure}
The second contribution to the current that we want to calculate
is
\begin{equation}\label{emissioneb}
{\cal{M}}_{bb} = -4\pi\alpha_S {m^2_b \over (p_b \cdot p_g)^2}.
\end{equation}
After the integration we get
\begin{equation}
{1 \over \Gamma_0} \left.{d\Gamma \over dM^2_X}\right)_{bb} =
-\alpha_S {C_F \over \pi} {1 \over m^2_X}.
\end{equation}
Recalling (\ref{relazione})
\begin{equation}\label{integrb}
{1 \over \Gamma_0} \left.{d\Gamma \over dM^2_X}\right)_{bb} = -
\alpha_S {C_F \over \pi} {1\over m^2_b - m^2_s} {1 \over
1-X_\gamma}
\end{equation}
This result represents the usual soft contribution as expected
from the massless case: this is the only term not affected by the
presence of a massive quark in the final state.
\\
The last term in the current is new with respect to the massless
case, because it doesn't appear if $m_s = 0$
\begin{equation}\label{emissiones}
{\cal{M}}_{ss} = - 4\pi\alpha_S {m^2_s \over (p_s\cdot p_g)^2}.
\end{equation}
The integration gives
\begin{eqnarray}\label{integrs}
{1 \over \Gamma_0} {d\Gamma \over dM^2_X}\left.\right)_{bb}&=&
\alpha_S{C_F \over \pi} {m^2_X \over M^2_x  m^2_s} = \nonumber
\\ &=& \alpha_S{C_F \over \pi} {1 \over m_b^2 - m^2_s} \left[ \  {1 \over 1
- X_\gamma} -{1 \over 1 -X_\gamma + \mu}\ \right].
\end{eqnarray}
Let us note that (\ref{integrs}) is zero if $m_s = 0$ as we
expected from the current.
\\
The result shows a pure soft term, $1 \over  1- X_\gamma$, and a
term screened by the massive quark, $1 \over (1-X_\gamma) + \mu $:
for $(1-X_\gamma) \ll \mu$ the second term is no longer enhanced
and we can neglect it.
\\
The photon spectrum for soft emissions becomes
\begin{eqnarray}\label{softmassfin}
{1 \over \Gamma_0} {d\Gamma \over dX_\gamma}\left.\right)_{soft}
&=& \delta(1-X_\gamma) - \alpha_S A_1 {1 \over 1-X_\gamma} \log
\left[{1 \over (1 - X_\gamma) + \mu}\right] \nonumber\\ &-&
\alpha_S \ 2S_1 \ {1 \over 1 - X_\gamma} +\alpha_S \ S_1 \ {1
\over (1 - X_\gamma) + \mu}
\end{eqnarray}
Let us note that for $m^2_s \rightarrow 0$ we get the massless
result, while for $\mu \gg (1 - X_\gamma)$ the soft logarithm has
a coefficient twice with respect to the massless case.
\\
We defined
\begin{equation}\label{auno}
A_1 = {C_F \over \pi},
\end{equation}
\begin{equation}\label{suno}
S_1 = -{ C_F \over \pi},
\end{equation}
in analogy with the massless case.

\section{Quasi-Collinear Emissions}
The mass of the quark in the final state provides a natural
cut-off for real collinear emissions: gluon emissions have to
respect an angular restriction, known as dead cone effect, as
discussed in section (\ref{seclogs}).
\\
In this case we can tell two main behaviours: if the mass of the
final quark is comparable to the hard scale (as maybe it could be
for the transition $b\rightarrow c$) then the collinear
logarithmic enhancement, matter of fact, disappears. In fact
logarithmic terms will surely appear as in every case where
different energy scales are involved, but these logarithms can be
considered small, or to be more precise, not enhanced, because the
two scales are not very much different.
\\
Otherwise, if the final quark mass does not vanish, but is small
with respect to the hard scales, collinear logarithms turn into
less divergent terms, which we will call \it quasi-collinear\rm.
\\
The formal difference is that in this case the collinear
logarithms does not diverge when the variable we are considering
(for example $E_\gamma$) reaches the endpoint of the spectrum;
from a physical point of view this means that the collinear
emission is screened because of the dead cone effect.
\\
The emission is still described by the Altarelli-Parisi kernel
(\ref{apkernel}),
\begin{equation}
P_{qq}(\omega)=\frac{2-2\omega+\omega^2}{\omega}\stackrel{{\rm
leading}}{\rightarrow}\frac{2}{\omega},
\end{equation}
but the integration over the polar angle is restricted according
to the relation (\ref{deadcone}):
\begin{equation}
z>1-\frac{M^2_X}{Q^2}.
\end{equation}
The integration over the phase space gives the result in
(\ref{doppiologmassive}) which represents the leading logarithmic
approximation and therefore has both collinear and soft
enhancement.
\\
As stated above, while in the massless case the distribution
formally diverges as
\begin{equation}
\frac{\log (1-X_\gamma)}{1-X_\gamma}
\end{equation}
at the endpoint of the spectrum, in this case the divergence is
less dramatic and goes as
\begin{equation}
\frac{\log \mu}{1-X_\gamma}.
\end{equation}
As discussed above if the mass of the initial and final quark are
not much different, that is their masses does not different for
one or more scales of magnitude, then this enhancement is no
longer present: physically this means that if the masses are
similar the collinear emission is so inhibited to be unrelevant.

\section{Final Considerations}
When the mass of the final quark is considered a new parameter is
introduced in the calculation and this makes the structure of the
result richer  even at the logarithmic level. Obviously we expect
that a complete calculation would show an increasing number of
regular functions, with respect to the massless case, depending on
the kinematics and in particular from the ratio of the masses of
the quarks.
\\
Concerning the logarithmic structure, which we are interested in,
the introduction of the mass of the final quark introduces
different regimes, which mainly depend on the value of this mass.
\\
For real emissions we have studied the behaviour of the eikonal
current, showing that if the mass is negligible the result of the
massless case are obtained: however, studying a specific case, for
example the photon spectrum, it is shown that near the endpoint
the presence of the mass changes the spectrum giving a less
divergent behaviour, because the logarithm of collinear nature is
screened. In fact, due to the presence of the mass, the collinear
emission  under a minimum angle is forbidden (dead cone effect).
\\
Moreover the mass introduces new terms in the eikonal current
(which correctly disappear in the massless limit). The importance
of each term strongly depend on the value of the mass: if the mass
is small the effect is a correction to the massless case, if the
mass is large the structure changes sizeably. In this case (see
eq. (\ref{softmassfin})), for example, the double logarithmic
enhancement disappears, the soft contribution has a coefficient
twice the one of the massless case and the collinear emission can
be neglected at this level of accuracy.
\\
The results discussed in the previous sections are only
preliminary: they should be developed to study the resummation of
large logarithms when they appear and they should apply to
different phenomenologies to understand when the effects of the
final mass can be relevant.
\\
In fact the presence of the mass manifest itself, as discussed, in
different point of the spectrum: if the mass is small just at the
endpoint, if the mass is large in regions far from the endpoint.
If they manifest very near the endpoint it is very probable that
they are overwhelmed by non perturbative effects and will be
difficult to study and show. This implies that the application of
this formalism should show different behaviours in the cases of
the transition $b\rightarrow c$ (where the final mass is large)
and in $b\rightarrow s\gamma$ (where the final mass is reasonably
small).
\\
We expect that the calculation of virtual diagrams will have the
main effect to introduce plus distribution, for example
\begin{equation}
\frac{1}{1-X_\gamma}\rightarrow \frac{1}{(1-X_\gamma)}_+
\end{equation}
without changing the general structure of the result.
\chapter{Conclusions}\label{conclusionith}
The main topic of our work has been the calculation of the
transverse momentum distribution of the $strange$ quark with
respect to the photon direction, in the rare decay of the $beauty$
quark $b\rightarrow s\gamma$.
\\
The physical motivations which lead us to study this particular
process are outlined in chapter \ref{introtmd}: there a discussion
of the physical effects which lead to a generation of transverse
momentum are described. In particular we focused our attention on
the perturbative part, whose limits give informations about the
structure of non perturbative contributions: such a distribution
is a typical case where large logarithms appear near the border of
the phase space, for the emission of soft and collinear gluons.
\\
The appearance of logarithms always occurs in calculations where
different scales are involved: however if the scales have very
different sizes these logarithms becomes so large to spoil the
perturbative expansion.
\\
In the past years a technique to resum them was developed, in
order to obtain an improved perturbative formula, with a range of
applicability in a wider region of the phase space: this technique
was successfully applied to processes at high energies, namely of
the order of intermediate boson mass. Our aim was an application
of such a technique to processes at lower energies, in our case at
the energy of the $beauty$ quark mass.
\\
In this case theoretical predictions are quite involved even in
inclusive quantities: several poorly known parameters appear, such
as CKM matrix elements or the quark mass, so that a stringent
check of QCD at this scale of energies is not trivial. Matter of
fact there are not indications that QCD should fail at this scale
of energies nor a very precise check of its reliability.
\\
The strong coupling constant seems to be small enough to allow a
reliable perturbative expansions, and in distributions such as the
one we studied the technique of resummation seems to be necessary
and viable. However only a precise calculation, for example with
next-to-leading accuracy, as we performed, compared with equally
precise experimental data can suggest if the perturbative QCD is
reliable or other effects play a crucial role.
\\
This thesis is divided in a first part, where general tools to
deal with QCD corrections in $b$ physics are outlined and a second
part where the calculations we have performed are explicitely
described.
\\
After a description of the kinematical properties of the process
of interest (Chapter \ref{introtmd}), in Chapter \ref{resumtmd}
the resummation of large logarithms, performed with the purposes
indicated above, is shown.
\\
As usual even the resummed formula cannot be applied in the whole
phase space, since near the infrared region non perturbative
effects make their appearance and singularities appear in the
resummed formula. These singularities give information about the
non perturbative physics.
\\
In Chapter \ref{compare} the comparison with a complementary
distribution calculated for the same process, the threshold
distribution, is outlined: the comparison with the different
structure of singularities allows to conclude that in the
transverse momentum case the non perturbative effects, above all
the Fermi Motion, cannot be dealt with as in the threshold case.
There a new non perturbative function (shape function), based on
the HQET was introduced, here this approach is not viable, for the
motivations contained in Chapter \ref{compare}. The main reason is
that in the threshold case Fermi Motion effects and hadronization
can be separated by a scale, while in the transverse momentum case
this does not occur. One should try in the latter case an
alternative approach, possibly based on a different effective
theory, as outlined in the text.
\\
Finally the calculation is completed in Chapter \ref{fixedorder},
where a complete computation of real and virtual diagrams to order
$\alpha_S$ is performed: this is necessary to extract some
ingredients appearing in the general resummed formula, above all
the constants involved in the coefficient function. Moreover it
allows the extraction of the remainder function, the regular
function describing the behaviour of the distribution in the hard
region of the phase space: its knowledge is not strictly required
for a next-to-leading resummation, but can improved the quality of
the prediction far from the semi-inclusive region. However most of
the events are expected in the region of small transverse momenta
and therefore the phenomenological impact of the remainder
function is expected to be of a few percent.
\\
On the other side, the analytical computation of this term, which
could seem academical, is an interesting application of techniques
of advanced analytical calculation, as described in Chapter
\ref{fixedorder}, which can be used for future purposes to
approach the computation of other processes.
\\
In the last Chapter, \ref{masseffects}, a different topic is
sketched: the effect of the mass of the final quark in $b$ decays.
The result there presented are preliminary, since a resummation of
the logarithms has not been performed yet and only the logarithmic
structure to one loop is analyzed. However this is the first step
towards a more general approach, taking into account the
logarithmic enhancement in decays with a massive final quark. This
will have an important application for the transition
$b\rightarrow c$, where the mass of the $charm$ quark cannot be
surely neglected. This process is very interesting from a
phenomenological point of view: having a large rate, its study is
intense at the $b$-factories.
\\
To summarize, the main object of our work has been the application
of the logarithmic resummation to the decay of an heavy quark. The
calculation has been performed for the first time with
next-to-leading accuracy, including constants and remainder terms,
and the comparison with experimental data could be very useful to
verify informations about the reliability of perturbative QCD in
$b$ decays, the role played by non perturbative physics.
\\
As underlined during the text, the process of interest is a rare
decay, but it can be considered as a first step to face other
processes that are very interesting from the phenomenological
point of view.
\\
This work could have several and possibly interesting extensions:
first of all we hope that a systematic phenomenological study
could be carried on within an extended comparison with data
collected at the $b$ factories. Obviously this would be crucial to
argue the correctness of the approaches that are used in $b$
decays, such as perturbative QCD.
\\
A second point would be the extension of the shape function
approach to the $p_t$ case: as underlined during the text,
physical reasons forbid the application of this approach to the
$p_t$ case and one should try to use different tools. Let us
notice here that a good parametrization of non perturbative
effects is crucial, not only for this particular decay, but also
for other important decays, such as $b\rightarrow u$, because the
same non perturbative functions are involved.
\\
A further point would be the application of these techniques to
other $b$ quark decays and to other distributions, for example the
jet broadening, in order to reach a more complete knowledge of the
dynamics of QCD at this scale of energies.
\\
Finally the results of Chapter \ref{masseffects} should be
completed to face more general problems, such as the decay of a
$beauty$ quark into another massive quark, mainly a $charm$, but
also a $strange$. The study of mass effects in perturbative
calculation and in the resummation of logarithms can give higher
reliability of the predictions of the theory.

\part{Appendices and Bibliography}
\large
\appendix
\chapter{Feynman Rules and Numerator Algebra}\label{feynrules}
This appendix contains relevant Feynman rules, used in the
calculations performed in perturbative QCD. Feynman rules are
extracted by the lagrangian of the theory, which reads:
\begin{eqnarray}
{\cal{L}}_{QCD}&=&\overline{\Psi}(i\gamma^\mu\partial_\mu-m)\Psi-\frac{1}{4}(\partial_\mu
A^a_\nu-\partial_\nu A^a_\nu)^2+g_S
A^a_\mu\overline{\Psi}\gamma^\mu t^a \Psi\nonumber\\&-&g
f^{abc}(\partial_\mu A^a_\nu)^2 A^{b\mu}
A^{c\nu}-\frac{1}{4}(f^{eab}A^a_\mu
A^b_\nu)(f^{ecd}A^{b\mu}A^{d\nu}).
\end{eqnarray}
\begin{minipage}{15cm}
\parbox{7cm}{
{Ingoing quark:}
\\
\mbox{\epsfig{file=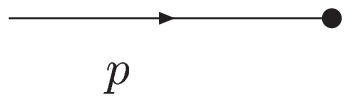,width=6cm}} }
\parbox{7cm}{
\begin{equation}
=u(p,s) \ {\rm ( for \ quark )}, \overline{v}(p,s)\  {\rm ( for\
antiquark )}\end{equation}}
\end{minipage}
\begin{minipage}{15cm}
\parbox{7cm}{
{Outgoing quark:}
\\
\mbox{\epsfig{file=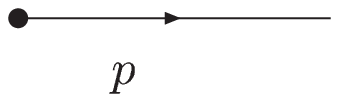,width=6cm}}}
\parbox{7cm}{
\begin{equation}
=\overline{u}(p,s)\  {\rm ( for \ quark )}, v(p,s) \ {\rm ( for\
antiquark )}\end{equation} }
\end{minipage}
\begin{minipage}{15cm}
\parbox{7cm}{
{Ingoing gluon:}
\\
\mbox{\epsfig{file=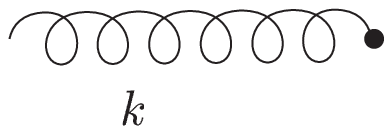,width=6cm}} }
\parbox{7cm}{
\begin{equation}
=\epsilon(k,\lambda)\end{equation}}
\end{minipage}
\begin{minipage}{15cm}
\parbox{7cm}{
{Outgoing gluon:}
\\
\mbox{\epsfig{file=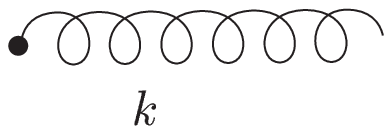,width=6cm}} }
\parbox{7cm}{
\begin{equation} \epsilon^\ast(k,\lambda)\end{equation}}
\end{minipage}
\begin{minipage}{15cm}
\parbox{7cm}{
Quark propagator:
\\
\mbox{\epsfig{file=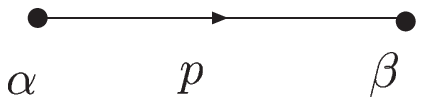,width=6cm}} }
\parbox{7cm}{
\begin{equation}
=i\frac{\gamma^\mu p_\mu-m}{p^2-m^2+i\epsilon}\end{equation}}
\end{minipage}
\begin{minipage}{15cm}
\parbox{7cm}{
{Gluon propagator:}
\\
\mbox{\epsfig{file=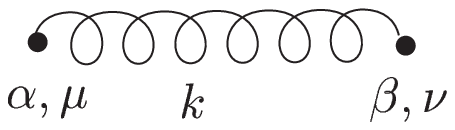,width=6cm}} }
\parbox{7cm}{
\begin{equation}\label{gluonprop} =-i\delta_{\alpha\beta}\left[\frac{g_{\mu\nu}+\eta
\frac{k_\mu k_\nu}{k^2}}{k^2+i\epsilon}\right]\end{equation}}
\end{minipage}
\begin{minipage}{15cm}
\parbox{7cm}{
Quark-Gluon vertex:
\\
\mbox{\epsfig{file=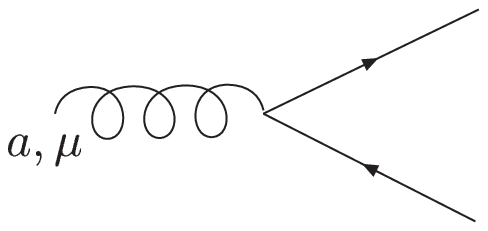,width=6cm}} }
\parbox{7cm}{
\begin{equation}
=-ig_S \gamma^\mu t^a_{ij}\end{equation}}
\end{minipage}
\begin{minipage}{15cm}
\parbox{7cm}{
3-gluon vertex:
\\
\mbox{\epsfig{file=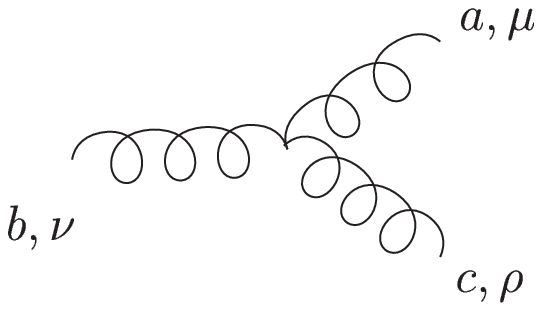,width=6cm}} }
\parbox{7cm}{
\begin{eqnarray}
=&-&g_S f^{abc}\left[ g^{\mu\nu}(k-p)^\rho+g^{\nu\rho}(p-q)^\mu
+\right. \nonumber\\
 &+& g^{\rho\mu}(q-k)^\nu ]
\end{eqnarray}
}
\end{minipage}
\begin{minipage}{15cm}
\parbox{7cm}{
{4-gluon vertex:}
\\
\mbox{\epsfig{file=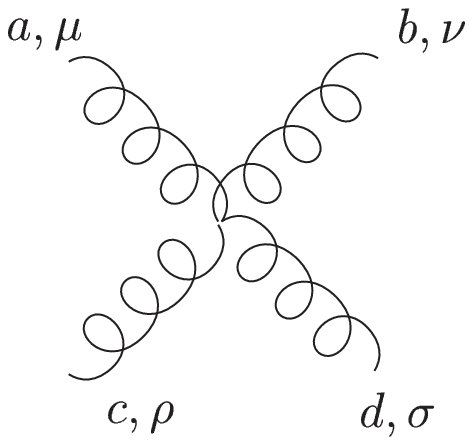,width=6cm}} }
\parbox{7cm}{
\begin{eqnarray}
=&-&ig_S [
f^{abe}f^{cde}(g^{\mu\rho}g^{\nu\sigma}-g^{\mu\sigma}g^{\nu\rho})\nonumber\\
&+&f^{ace}f^{bde}(g^{\mu\nu}g^{\rho\sigma}-g^{\mu\sigma}g^{\nu\rho})\nonumber\\
&+&f^{ade}f^{bce}(g^{\mu\nu}g^{\rho\sigma}-g^{\mu\rho}
g^{\nu\sigma}]\nonumber\\
\end{eqnarray}}
\end{minipage}
Ghost vertices and propagator are not important for our work and
will not be recalled.
\\
The factor $\eta$ in the gluon propagator selects the gauge: all
the calculation have been performed in the Feynman gauge which
corresponds to $\eta=0$.
\\
Sum over polarizations:
\begin{equation}
\sum_s u(p,s)\overline{u}(p,s)=\gamma^\mu p_\mu + m,
\end{equation}
\begin{equation}
\sum_s v(p,s)\overline{v}(p,s)=\gamma^\mu p_\mu - m.
\end{equation}
In the Feynman gauge the sum over the polarizations of external
gluons allows the following replacement
\begin{equation}
\sum_\lambda
\epsilon^\ast_\mu(\lambda)\epsilon_\nu(\lambda)\rightarrow
-g_{\mu\nu}.
\end{equation}
In the calculation of matrix elements in dimensional
regularization traces of gamma matrices arise. The following
properties have been used (n = dimension of the space-time):
\begin{eqnarray}
\{ \gamma^\mu,\gamma^\nu \}&=& 2g^{\mu\nu}\nonumber\\
\gamma^\mu \gamma_\mu &=& n \nonumber\\
{\rm Tr}(\gamma^\mu \gamma^\nu)&=&4g^{\mu\nu}\nonumber\\
\gamma^\mu \gamma^\nu \gamma_\mu &=& (2-n) \gamma^\nu\nonumber\\
\gamma^\mu \gamma^\nu \gamma^\rho
\gamma_\mu&=&4g^{\nu\rho}+(n-4)\gamma^\nu\gamma^\rho.
\end{eqnarray}
In $n$ dimension one can consistently introduce a matrix
$\gamma_5$ in a scheme where the property
\begin{equation}
\{\gamma_g,\gamma_\mu\}=0
\end{equation}
holds. $\gamma_5$ in 4 dimension reads:
\begin{equation}
\gamma_5=i\gamma_0 \gamma_1 \gamma_2 \gamma_3.
\end{equation}
Traces of an even number of gamma matrices and a $\gamma_5$
vanish:
\begin{equation}
{\rm Tr} [\gamma_5 \gamma_{\mu_1}\dots\gamma_{\mu_{2n}}]=0.
\end{equation}
The metric chosen to perform the calculations is
\begin{equation}
g_{\mu\nu}=\left(
\begin{array}{cccc}
1 & 0 & 0 & 0\\
0 & -1 & 0 & 0 \\
0 & 0 & -1 & 0 \\
0 & 0 & 0 & -1
\end{array}
\right).
\end{equation}

\chapter{Phase Space in Dimensional
Regularization}\label{phasespace} The formula used to calculate
the decay rate of a particle in its rest frame is
\begin{equation}
d\Gamma=\frac{1}{2m} \ d\Phi_n \ \delta(\sum_i p_i^\mu) \
|\overline{M}|^2.
\end{equation}
$m$ is the mass of the particle, $\Phi_n$ is the $n$-body phase
space, $\delta(\sum_i p_i^\mu)$ assures the momentum conservation
and $|\overline{M}|^2$ is the squared matrix element of the
process.
\\
For the distribution of the transverse momentum in $b\rightarrow
s\gamma$, $|\overline{M}|^2$ has been calculated in Chapter
\ref{fixedorder} , in this appendix the results for the 2-body and
3-body phase space in dimensional regularization are shown: these
terms appears in the calculations of the Born amplitude and in
radiative corrections.
\\
The lowest order process is $b\rightarrow s\gamma$.
\\
The general expression for the phase space is
\begin{equation}\label{phi2}
d\Phi_2=\frac{d^n p}{(2\pi)^n}\frac{d^n
q}{(2\pi)^n}(2\pi)\delta_+(p^2)(2\pi)\delta_+(q^2)(2\pi)^n\delta(P^\mu-p^\mu-q^\mu),
\end{equation}
where $p^\mu$ is the strange quark momentum, $q^\mu$ the photon
momentum and $P^\mu$ the $beauty$ quark momentum and
\begin{equation}
\delta_+(p^2)=\delta(p^2)\theta(p^0)
\end{equation}
$n$ is the dimension of the space-time: for the regularization of
infrared divergent digrams $n=4+\epsilon$.
\\
(\ref{phi2}) can be integrated over all the variable on which the
matrix element do not depend explicitely: for example the matrix
element for $b\rightarrow s\gamma$ can be rearranged to depend
only on the photon energy $E_\gamma$ and all the other variables
can be integrated out.
\\
At first the integration over the strange quark momentum can be
performed, so that (\ref{phi2}) turns out to be
\begin{equation}\label{phi2bis}
d\Phi_2=\frac{d^n
q}{(2\pi)^{n-2}}\delta_+\left[(P-q)^2\right]\delta_+(q^2).
\end{equation}
Let us recall that
\begin{equation}\label{prop1}
\int d^n q \ \delta_+(q^2)=\int \ d^n q \
\delta_+(q^2)=\frac{d^{n-1} q}{(2E_\gamma)}.
\end{equation}
and
\begin{equation}
\frac{d^{n-1}
q}{(2E_\gamma)}=\frac{1}{2}E_\gamma^{n-3}\sin^{n-3}\theta_1
\sin^{n-4}\theta_2 \dots \sin\theta_{n-2} d\theta_1 d\theta_2
\dots d\theta_{n-2} dE_\gamma.
\end{equation}
$\theta_i$ are the angles with respect of the axes in the $n-1$
dimensional space, for $n=4$ the formulae reduce to the well known
one in physical phase space. Since the matrix element does not
depend on the angles $\theta_i$ they can be integrated, using
\begin{equation}
\int_0^\pi \sin^n \theta d\theta=\sqrt{\pi}\
\frac{\Gamma(n/2+1/2)}{\Gamma(n/2+1)}.
\end{equation}
Using the properties of the gamma function in appendix
(\ref{gammadilog}), finally
\begin{equation}\label{prop2}
\int \ \frac{d^{n-1}q}{2E_\gamma}=2^{n-3}\pi^{(n-2)/2}\
\frac{\Gamma(n/2-1)}{\Gamma(n-2)}E_\gamma^{n-3}dE_\gamma.
\end{equation}
Substituting (\ref{prop1}) and (\ref{prop2}) in (\ref{phi2bis})
finally one gets
\begin{equation}
d\Phi_2=\frac{1}{2^{n-1}\pi^{n/2-1}}\frac{\Gamma(n/2-1)}{\Gamma(n-2)}y
\ m^{n-4}\delta(1-y),
\end{equation}
where the adimensional variable $y=2E_\gamma/m$ has been defined
and the kinematical property $(P-q)^2=m^2 (1-y)$ has been used.
\\
\\
The 3-body phase space can be calculated in a similar way. It
appears in the calculation of QCD corrections and corresponds to
the process $b\rightarrow s\gamma g$.
\\
In general it takes the look
\begin{equation}\label{phi3}
d\Phi_2=\frac{d^n p}{(2\pi)^n}\frac{d^n q}{(2\pi)^n}\frac{d^n
k}{(2\pi)^n}(2\pi)\delta_+(p^2)(2\pi)\delta_+(q^2)(2\pi)\delta_+(k^2)(2\pi)^n\delta(P^\mu-p^\mu-q^\mu-k^\mu),
\end{equation}
where $k^\mu$ is the gluon momentum.
\\
The matrix element turns out to depend only on three variables:
for example the gluon energy, the photon energy and the angle
between the photon and the gluon and all the other variables can
be integrated out.
\\
After the integration over the strange quark momentum $p^\mu$, by
using the momentum conservation and  (\ref{prop1}), (\ref{phi3})
reads
\begin{equation}\label{phi3bis}
d\Phi_3=\frac{d^{n-1}q}{(2\pi)^{n-1}2E_\gamma}\frac{d^{n-1}k}{(2\pi)^{n-1}2E_g}\delta\left[m^2-2mE_\gamma
-2m E_g + 2 E_\gamma E_g (1-cos\ \theta)\right],
\end{equation}
being $\theta$ the angle between the gluon and the photon.
\\
Moreover
\begin{equation}\label{prop3}
\int\int\frac{d^{n-1}q}{(2\pi)^{n-1}2E_\gamma}\frac{d^{n-1}k}{(2\pi)^{n-1}2E_g}
= \frac{2^{n-3}\pi^{n-2}}{\Gamma(n-2)} E_\gamma^{n-3}dE_\gamma
E_g^{n-3} dE_g dt t^{n/2-2} (1-t)^{n/2-2} dE_g dE_\gamma dt,
\end{equation}
where the angular variable $t$ is defined as $t=(1+\cos\
\theta)/2$.
\\
Substituting (\ref{prop3}) in (\ref{phi3bis}) and defining the
adimensional energy fractions
\begin{equation}
y=\frac{2E_\gamma}{m} \ \ \ , \ \ \ z=\frac{2E_g}{m},
\end{equation}
the final result for the 3-body phase space turns out to be
\begin{equation}
d\Phi_3=\frac{m^{2n-7}}{2^{2n}\pi^{n-1}\Gamma(n-2)}\ y^{n-3} \
z^{n-3} \ t^{(n-4)/2} \ (1-t)^{(n-4)/2}\ \delta\left[1-y-z+y z
(1-t)\right]\ dz \ dy \ dt.
\end{equation}

\chapter{Loop Integrals in Dimensional
Regularization}\label{feynpar} A standard technique for the
calculation of loop integrals consists in the introduction of
Feynman parameters. This technique has been used to calculate the
self energy corrections in virtual diagrams.
\\
It lies on the following identity:
\begin{equation}
\frac{1}{a_1 a_2 \dots a_n}=\Gamma(n)\int_0^1 \ dx_1 \ \int_0^1 \
dx_2 \ \int_0^1 \ dx_n \ \frac{\delta(1-\sum_1^n
x_n)}{\left[\sum_i a_i x_i\right]^n}.
\end{equation}
\\
In Feynman gauge at one loop it can be used to parametrize the
denominators:
\begin{eqnarray}
\frac{1}{abc}&=&2 \int_0^1 \ dx \int_0^1 \ dy \
\frac{1}{\left[ax+by+c(1-x-y)\right]^3}\nonumber\\
\frac{1}{ab}&=&\int_0^1 \ dx \frac{1}{\left[ax+b(1-x)\right]^2}.
\end{eqnarray}
In Landau gauge other denominators can appear, such as:
\begin{eqnarray}
\frac{1}{a^2b}&=& 2 \int_0^1 \ dx
\frac{x}{\left[ax+b(1-x)\right]^3}\nonumber\\
\frac{1}{a^3b}&=&6\int_0^1 \ dx
\frac{x^2}{\left[ax+b(1-x)\right]^4}\nonumber\\
\frac{1}{a^2bc}&=& 12 \int_0^1 \ dx \int_0^1 \ dy \frac{x^2
y}{\left[a x y+bx(1-y)+c(1-x)\right]^4}.
\end{eqnarray}
Let us recall that Feynman gauge corresponds to choose $\eta=0$ in
(\ref{gluonprop}) and Landau gauge to $\eta=1$.
\\
These parametrization are useful to calculate loop integrals in
every regularization: in particular in dimensional regularization,
after the introduction of Feynman parameters, the integral over
the loop momentum can be performed using
\begin{equation}\label{intn}
\int\frac{d^n k}{(2\pi)^n} \
\frac{1}{\left[k^2-C\right]^\alpha}=i\frac{(-1)^\alpha}{(16\pi^2)^{n/4}}
\frac{\Gamma(\alpha-n/2)}{\Gamma(\alpha)}\frac{1}{C^{\alpha-n/2}}
\end{equation}
where $n$ is the dimension of the space-time and $k^\mu$ the loop
momentum to integrate. $C$ is a constant function of the external
momenta in the loop: usually the denominator does not appear as in
(\ref{intn}), but with scalar product of $k^\mu$ with external
momenta $(k^2+2k\cdot q+\lambda)$. Such linear terms can be
eliminated by adding and subtracting terms which make the
denominator a square and by redefining the integration momentum
with a translation:
\begin{equation}\label{ridden}
(k^2+2k\cdot q+\lambda)=(k^2+2k\cdot q
+\lambda+q^2-q^2)=(\left[k+q\right]^2-\left[q^2-\lambda\right])\rightarrow
(k^\prime-C).
\end{equation}
Another kind of integral appearing in ultraviolet divergent
integrals is:
\begin{equation}
\int\frac{d^n k}{(2\pi)^n} \
\frac{k^2}{\left[k^2-C\right]^\alpha}=i\frac{(-1)^\alpha}{(16\pi^2)^{n/4}}
\frac{n}{2}\frac{\Gamma(\alpha-1-n/2)}{\Gamma(\alpha)}\frac{1}{C^{\alpha-n/2-1}}.
\end{equation}
Integrals with linear terms in $k^\mu$ in the numerator vanish
once the reduction in (\ref{ridden}) is performed.

\chapter{Gamma Function and Dilogarithms}\label{gammadilog}
\begin{figure}[h]
\begin{minipage}{15cm}
\parbox{7cm}{
\mbox{\epsfig{file=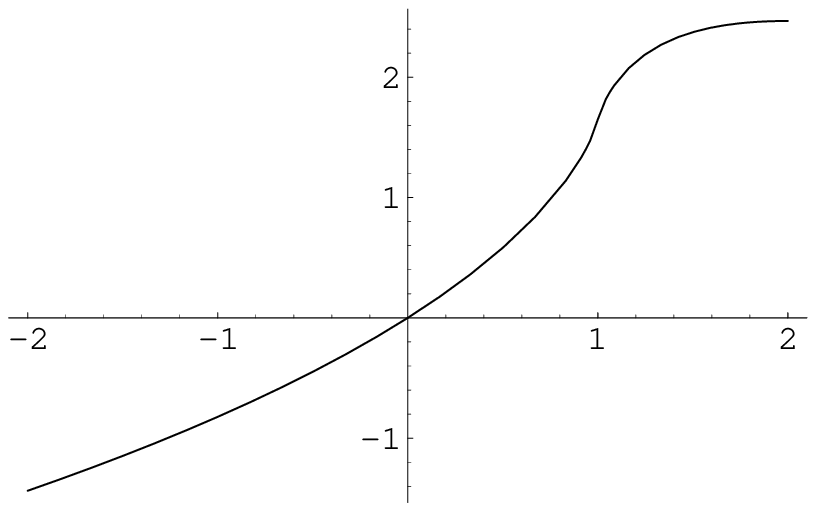, width=6.5cm}}\caption{Real part
of the dilogarithms.\label{polylogre}}}
\parbox{7cm}{
\mbox{\epsfig{file=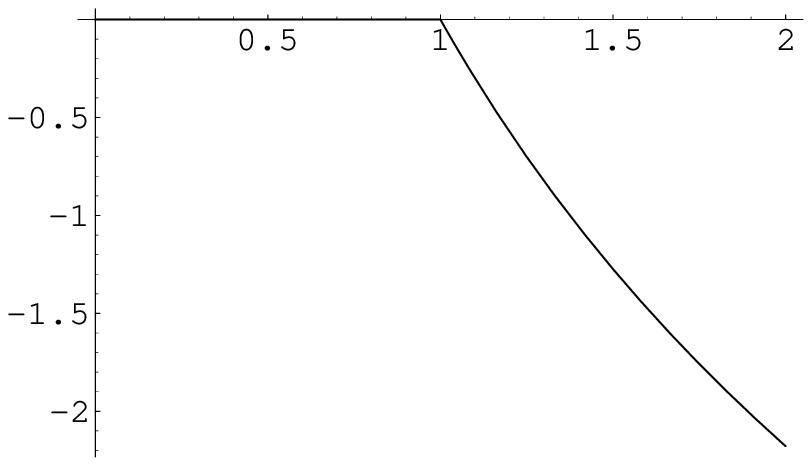,
width=6.5cm}}\caption{\label{polylogim}Imaginary part of the
dilogarithm.}}
\end{minipage}
\end{figure}

In the analytical evaluation of processes involving QCD
corrections a wide range of functions appears: multi-loop
calculations are very difficult to perform in analytical form and
up to now people try to face complete calculations at two loops.
\\
For one loop calculations the range of function which can be
encountered is well known: apart regular and elementary functions
(powers of a variable basically), trascendental functions may
appears, above all in semi-inclusive calculations.
\\
Beside logarithms, which are widely treated during the thesis and
are relevant in the endpoint region of the spectrum, other special
functions appear: in particular $\Gamma$ function and
dilogarithms.
\\
The $\Gamma$ function basically is involved in the measure of
integrals in dimensional regularization and is defined as:
\begin{equation}
\Gamma(x)=\int_0^1 \ e^{-y} \ y^{x-1} \ dy, \ \ \ {\rm x>0}.
\end{equation}
It provides a generalization of the factorial for non integer
number, since
\begin{equation}
\Gamma(n+1)=n\Gamma(n)=n!
\end{equation}
which can be extended for real positive numbers as
\begin{equation}\label{fattoriale}
\Gamma(x+1)=x\Gamma(x).
\end{equation}
In dimensional regularization $\Gamma$ functions arise from one
loop integrations and may give rise to poles in the regulator
$\epsilon$.
\\
For example it is usual to perform the expansion
\begin{equation}
\Gamma(\epsilon)=\frac{1}{\epsilon}-\gamma+\frac{1}{2}\left(\gamma^2+\frac{\pi^2}{6}\right)\epsilon+O(\epsilon^2),
\end{equation}
valid for $\epsilon \rightarrow 0$, in case after the application
of (\ref{fattoriale}).
\\
$\gamma$ is the \it Euler constant \rm, which satisfies the
property
\begin{equation}
\gamma=\lim_{n\rightarrow\infty} \left[-\log
n+1+\frac{1}{2}+\frac{1}{3}+\dots+\frac{1}{n}\right]=0.5772\dots
\end{equation}
and
\begin{equation}
\gamma=-\int_0^\infty \ e^{-x} \ \log x \ dx.
\end{equation}
Finally it turns out that
\begin{eqnarray}
\frac{d\Gamma(x)}{dx}&=&-\gamma\nonumber\\
\frac{d^2\Gamma(x)}{dx^2}&=&\gamma^2+\frac{\pi^2}{6}.
\end{eqnarray}
In one loop calculations (for example in virtual corrections)
often the following integral appears:
\begin{equation}
\int_0^1dx \ x^{\alpha-1} \ (1-x)^{\beta-1} =
\frac{\Gamma(\alpha)\Gamma(\beta)}{\Gamma(\alpha+\beta)}.
\end{equation}
A useful property of the $\Gamma$ function is
\begin{equation}
\Gamma(x)\Gamma(1-x)=\frac{\pi}{\sin \pi x},
\end{equation}
from which it follows
\begin{eqnarray}
\Gamma(1)&=&1\nonumber\\
\Gamma(\frac{1}{2})&=&\sqrt{\pi}.
\end{eqnarray}
Finally let us notice that the $\Gamma$ function can be extended
for $x<0$, but in this region singularities appears for $x \in
Z^-$, that is for negative integer numbers (and for $x=0$).
\\
\\
While the $\Gamma$ function appear in the measure of integrals in
dimensional regularization, other special functions appears in the
final analytical results: they are dilogarithms.
\\
The dilogarithm is defined as:
\begin{equation}
{\rm Li_2}(x)=\sum_{n=1}^\infty\frac{x^2}{n^2} \ \ \ {\rm for}\
|x|<1.
\end{equation}
and has a branch cut discontinuity in the complex plane from 1 to
$\infty$.
\\
Equivalent definitions can be stated in integral form:
\begin{eqnarray}
{\rm Li_2}(x)&=&-\int_0^x \frac{\log(1-t)}{t}dt\nonumber\\
{\rm Li_2}(x)&=&-\int_0^1 \frac{\log(1-xt)}{t}dt\nonumber\\
{\rm Li_2}(x)&=&-\int_{1-x}^1\frac{\log t}{1-t}dt\nonumber\\
{\rm Li_2}(x)&=&\int_0^1 \frac{\log t}{t-1/x}dt.
\end{eqnarray}
In figure (\ref{polylogre}) and (\ref{polylogim}) the real and
imaginary part of the dilogarithm are depicted.
The dilogarithm has the properties:
\begin{eqnarray}
{\rm Li_2}(0)&=&0\nonumber\\
{\rm Li_2}(1)&=&\frac{\pi^2}{6}\nonumber\\
{\rm Li_2}(-1)&=&-\frac{\pi^2}{12}\nonumber\\
{\rm Li_2}(1/2)&=&\frac{\pi^2}{12}-\frac{1}{2}\log^2 2.
\end{eqnarray}
Other relevant properties are:
\begin{eqnarray}
{\rm Li_2}(-x)+{\rm Li_2}(-1/x)&=&
-\frac{\pi^2}{6}-\frac{1}{2}\log^2 x \ \ \ {x>0}\nonumber\\
{\rm Li_2}(x)+{\rm Li_2}(1/x)&=&\frac{\pi^2}{3}-\frac{1}{2}\log^2
x-i\pi\log x \ \ \ x>1.
\end{eqnarray}
Sometimes ${\rm Li_2}(1-x)$ is referred as \it Spence \rm
function.
\\
Dilogarithms are a special case of the class of polylogarithms,
defined as:
\begin{equation}
{\rm
Li_n}(x)=\sum_{k=1}^{\infty}\frac{x^k}{k!}=\frac{(-1)^{n-1}}{(n-1)!}\int_0^1
\ dt \frac{\log^{n-1} t}{t}.
\end{equation}

\chapter{Harmonic Polylogarithms}\label{apphpl}
The occurrence of dilogarithms in one loop calculation in QED and
QCD is well known since a long time. It is also well known that
two loops calculations require the introduction of the Nielsen's
generalization of dilogarithms, also known as polylogarithms.
\\
Nielsen's polylogarithms are defined as
\begin{equation}
S_{n,p}(z)=\frac{(-1)^{n+p-1}}{(n-1)!p!}\ \int_0^1 \ \frac{(\log \
t)^{n-1} \ \log^p(1-z t)}{t} \ dt
\end{equation}
It is known that, while the class of Nielsen's polylogarithms is
sufficient for a complete analytical calculation of two loops
amplitudes, it is not when the number of loops becomes higher or
several different scales are involved in the calculations.
\\
In these case a generalization of Nielsen's polylogarithms which
allows a complete analytical calculations is provided by the class
of harmonic polylogarithms (hpl's) introduced in \cite{remiddi},
by Remiddi and Vermaseren.
\\
Following the formalism in \cite{remiddi}, the harmonic
polylogarithms are labelled by a set of arguments contained in the
$w$-dimensional vector $\vec{m}_w$ and are indicated as
$H(\vec{m}_w;x)$.
\\
For example for $w=1$ they read
\begin{eqnarray}
H(0;x)&=& \log \ x,\nonumber\\
H(1;x)&=& \int_0^x \ \frac{dx^\prime}{1-x^\prime} \ =-\log
(1-x),\nonumber\\
H(-1;x)&=& \int_0^x \ \frac{dx^\prime}{1+x^\prime} \ =\log (1+x).
\end{eqnarray}
Their derivatives are:
\begin{equation}
\frac{d}{dx}H(a;x)=f(a;x),
\end{equation}
where $a=0,1,-1$ and the functions $f(a;x)$ are
\begin{eqnarray}\label{basef}
f(0;x)&=&\frac{1}{x}\nonumber\\
f(1;x)&=&\frac{1}{1-x}\nonumber\\
f(-1;x)&=&\frac{1}{1+x}.
\end{eqnarray}
One could start by defining the functions in (\ref{basef}) as the
basis to build harmonic polylogarithms of weight 1 or higher.
\\
In fact higher weight hpl's are defined as
\begin{equation}\label{defhpl}
H(\vec{m}_w;x)=\int_0^x \ dx^\prime \
f(a;x^\prime)H(\vec{m}_{w-1};x^\prime),
\end{equation}
where
\begin{equation}
\vec{m}_w=(a,\vec{m}_{w-1}),
\end{equation}
being $a$ the leftmost index of $\vec{m}_{w}$. The definition in
(\ref{defhpl}) is valid for $\vec{m}_w\not= \vec{0}$, that is if
almost one index is different from zero.
\\
For $\vec{m}_w=\vec{0}_w$
\begin{equation}\label{logpolylog}
H(\vec{0}_w;x)=\frac{1}{w!} \ \log^w \ x.
\end{equation}
In compact form the derivative of a general harmonic polylogarithm
of weight $w$ is
\begin{equation}
\frac{d}{dx}H(\vec{m}_w;x)=f(a;x)H(\vec{m}_{w-1};x).
\end{equation}
For $w=2$ the harmonic polylogarithms can be expressed in terms of
well known functions as logarithms and dilogarithms:
\begin{eqnarray}
H(0,0;x)&=&\frac{1}{2}\log^2 x,\nonumber\\
H(0,1;x)&=&Li_2(x),\nonumber\\
H(0,-1;x)&=&-Li_2(-x),\nonumber\\
H(1,0;x)&=&-\log \ x \log(1-x)+Li_2(x),\nonumber\\
H(1,1;x)&=&\frac{1}{2}\log^2(1-x),\nonumber\\
H(1,-1;x)&=&Li_2(\frac{1-x}{2})-\log 2
\log(1-x)-Li_2\left(\frac{1}{2}\right),\nonumber\\
H(-1,0;x)&=&\log x \log(1-x)+Li_2(-x),\nonumber\\
H(-1,1;x)&=&Li_2\left(\frac{1+x}{2}\right)-\log
2\log(1+x)-Li_2\left(\frac{1}{2}\right),\nonumber\\
H(-1,-1;x)&=&\frac{1}{2}\log^2(1+x).
\end{eqnarray}
For polylogarithms of weight $w=3$ the same thing happens, that is
they can be all expressed in terms of logarithms, dilogarithms and
Nielsen's polylogarithms of weight 3. For higher weights this is
no longer true: the class of harmonic polylogarithms in this case
is a wider class with respect to Nielsen's polylogarithms.
\\
The behaviour at the border of the integration domains is the
following:
\begin{itemize}
\item
if $\vec{m}_w=0$, trivially $H(\vec{0}_w;x)$ has a logarithmic
divergence for $x\rightarrow 0$ (see (\ref{logpolylog}));
\item
if $\vec{m}_w\not=0$, $H(\vec{m}_w;0)=0$ for $x=0$;
\item
if $m_w\not=1$, $H(\vec{m}_w;1)$ is finite;
\item
if $m_w=1$ and $\vec{m}_{w-1}=\vec{0}_{w-1} $, $H(\vec{m}_w;1)$ is
finite;
\item
if $m_w=1$ and $\vec{m}_{w-1}\not=\vec{0}_{w-1}$, $H(\vec{m}_w;1)$
has a logarithmic divergence in $x=1$, and the term with the
highest divergence is $\log^p(1-x)$, where $p$ is the number of
the leftmost indices equal to 1.
\end{itemize}

C.Greub, T.Hurth, \it Two-loop Matching of the Dipole Operators
for $b \to s \gamma$ and $b \to  s g$, \rm Phys. Rev. D56, 2934
(1997);

\bibitem{ali2}A.Ali, C.Greub, \it Photon Energy Spectrum in $B\rightarrow X_s\gamma$
and Comparison with Data\rm, Phys. Lett. B361 C49, 431 (1991)

\bibitem{pott} N.Pott, \it Brehmsstrahlung Corrections to the
Decay $b\rightarrow s\gamma$, \rm  Phys. Rev. D54, 938 (1996)

\bibitem{greub} C.Greub, T.Hurth, D.Wyler, \it Virtual Corrections
to the Decay $b\rightarrow s\gamma$, \rm Phys. Lett. B380, 385
(1996)

\bibitem{greub2} C.Greub, T.Hurth, D.Wyler, \it Virtual $O(\alpha_S)$ Corrections
to the Inclusive Decay $b\rightarrow s\gamma$, \rm Phys. Rev. D54,
3350 (1996)

\bibitem{ali} A.Ali, C.Greub, \it Inclusive Photon Energy Spectrum
in Rare B Decays\rm, Z. Phys. C49, 431 (1991)

\bibitem{cleo} CLEO Collaboration (S.Chen et al.), \it Branching
Fraction and Photon Energy Spectrum for $b\rightarrow s\gamma$\rm,
Phys. Rev. Lett. 87, 251807 (2001)

\bibitem{altpet} G.Altarelli, S.Petrarca, \it Inclusive beauty decays and the spectator model, \rm Phys. Lett. B261, 303 (1991)

\bibitem{generale}  I.Bigi, M.Shifman, N.Uraltsev, A.Vainshtein, \it QCD predictions for lepton spectra in inclusive
heavy flavor decays, \rm Phys.Rev.Lett. 71, 496 (1993)
\\
\\
I.Bigi, M.Shifman, N.Uraltsev, A.Vainshtein, \it On the motion of
heavy quarks inside hadrons: Universal distributions and inclusive
decays, \rm Int. J. Mod. Phys. A 9, 2467 (1994)
\\
\\
A.Manohar, M.Wise, \it Inclusive Semileptonic B and polarized
Lambda(b) decays from QCD, \rm Phys.Rev. D49, 1310 (1994)
\\
\\
M.Neubert, \it QCD based interpretation of the lepton spectrum in
inclusive anti-B $\to$ X(u) lepton anti-neutrino decays, \rm
Phys.Rev. D49, 3392 and \it Analysis of the photon spectrum in
inclusive B $\to$ X(s) gamma decays, \rm 4623 (1994)
\\
\\
T.Mannel, M.Neubert, \it Resummation of non-perturbative
corrections to the lepton spectrum in inclusive B $\to$ X lepton
anti-neutrino decays, \rm Phys.Rev. D50, 2037 (1994)

\bibitem{ugo}
U.Aglietti, \it Resummed coefficient function for the shape
function, \rm arXiv:hep-ph/0102138
\\
\\
U.Aglietti, \it Next-to-leading resummed coefficient function for
the shape function, \rm  Phys.Lett. B515, 308 (2001)

\bibitem{hqet}
For a detailed description of the Heavy Quark Effective Theory see
M.Neubert, \it Heavy quark symmetry, \rm Phys.Rept. 245, 259
(1994) and references thereinn

\bibitem{americani}
R.Akhoury, I.Rothstein, \it The Extraction of $V_{ub}$ from
Inclusive B Decays and the Resummation of End Point Logs, \rm
Phys.Rev. D54, 2349 (1996)

\bibitem{ks}
G.Korchemsky, G.Sterman, \it Infrared factorization in inclusive B
meson decays, \rm Phys.Lett. B340, 96 (1994)

\bibitem{montp}  U.Aglietti, \it The shape function in field theory, \rm Nucl.Phys.B Proc.Suppl. 96, 453 (2001).

\bibitem{lep3proc}  U.Aglietti, \it A new sum rule to determine $|$V(ub)$|$/$|$V(cb$|$ \rm, talk given at the
LEP3 conference, Rome 18-20 April 2001, Published in *Rome 2001,
LEP physics* 217, hep-ph/0105168.

\bibitem{mangano}
S.Catani, M.Mangano, P.Nason, L.Trentadue, \it The Resummation of
Soft Gluon in Hadronic Collisions, \rm Nucl.Phys. B478, 273 (1996)

\bibitem{nostri}  G.Altarelli, N.Cabibbo, G.Corb\'{o}, L.Maiani, G.Martinelli, \it Leptonic Decay Of Heavy Flavors: A Theoretical Update,
\rm Nucl.Phys. B208, 365 (1982)

\bibitem{congiulia2} U.Aglietti, G.Ricciardi, \it The structure function of semi-inclusive heavy flavor decays in field
theory, \rm Nucl.Phys. B587, 363 (2000)

\bibitem{UC}  U.Aglietti, G.Corbo, \it Factorization and effective theories, \rm Phys.Lett. B431, 166 (1998)
and Int. J. Mod. Phys. A 15, 363 (2000)
\\
\\
U.Aglietti, G.Corbo, \it Factorization in exclusive and
semi-inclusive decays and effective  theories for massless
particles, \rm Int.J.Mod.Phys. A15, 363 (2000)

\bibitem{separati}  C.Bauer, S.Fleming, M.Luke, \it Summing Sudakov Logarithms in B $\to$ X/s gamma in Effective
Field Theory, \rm Phys.Rev. D63, 014006 (2001)
\\
\\
C. Bauer, S. Fleming, D. Pirjol and W. Stewart, \it An Effective
Field Theory for Collinear and Soft Gluons: Heavy to Light Decays,
\rm Phys.Rev. D63, 114020 (2001)

\bibitem{shapelattice} U.Aglietti, M.Ciuchini, G.Corb\'o,
E.Franco, G.Martinelli, L.Silvestrini, \it Model Independent
Determination of the Shape Function for Inclusive $B$ Decays and
of the Structure Function in DIS, \rm Phys.Lett. B432, 411 (1998)

\bibitem{remiddi} E.Remiddi, J.A.M.Vermaseren, \it Harmonic
Polylogarithms, \rm Int.J.Mod.Phys.A15, 725 (2000)

\bibitem{chettak} K.G.Chetyrkin, F.V.Tkachov, \it Integration By Parts: The Algorithm To Calculate Beta Functions In 4
Loops, \rm Nucl.Phys. B192, 159 (1981)

\bibitem{acg}
U.Aglietti, M.Ciuchini, P.Gambino, \it A new model-independent way
of extracting $|$V(ub)/V(cb)$|$, \rm Nucl.Phys. B637, 427 (2002)

\bibitem{ugoscala} U.Aglietti, \it Resummed $B\rightarrow X_u$ Lepton Neutrino Decay
Distributions to Next-to-Leading Order, \rm Nucl.Phys.B610,293
(2001) [HEP-PH 0104020]

\end{document}